\documentclass[preprint,notoc]{JHEP3}
\usepackage{amssymb,amsmath,url}
\usepackage{graphicx}
\usepackage{psfrag}
\renewcommand{\bottomfraction}{0.999}
\renewcommand{\topfraction}{0.999}
\renewcommand{\textfraction}{0.001}

\def\tl{\tilde l}

\def\tb{\tilde b}

\def\tst{\tilde t}
\def\ttau{\tilde \tau}
\def\tmu{\tilde \mu}
\def\tg{\tilde g}
\def\tw{\widetilde W}
\def\tz{\widetilde Z}
\def\tnu{\tilde\nu}

\def\tq{\tilde q}
\def\stac{$\ttau \tz $}
\def\stoc{$\tst \tz $}
\def\fp{HB/FP}

\newcommand{\beq}{\begin{equation}}
\newcommand{\eeq}{\end{equation}}
\newcommand{\bea}{\begin{eqnarray}}
\newcommand{\eea}{\end{eqnarray}}

\title{Interplay of Higgs and Sparticle Masses  in the CMSSM 
     with updated SUSY constraints }
\author{Alexander Belyaev$^a$,
Shahida Dar$^b$,Ilia Gogoladze$^b$, Azar Mustafayev$^c$ and Qaisar Shafi$^b$\\
$^a$School of Physics and Astronomy, University of Southampton,
Highfield, Southampton SO17 1BJ, UK; E-mail:~\email{a.belyaev@phys.soton.ac.uk}\\
$^b$Bartol Research Institute, Department of Physics \& Astronomy,
University of Delaware, Newark, DE 19716, USA;  
E-mail:~\email{dars@physics.udel.edu,ilia@physics.udel.edu,shafi@bartol.udel.edu}\\
$^c$Department of Physics and Astronomy, University
of Kansas, Lawrence, KS 66045, USA; E-mail:~\email{amustaf@ku.edu} }
\preprint{\vbox{BA-07-029}}

\abstract{ 
We estimate the bounds on Higgs and sparticle masses and discuss 
their correlations in the constrained minimal supersymmetric 
standard model (CMSSM). In our analysis we have applied the present 
constraints from collider and low energy experiments, as well as the 
experimental bound on cold dark matter from WMAP. For a given lightest
Higgs boson mass, which is expected to be measured with good precision at the LHC, 
we find important correlations between the Higgs and sparticle masses which allows one to 
delineate the MSSM model parameters and particle spectra. 
We have also  demonstrated an important complementarity between 
the LHC and direct dark matter detection experiments
emphasizing that by including the experimental
input both from collider physics and from dark matter detection experiments, 
one would significantly improve the measurement of the SUSY spectrum and the underlying
parameter space.
}

\begin{document}
\renewcommand{\bottomfraction}{0.999}
\renewcommand{\topfraction}{0.999}
\renewcommand{\textfraction}{0.001}

\section{Introduction}

In recent years the measured value of the top quark mass has 
gradually inched its way down to its current value of $170.9\pm 
1.8$~GeV~\cite{:2007bxa}, which is significantly below the central 
value of around 178 GeV used in various studies just a few years 
ago. This downward shift of the top quark mass has important 
implications for the particle spectrum of the minimal supersymmetric 
standard model (MSSM), in particular for the lightest CP-even Higgs 
boson mass, $m_h$, since the leading radiative corrections to $m_h$ 
are proportional to the square of the top mass, $m_t^2$, while  $m_h \geq 114.4$~GeV 
from LEP2 in the decoupling limit~\cite{Schael:2006cr}. At tree 
level $m_h \leq M_Z |\cos2\beta|$, and so significant radiative 
corrections are necessary to overcome the LEP2 bound. In the 
absence of large trilinear couplings, significant radiative 
corrections can be generated only if the stop masses are in the TeV 
range. Alternatively, significant stop mixing in the presence of 
large trilinear couplings can drive $m_h$ up without excessively 
heavy stops. In this paper, we investigate both possibilities and 
study the correlations of the Higgs and sparticle mass spectrum in 
the constrained MSSM 
(CMSSM)~\cite{Chamseddine:1982jx, Barbieri:1982eh, Hall:1983iz, Cremmer:1982vy, Ohta:1982wn} 
under the plausible assumption that the neutralino constitutes 
the cold dark matter in the universe.

The CMSSM has received a great deal of attention in recent years, especially 
for $m_t$ close to 178~GeV~\cite{Baer:2002gm, Baer:2003yh, 
Chattopadhyay:2003xi, Ellis:2003cw, Battaglia:2003ab, Arnowitt:2003vw, 
Ellis:2003si, Baer:2004xx, Gomez:2004eka, Ellis:2004tc, deBoer:1995fr}. 
When we started this analysis, the central value was $m_t=171.4\pm 2.1$~GeV
~\cite{Brubaker:2006xn}, which has recently drifted slightly lower 
to $170.9\pm 1.8$~GeV~\cite{:2007bxa}. Our results are based on $m_t=171.4$~GeV. 
Note that we have found agreement between our results with partially overlapping 
results of some recent papers which also use an updated value for the top-quark mass
~\cite{Ellis:2005tu, Djouadi:2006be, Ellis:2007ka, Roszkowski:2007fd}.

The CMSSM contains the following five fundamental parameters 
(we follow notations and conventions of Ref.~\cite{Baer:2006rs})
\beq 
\label{paras}
m_0, \ m_{1/2}, \ A_0,  \tan \beta, \ {\rm sign}(\mu)\,,
\eeq 
where $m_0$ is the universal soft supersymmetry breaking (SSB) 
scalar mass, $m_{1/2}$ the universal SSB gaugino mass, and $A_0$ the 
universal SSB trilinear scalar interaction (with the corresponding 
Yukawa coupling factored out). The values for these three 
parameters are prescribed at the GUT scale, $M_{{GUT}}\simeq 2 
\times 10^{16}$~GeV. $\tan\beta$ is the ratio of the vacuum 
expectation values (VEV) of the two Higgs doublets at the weak 
scale, and $\mu$ is the supersymmetric bilinear Higgs parameter 
whose magnitude, but not the sign, is determined by the radiative 
electroweak breaking conditions\footnote{Historically this framework 
is also called mSUGRA and is associated with supergravity. However, 
supergravity does not necessarily lead to high scale universality 
as originally thought. Occasionally, the abbreviation mSUGRA is used 
only for a subset of (\ref{paras}) where some additional relations coming from 
a particular form of K\"ahler potential are assumed (see e.g. \cite{Ellis:2003pz, Ellis:2004qe}). 
To avoid possible confusion, we decided to use the term CMSSM and refer interested 
readers to Ref.~\cite{Baer:2006rs,Drees:2004jm} for textbook discussions.}.

The results for CMSSM studies are usually  presented in the 
$(m_0,m_{1/2})$ plane. However, anticipating that the lightest MSSM 
Higgs will be found at the LHC and that its mass according to 
ATLAS~\cite{ATLAS_TDR} and CMS~\cite{CMS_TDR} technical design 
reports, will be determined to a precision which is better then 
$1\%$,  it is worthwhile to understand possible correlations between 
such a precisely measured Higgs mass and sparticles masses as well 
as the underlying  model parameters. Such correlations would allow 
one to delineate the underlying SUSY model and, moreover, to 
predict some model parameters and masses, such as the trilinear 
coupling, $A_t$, whose measurement at the 
LHC may be  problematic. The mass $m_h$ may be experimentally determined to 
an accuracy better than a GeV, which is better than the theoretical calculation of 
$m_h$ at present~\cite{Martin:2007pg}.
In this study, we 
assumed a theoretical uncertainty $\Delta m_h$ of 1~GeV in estimating the precision 
with which the sparticles masses and 
model parameters are predicted as a function of $m_h$, keeping in mind 
the optimistic hope that the theoretical uncertainty for the Higgs mass 
calculation will be improved by the time of the actual Higgs boson 
search at the LHC. In our  analysis we require that the magnitudes of 
$m_{1/2}$ and $A_0$ should not exceed 2~TeV, while $m_0$ is 
taken to be $\lesssim 5$~TeV. This choice covers the whole dark 
matter-motivated  CMSSM parameter space except the focus point 
region (see discussion below), which can go with $m_0$ all the way 
up to $M_{GUT}$, and turning the CMSSM scenario into a split-SUSY 
one~\cite{Giudice:2004tc}.

The sign of the MSSM parameter $\mu$ is taken to be positive, as dictated by 
the experimental measurements of the anomalous magnetic moment of 
the muon, $\Delta a_\mu$.  
We take into account the LEP2 constraints 
on the chargino mass, experimental data on  rare B decays, and the WMAP relic density constraints.

The paper is organized as follows. In Section~\ref{ch:constraints} we review details 
of the MSSM parameter space scan and implementation of the various 
constraints.  In Section~\ref{ch:results} we present our results on various 
correlations of the MSSM spectra which would be crucial for LHC 
collider phenomenology, and in Section~\ref{ch:conclusions} we summarize our conclusions.

\section{SUSY constraints and scanning procedure}
\label{ch:constraints}

For the MSSM spectrum calculation we use the IsaSUGRA program, 
which is a part of ISAJET~7.74 package~\cite{Baer:1999sp}. In this 
package, the weak scale values of gauge and third generation Yukawa 
couplings are evolved to $M_{GUT}$ via the MSSM 
renormalization group equations (RGEs) in the $\overline{DR}$ 
regularization scheme, where $M_{GUT}$ is defined to be the 
scale at which $g_1=g_2$. We do not enforce an exact unification of the 
strong coupling $g_3=g_1=g_2$ at $M_{GUT}$, since a few percent 
deviation from unification can be assigned to unknown GUT-scale 
threshold corrections~\cite{Hisano:1992jj,Yamada:1992kv}. At $M_{GUT}$, 
the universal soft supersymmetry breaking (SSB) boundary 
conditions are imposed and all SSB parameters along with the gauge 
and Yukawa couplings are evolved back to the weak scale $M_Z$. In 
the evaluation of  Yukawa couplings the SUSY threshold corrections~\cite{Pierce:1996zz} 
are taken into  account at the common scale $M_{SUSY}= \sqrt{m_{\tst_L}m_{\tst_R}}$, 
and the entire parameter set is iteratively run between $M_Z$ and $M_{GUT}$ using 
full 2-loop RGEs until a stable solution is obtained. To better 
account for leading-log corrections, one-loop step-beta functions 
are adopted for gauge and Yukawa couplings and the SSB parameters 
$m_i$ are extracted from RGEs at multiple scales $m_i=m_i(m_i)$.~
\footnote{Detailed discussion of SUSY threshold effects in ISAJET and other public codes can be 
found in ref.~\cite{Baer:2005pv}.} 
The RGE-improved 1-loop effective potential is minimized at an optimized 
scale $M_{SUSY}$, which effectively accounts for the leading 2-loop 
corrections. Full 1-loop radiative corrections are incorporated for all 
sparticle masses according to ref.~\cite{Pierce:1996zz}. The numerical results 
are in close accord with those generated by the FeynHiggsFast program~\cite{Heinemeyer:2000nz}.

The requirement of radiative electroweak symmetry breaking
(REWSB)~\cite{Ibanez:1982fr, Ibanez:1982ee, Ellis:1982wr, AlvarezGaume:1983gj} 
puts an important theoretical constraint on the parameter space. 
Another important constraint comes from limits on the cosmological 
abundance of stable charged particles~\cite{Yao:2006px} -- this 
excludes the parameter points where charged SUSY particles, such as 
$\ttau_1$ or $\tst_1$, become the lightest supersymmetric particle (LSP).

We also impose bounds on sparticle masses resulting from direct 
searches at colliders. The LEP2 search for pair production of 
sleptons~\cite{LEP2:slepton} puts the lower limit $m_{\tl} \geq 
99$~GeV provided the mass difference $m_{\tl}-m_{\tz_1}>10$~GeV. 
Another important LEP2 constraint comes from the  lightest 
chargino~\cite{LEP2:chargino},  $m_{\tw_1} \geq 103.5$~GeV.

Recent search results for squark/gluino production at the 
Tevatron~\cite{CDF:gluino,D0:gluino} put bounds on the gluino mass 
$m_{\tg} >370$~GeV for $m_{\tq}\simeq m_{\tg}$, and $m_{\tg} >220$~GeV for $m_{\tq}<
m_{\tg}$. In models with gaugino mass universality, like CMSSM, 
$m_{\tg} \sim 3.5 m_{\tw_1}$. Thus, the LEP2 chargino bound 
translates into a bound on the gluino mass, $m_{\tg} \geq 350$~GeV, 
to which Tevatron comes close only for  $m_{\tq}\simeq m_{\tg}$.

Tevatron searches for direct pair production of stops~\cite{CDF:stop,D0:stop} 
impose the  bound $m_{\tst_1} > 132$~GeV for $m_{\tz_1} \leq 48$~GeV. 
However, the requirements of GUT-scale gaugino mass universality 
tell us that $m_{\tz_1} \simeq 0.5m_{\tw_1} \geq 51$~GeV for points satisfying 
the LEP2 chargino bound. This chargino-neutralino mass ratio gets smaller 
in the HB/FP region at large $m_0$, but here the stops are always 
heavier than about 500~GeV,  as we will show later.

Similar searches for direct pair production by D0 collaboration~\cite{D0:sbottom} 
put a bound $m_{\tb_1} > 222$~GeV for $m_{\tz_1} \leq 90$~GeV; 
the CDF limit is slightly lower, $m_{\tb_1} > 193$~GeV for $m_{\tz_1} 
\leq 80$~GeV~\cite{CDF:stop}. In CMSSM, the lightest sbottom is 
dominantly left-handed whose mass is given by the approximate relation 
$m^2_{\tb_L} \sim 0.52 m_0^2+ 5 m_{1/2}^2$. The lightest chargino is 
mostly wino and obeys the approximate relation $m_{\tw_1} \sim \frac{2}{3} m_{1/2}$. 
Therefore, the LEP2 chargino bound leads to $m_{\tb_L} > 350$~GeV, 
significantly above the Tevatron limit.

The LEP2 Higgs bound~\cite{Schael:2006cr} further constrains the 
MSSM parameter space. After the LEP2 direct SUSY searches 
constraints are  applied, the pseudoscalar mass  $m_A$  is limited 
to be above 200~GeV (as we present later on), which ensures the 
decoupling regime and SM-like coupling of $h$ to gauge bosons. 
Therefore, the SM-Higgs LEP2 bound $m_h \geq 114.4$~GeV is applicable to the 
CMSSM scenario. 
This bound can be very different in the MSSM light Higgs 
boson scenario~\cite{Belyaev:2006rf}, when the LEP2 bound 
allows a higgs scalar as light as about $60$~GeV.
We would like to point out that 
in our  plots  we do not remove points with the Higgs boson mass below 
the SM-higgs LEP2 bound, but indicate its position whenever possible for the sake of clarity.

A combination of the recent WMAP data with other cosmological 
observations leads to tight constraints on the cold dark 
matter (CDM) relic abundance, and the most important and dramatic 
constraint on the CMSSM parameter space. The exact numerical value 
depends on a number of assumptions about the history of the early 
Universe~\cite{Drees:2005bx} as well as on a combination of data 
sets chosen. In our analysis we use the constraints arising  from a 
combination of WMAP and the Sloan Digital Sky Survey 
data~\cite{Spergel:2006hy}
\beq 
\Omega_{\rm CDM}h^2 = 0.111^{+0.011}_{-0.015} \qquad (2\sigma)\,, 
\label{eq:Oh2} 
\eeq
which assumes standard $\Lambda$CDM cosmology. We further assume that the 
bulk of CDM is composed of the lightest neutralino 
$\tz_1$ which was in thermal equilibrium in the early 
Universe~\footnote{Non-thermalized neutralinos were recently 
discussed in ref.~\cite{Drees:2006vh}}. To evaluate the neutralino 
relic density we employed the IsaReD code~\cite{Baer:2002fv} (part 
of IsaTools package), which uses several thousands $2\rightarrow 2$ 
annihilation and co-annihilation Feynman diagrams generated by the 
CompHEP package~\cite{Pukhov:1999gg,Boos:2004kh}.

An important constraint comes from the  measurement of the muon 
anomalous magnetic moment $a_\mu = \frac{(g-2)_\mu}{2}$ by the E821 
collaboration~\cite{Bennett:2006fi}. The current experimental value, 
$a^{exp}_{\mu}=(11659208.0 \pm 6.3)\times 10^{-10}$ deviates by 
$3.4\sigma$ from the SM prediction which uses hadronic vacuum 
polarization determined from $e^+e^-$ annihilation data, but there is no 
significant deviation from the SM predictions if the hadronic 
contribution  is calculated using $\tau$-decay 
data~\cite{Davier:2007ua, Hagiwara:2006jt, Hertzog:2007hz}. 
There is a stronger and stronger tendency at present to 
prefer  $e^+e^-$  data to evaluate the lowest-order hadronic 
contribution to $a_\mu$. The most recent analysis performed in 
ref.~\cite{Davier:2007ua}, based on $e^+e^-$ data, yields  a 
deviation  $\Delta a_{\mu}=(27.5 \pm 8.4)\times 10^{-10}$ from the 
experimental value. In view of the lack of consensus on the computation 
of the SM part~\cite{Davier:2007ua, Hagiwara:2006jt, Hertzog:2007hz}, 
we have decided not to apply the $(g-2)_\mu$ constraint but to indicate 
the $3\sigma$ allowed region for $(g-2)_\mu$,
\beq 
3.4 \times 10^{-10}\leq \Delta a_{\mu} \leq 55.6 \times 10^{-10}\,, 
\label{eq:g-2}
\eeq
with special colors to allow readers to make their own judgment 
on this constraint. This  $3\sigma$ interval uses $(g-2)_\mu$ 
predictions based on $e^+e^-$  data and was taken from the latest 
summary of $(g-2)_\mu$ status~\cite{Miller:2007kk}.

We used IsaAMU subroutine described in ref.~\cite{Baer:2001kn} to 
evaluate the SUSY contribution to $(g-2)_\mu$. The contribution of 
SUSY particles comes from $\tw_i \tnu_{\mu}$ and $\tz_i \tmu_{1,2}$ 
loops, where the total chargino contributions dominates in CMSSM. 
Therefore, $(g-2)_\mu$ puts an upper limit on the masses of 
charginos, neutralinos and second generation sleptons. For 
$\tan\beta$ not too small,  $\Delta a_{\mu}^{SUSY} \propto \mu M_2 
\tan\beta / \tilde{m}^4$, where $M_2$ is the $SU(2)$ gaugino mass and 
$\tilde{m}$ is the heaviest mass in the loop. 
Thus, it is particularly important for large $\tan\beta$ 
values and light sleptons and -inos ({\it i.e.}, small $m_0$ and $m_{1/2}$ 
region).  Note also that for models with a positive gaugino mass, 
like CMSSM, the sign of $\mu$ has to be positive.

The good agreement of the experimental value of the flavor changing 
$b\rightarrow s\gamma$ decay rate with the SM result imposes an 
important constraint on any beyond the SM physics. In MSSM, the most 
important SUSY contributions come from $t H^{\pm}$ and $\tst \, 
\tw^{\pm}$ loops. To evaluate the SM+MSSM branching fraction, we 
employed the IsaBSG subroutine~\cite{Baer:1996kv,Baer:1997jq}, which 
uses the effective theory approach. Namely, when any sparticle 
threshold is crossed, the sparticle is integrated out,  thereby inducing a 
new effective-operator basis. The resulting Wilson 
coefficients are evolved down to $Q=m_b$, thus summing large logarithms 
that can occur from the disparity between the scales at which 
various particles enter the loop corrections. In the CMSSM, the 
assumed universality of soft SUSY breaking (SSB) masses leads to a 
particularly simple framework known as minimal flavor violation 
(MFV) in which flavor mixing arises only from the CKM matrix. The 
experimental world average measurement for $Br(b \rightarrow s 
\gamma)$ is known to be $(3.55\pm 0.26) \times 10^{-4}$~\cite{Barberio:2006bi}, 
while its updated SM prediction has been recently shifted down to 
$(3.15\pm0.23)\times10^{-4}$ value~\cite{Misiak:2006bw,Misiak:2006ab}. 
Combining the experimental and theoretical errors in quadratures we 
apply the following constraints at $2\sigma$ level in our study:
\beq
2.85 \times 10^{-4} \leq Br(b \rightarrow s \gamma)
\leq 4.24 \times 10^{-4}.
\label{eq:bsg}
\eeq
This conservative approach has two advantages. Firstly, it 
allows one to accommodate the theoretical uncertainties that mainly 
come from the residual scale dependence and are $\sim 
12\%$~\cite{Gambino:2001ew}. Secondly, the SSB terms might have small 
off-diagonal entries at $M_{GUT}$ that will change $Br(b 
\rightarrow s \gamma)$ without significantly affecting other quantities.

Another B-physics constraint comes from the upper limit on $B_s \rightarrow \mu^+ \mu^-$ 
decay branching fraction,
\beq
 BF(B_s \rightarrow \mu^+ \mu^-)< 1.0 \times 10^{-7},
\label{eq:bmumu}
\eeq
as was reported by the CDF collaboration~\cite{bs2mumu}. We compute the 
SUSY contribution using IsaBMU code~\cite{Mizukoshi:2002gs} which 
assumes MFV. The SM predicts a very small value for this branching 
fraction, namely $BF_{SM}(B_s\rightarrow \mu^+ \mu^-) \simeq 3.4\times 
10^{-9}$, while the SUSY contribution behaves as $\tan ^6 \beta 
/m^4_A$ and hence is particularly important at large $\tan\beta$. 
However, it turns out that this constraint is always superseded in 
the CMSSM parameter space under study by the  one from $b \rightarrow s\gamma$.

To find bounds and correlations for the Higgs and sparticle masses, 
we performed random scans for the following range of the CMSSM parameters:
\begin{align}
0\leq m_0 \leq 5\, \textrm{TeV}, &\quad & 0\leq m_{1/2} \leq 2 \textrm{TeV} \nonumber \\
A_0= 0.5\, \textrm{TeV},\ 0,\ -1\, \textrm{TeV},\  -2\, \textrm{TeV}, &\quad & \tan \beta =5,\ 10,\ 50\ \textrm{and}\ 53
\label{ppp1}
\end{align}
with $\mu >0$ and $m_t = 171.4$~GeV. We have also performed scans for intermediate values 
of $\tan\beta =30,\ 45$, but found results to be similar to $\tan\beta =10,\ 50$ cases 
respectively and decided to omit them.
The use of IsaTools package for  implementation of the various 
constraints mentioned above was crucial in our study.

\section{Results}
\label{ch:results}

Let us first demonstrate our procedure by consecutively applying the 
constraints mentioned in the previous section to the case with $A_0=-2$~TeV and $\tan\beta =10$.
\FIGURE{
\includegraphics[width=1.0\textwidth]{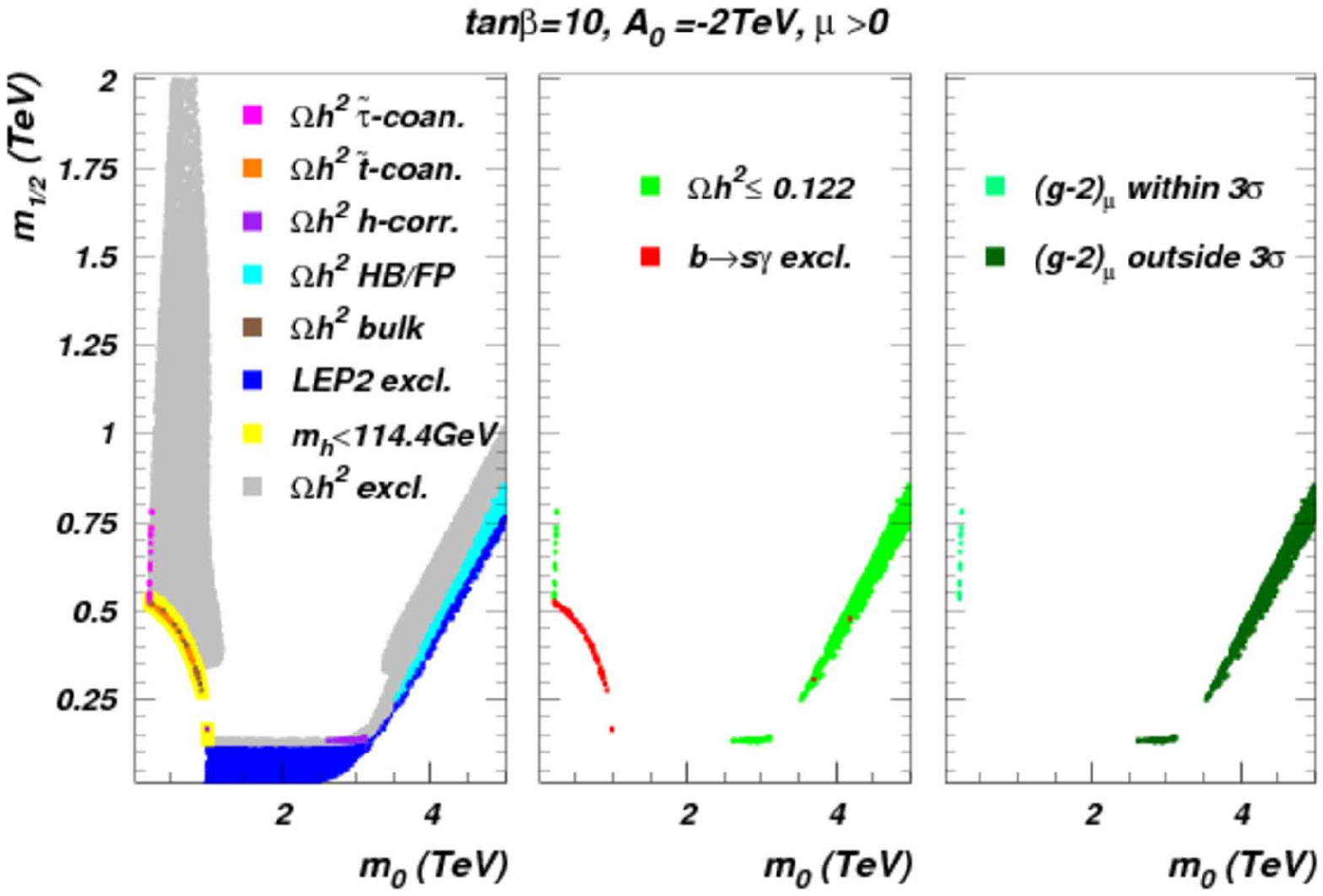}
\vspace*{-1cm}
\caption{
$(m_0,m_{1/2})$ plane for $\tan\beta =10,\ A_0 =-2$~TeV. In frame
a),  the gray regions are excluded by WMAP, blue ones are excluded by LEP2 chargino bound, 
and the
yellow  region has $m_h$ below 114.4~GeV; the  remaining colors denote various WMAP-allowed
regions described in the text. In frame b), the green regions satisfy both the WMAP and
$b\rightarrow s \gamma$ constraints, while the red region satisfies the WMAP bound but fails the
$b\rightarrow s \gamma$ constraint. In frame c), regions surviving the WMAP and $b\rightarrow s
\gamma$ constraint are presented; light  (dark) 
green region corresponds to $(g-2)_{\mu}$ within (outside) 
the $3\sigma$ range (\ref{eq:g-2}).}
\label{fig:m0-mhf-tb10-n2}
}
 We start by presenting our results in the conventional $(m_0,\ m_{1/2})$ plane illustrated 
in Fig.~\ref{fig:m0-mhf-tb10-n2}.  Frame a) displays the colored points 
allowed by radiative electroweak symmetry  breaking (REWSB) and 
from the requirement that the neutralino is the LSP~\footnote{In fact, to keep the data 
set size manageable, we show only points where $\Omega h^2 \leq 10$. 
The CMSSM points with too high a relic density, denoted by gray 
 or white color above the gray-colored area, are not relevant to the 
current discussion.}. 
 
The gray region at low $m_0$ and large-to-medium $m_{1/2}$ values   at
the left is excluded because $m_{\ttau_1} < m_{\tz_1}$,  
while the white  bulge at low $m_0$ and low $m_{1/2}$ 
under the $\tilde{t}$-coannihilation contour 
is excluded because the top-squark is LSP in this region
(typical for such a large negative $A_0$). 
Finally, the white corner region at the bottom-right
below the colored band is excluded because of the failure
of the  REWSB condition. The
LEP2 constraint on chargino mass removes the points shown in blue
color.  After application of the WMAP upper bound  on LSP DM, the allowed region is
drastically  shrunk to several small regions of the  parameter space,
with the remaining gray points associated with  unacceptably high  DM relic density.
This exhibits a well known fact that in CMSSM  the neutralinos do not
efficiently annihilate in the early universe except in a few very 
special narrow regions of the parameter
space\footnote{ These regions sometimes appear discontinuous on some plots 
only due to the finite resolution of our scan.}:
\begin{itemize}
\item the bulk region at low $m_0$ and low $m_{1/2}$, shown in brown color, where 
neutralinos efficiently annihilate via $t-$channel slepton exchange. This area is 
almost completely ruled out by LEP2 searches~\cite{Ellis:1983wd, Ellis:1983ew, Baer:1995nc,
Barger:1997kb}.
\item the stau co-annihilation region (\stac)~\cite{Ellis:1998kh, Ellis:1999mm, Gomez:1999dk, Gomez:2000sj, 
Ananthanarayan:1992cd, Ananthanarayan:1994qt, Lahanas:1999uy, Arnowitt:2001yh, Baer:2002fv} 
at low $m_0$, shown in magenta color, where $m_{\tz_1}\simeq m_{\ttau_1}$ and rapid reactions 
$\tz_1 \ttau_1 \rightarrow X$ and $\ttau_1 \ttau_1 \rightarrow X$ (here $X$ stands for any allowed 
final state of SM and/or higgs particles) in the early 
universe lower  the neutralino relic density to the WMAP range (\ref{eq:Oh2});
\item the stop coannihilation region  (\stoc) appearing at low $m_0$ and low $m_{1/2}$, but for 
particular values of $A_0$, shown in orange color, 
where $m_{\tz_1}\simeq m_{\tst_1}$ ~\cite{Boehm:1999bj, Ellis:2001nx, Edsjo:2003us};
\item the hyperbolic branch/focus point region (HB/FP) at large $m_0$, shown in 
turquoise, where $|\mu|$ is  small and the  neutralino develops a substantial higgsino 
component that enhances its annihilation into $WW,\ ZZ$ and $Zh$ pairs in the early 
universe~\cite{Chan:1997bi,Feng:1999mn,Baer:1995nq,Baer:1995va,Baer:1998sz}; 
\item the $h-$corridor (HF) at low $m_{1/2}$, shown in purple color, where $2m_{\tz_1} \simeq m_h$ 
and efficient annihilation through the light MSSM Higgs boson resonance occurs in the early 
universe~\cite{Nath:1992ty,Baer:1995nc};
\item the $A-$annihilation funnel (AF) that appears only at suitably large values 
of $\tan\beta$ and is therefore absent in the current plot. In this case, 
$2m_{\tz_1} \simeq m_A$ and resonance annihilation of neutralinos via the 
broad $A$ and $H$ Higgs bosons becomes possible
~\cite{Drees:1992am, Baer:1997ai, Baer:2000jj, Ellis:2001msa, Roszkowski:2001sb, Lahanas:2001yr}.
\end{itemize}
Next, in frame b), we show the effect of the $b\rightarrow s \gamma$ 
constraint on the WMAP allowed region: red points denote the excluded region where 
the branching fraction exceeds the range (\ref{eq:bsg}) which covers 
the bulk region, stop coannihilation and the low $m_{1/2}$ part of the stau 
coannihilation regions. The green points survive the combination of the WMAP and $\ b\rightarrow s \gamma$ 
and are the  subject of our analysis in this paper. 
Finally, in frame c) we illustrate the effect of $(g-2)_\mu$ constraint (\ref{eq:g-2}) 
which, at 3$\sigma$ level, excludes the h-corridor and HB/FP regions 
at medium and high $m_0$, denoted by dark green color. This occurs 
since sleptons are heavy in these regions and their loop 
contributions do not make significant contributions to $(g-2)_\mu$. On 
the contrary, in the \stac ~region, shown in light 
green color, the sleptons are light and can account for the observed deviation (\ref{eq:g-2}).

\FIGURE{
\vspace*{-1.0cm}
\includegraphics[width=\textwidth]{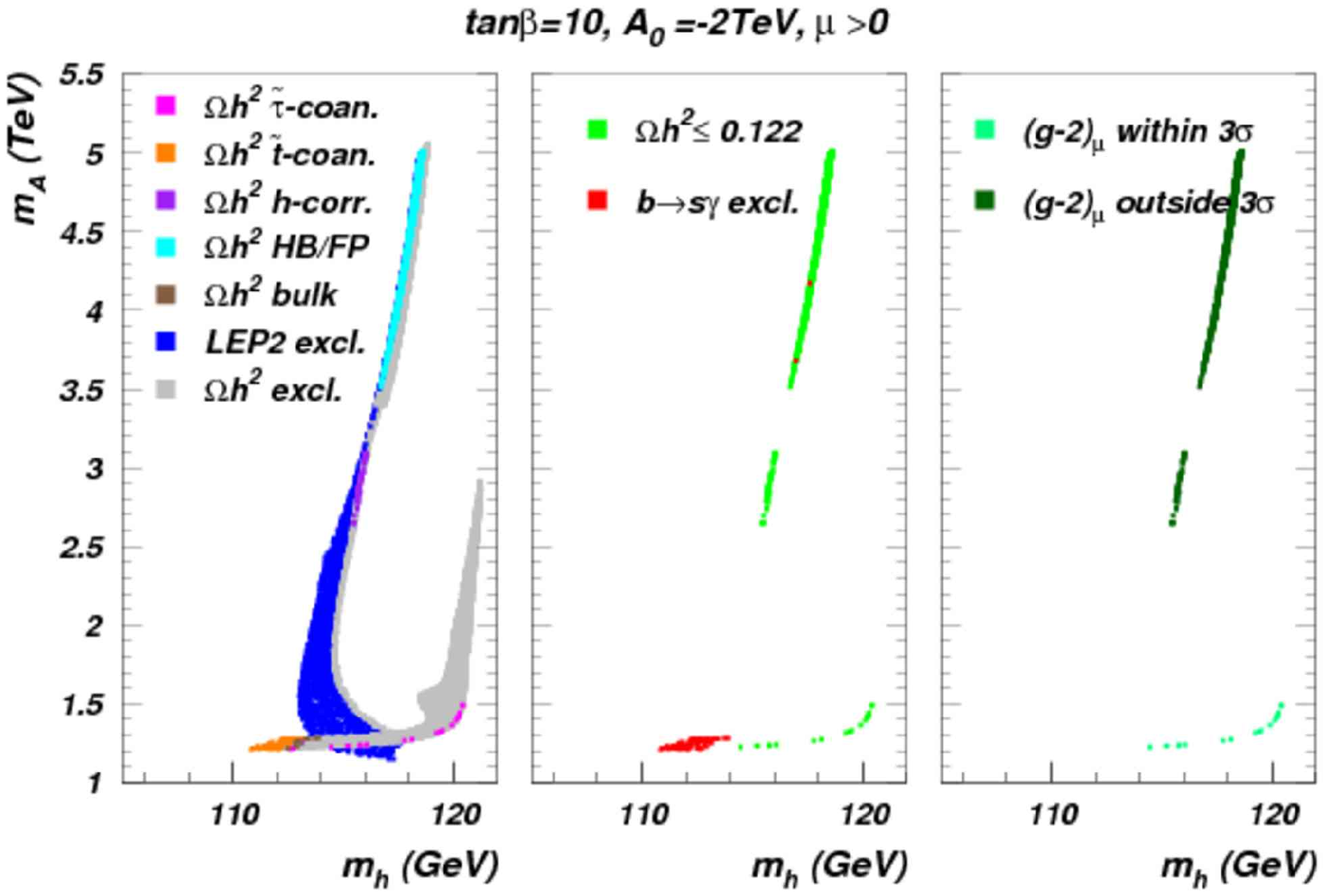}
\caption{
 $(m_h,m_A)$ plane for $\tan\beta =10,\ A_0 =-2$~TeV. 
 The color code is  the same as in Fig.~\ref{fig:m0-mhf-tb10-n2}.}
\label{fig:ma-mh-tb10-n2}
}

In Fig.~\ref{fig:ma-mh-tb10-n2} we present results for the same 
parameter space, but in the $(m_h,m_A)$ plane with the color 
coding as in Fig.~\ref{fig:m0-mhf-tb10-n2}. 
Note that the LEP2 Higgs boson limit excludes the 
\stoc ~region, leaving \stac, HF and HB/FP 
regions respecting the WMAP dark matter constraint.
It is very noticeable, that 
the regions allowed by SUSY constraints appear as very narrow bands 
exhibiting an obvious correlation of $m_A$ with $m_h$. 
This correlation is related to the two-fold solutions mainly defined by 
the DM constraints:  
the lower band corresponding to 
\stac ~region (connected to \stoc ~and AF regions at high values of $\tan\beta$)
and the upper band 
corresponding to the \fp ~region 
(and connected to HF region at low to intermediate values of $\tan\beta$) . 
\FIGURE{
\vspace*{-0.3cm}
\psfrag{q3}{{\normalsize {\boldmath $m_h~(GeV)$}}}
\psfrag{q4}{{\normalsize {\boldmath $m_{A}~(TeV)$}}}
\psfrag{q1}{{\small {\boldmath {\it (a)} $A_0=0.5~TeV$}}}
\psfrag{q2}{{\small {\boldmath {\it (b)} $A_0=0$}}}
\psfrag{q5}{{\small {\boldmath {\it (c)} $A_0=-1~TeV$}}}
\psfrag{q6}{{\small {\boldmath {\it (d)} $A_0=-2~TeV$}}}
\includegraphics[height=0.44\textwidth,angle=270]{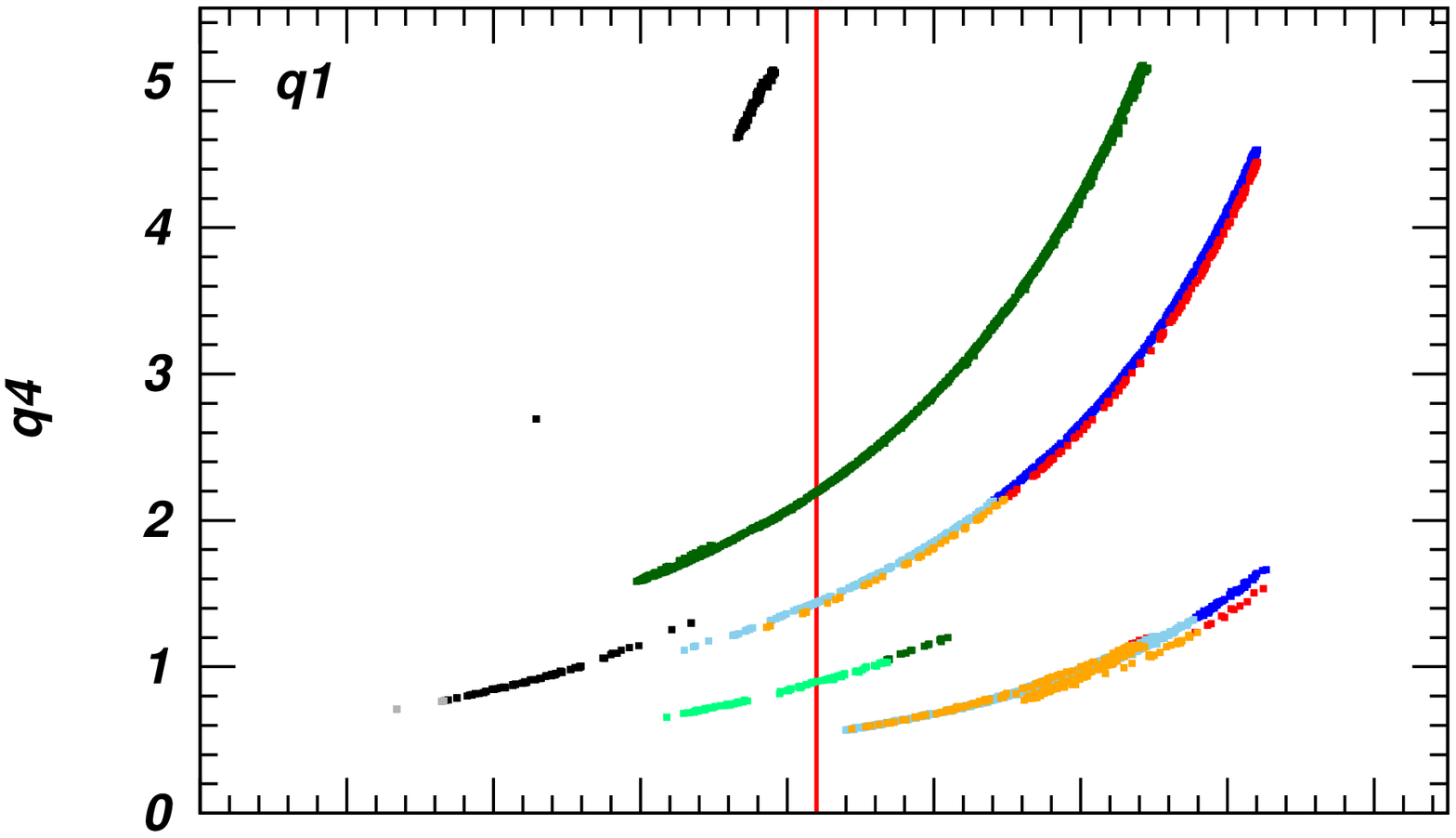}%
\includegraphics[height=0.44\textwidth,angle=270]{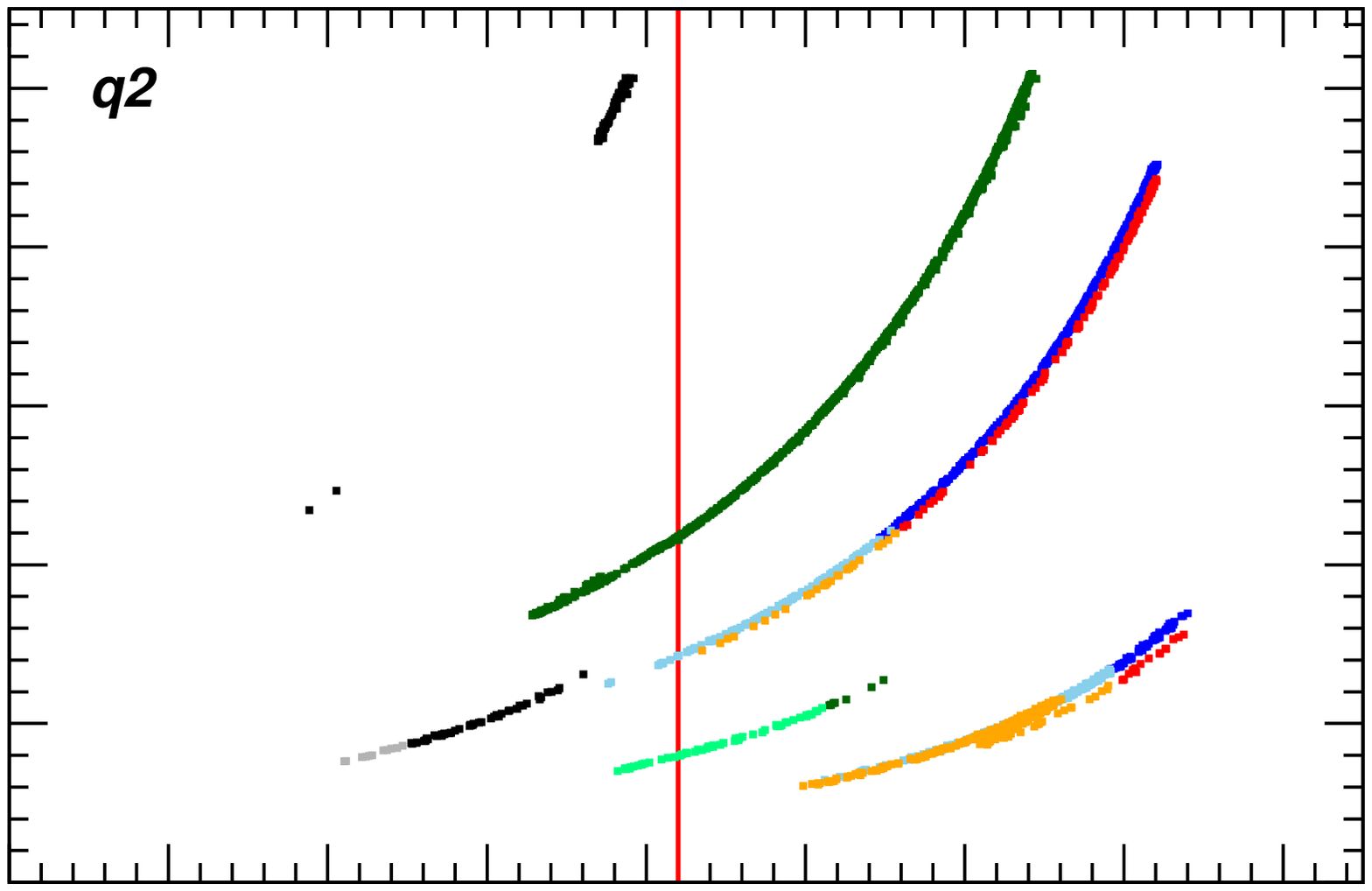}\vspace{-0.9cm}\\\hspace*{-0.2cm}%
\includegraphics[height=0.44\textwidth,angle=270]{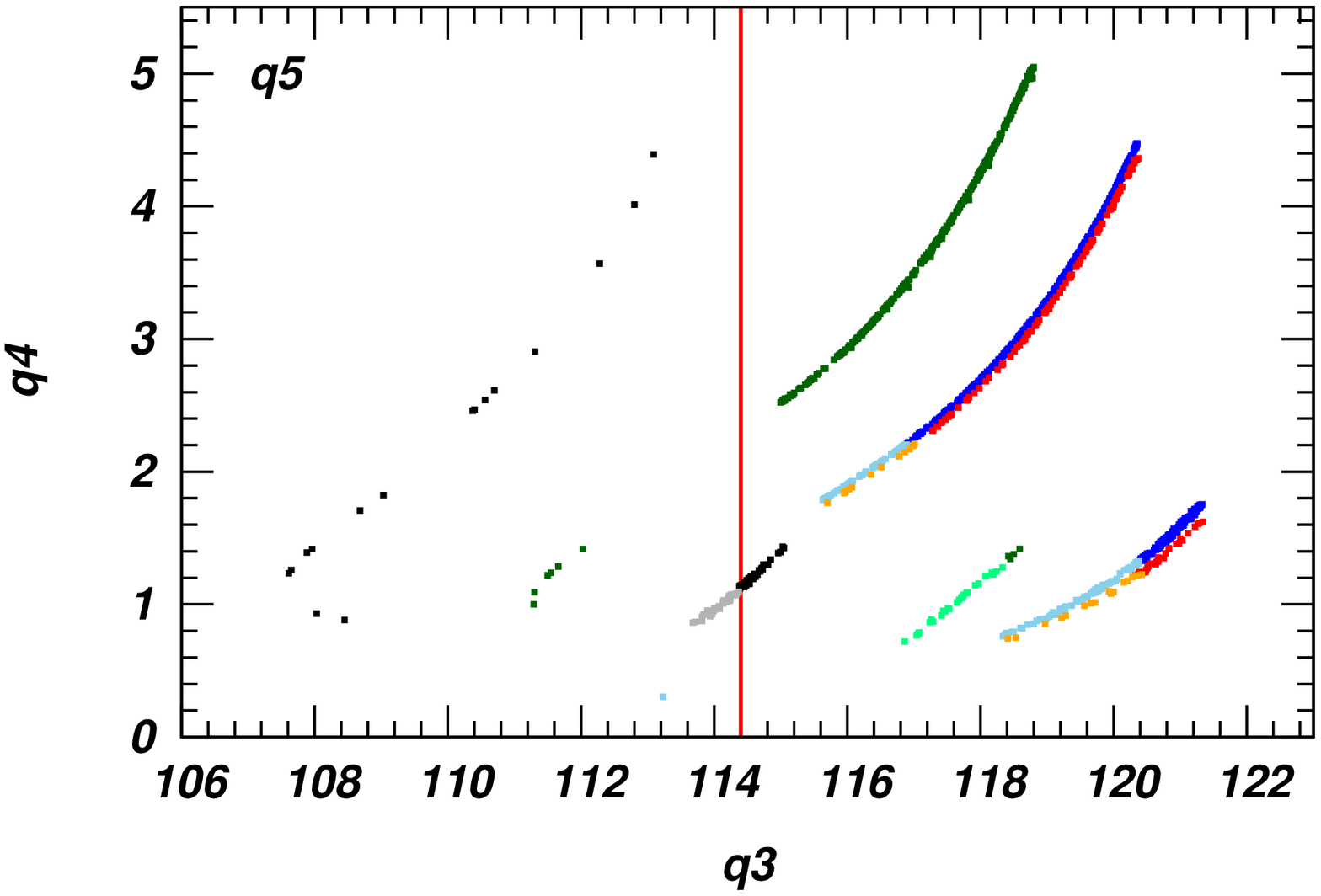}%
\includegraphics[height=0.44\textwidth,angle=270]{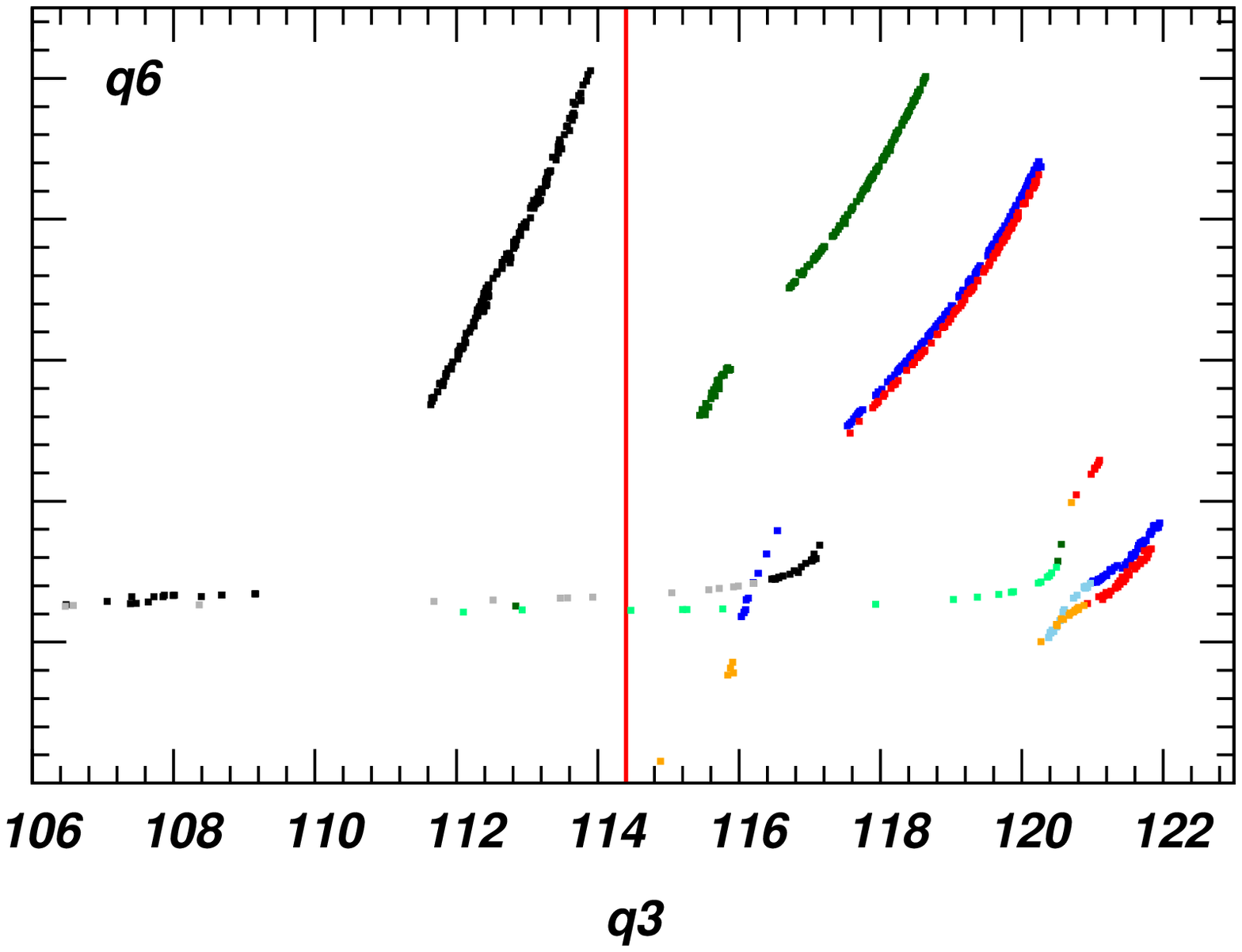}
\caption{
Allowed region for CP-odd Higgs boson mass $m_A$ versus $m_h$.
Gray, light green, light blue, and orange correspond to $\tan\beta =$5, 10, 50 and 53
 respectively and satisfy the $3 \sigma$ bound on $\Delta a_{\mu}$
(\ref{eq:g-2}). Black, dark green, blue and
 red correspond to $\tan\beta =$5, 10, 50 and 53, with $\Delta a_{\mu}$ 
 outside the $3 \sigma$ range.\label{fig:ma-mh}}
\vspace*{-0.1cm}
}

\FIGURE{
\vspace*{-0.3cm}
\psfrag{q3}{{\normalsize {\boldmath $m_h~(GeV)$}}}
\psfrag{q4}{{\normalsize {\boldmath $m_{\tg}~(TeV)$}}}
\psfrag{q1}{{\small {\boldmath {\it (a)} $A_0=0.5~TeV$}}}
\psfrag{q2}{{\small {\boldmath {\it (b)} $A_0=0$}}}
\psfrag{q5}{{\small {\boldmath {\it (c)} $A_0=-1~TeV$}}}
\psfrag{q6}{{\small {\boldmath {\it (d)} $A_0=-2~TeV$}}}
\includegraphics[height=0.44\textwidth,angle=270]{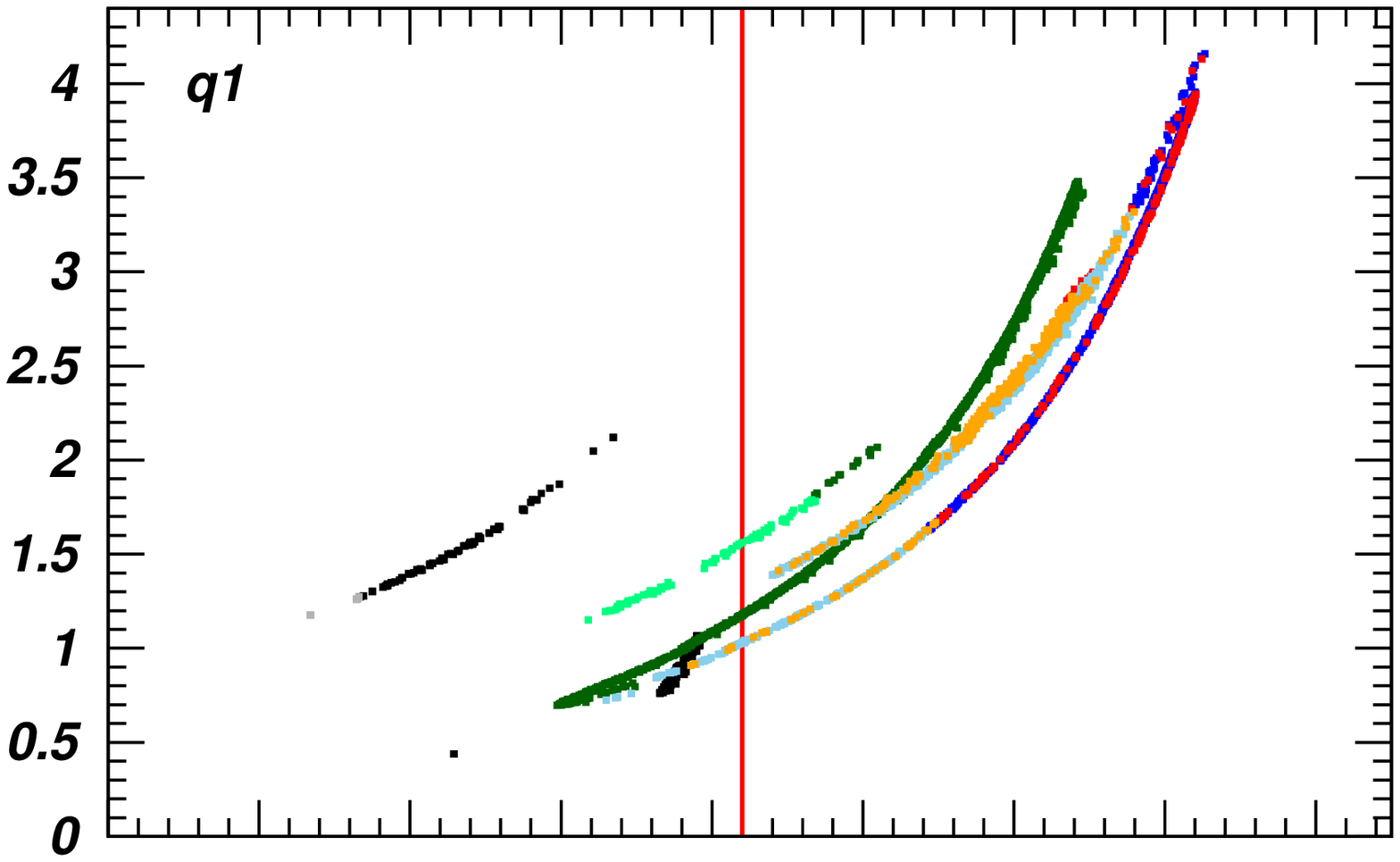}%
\includegraphics[height=0.44\textwidth,angle=270]{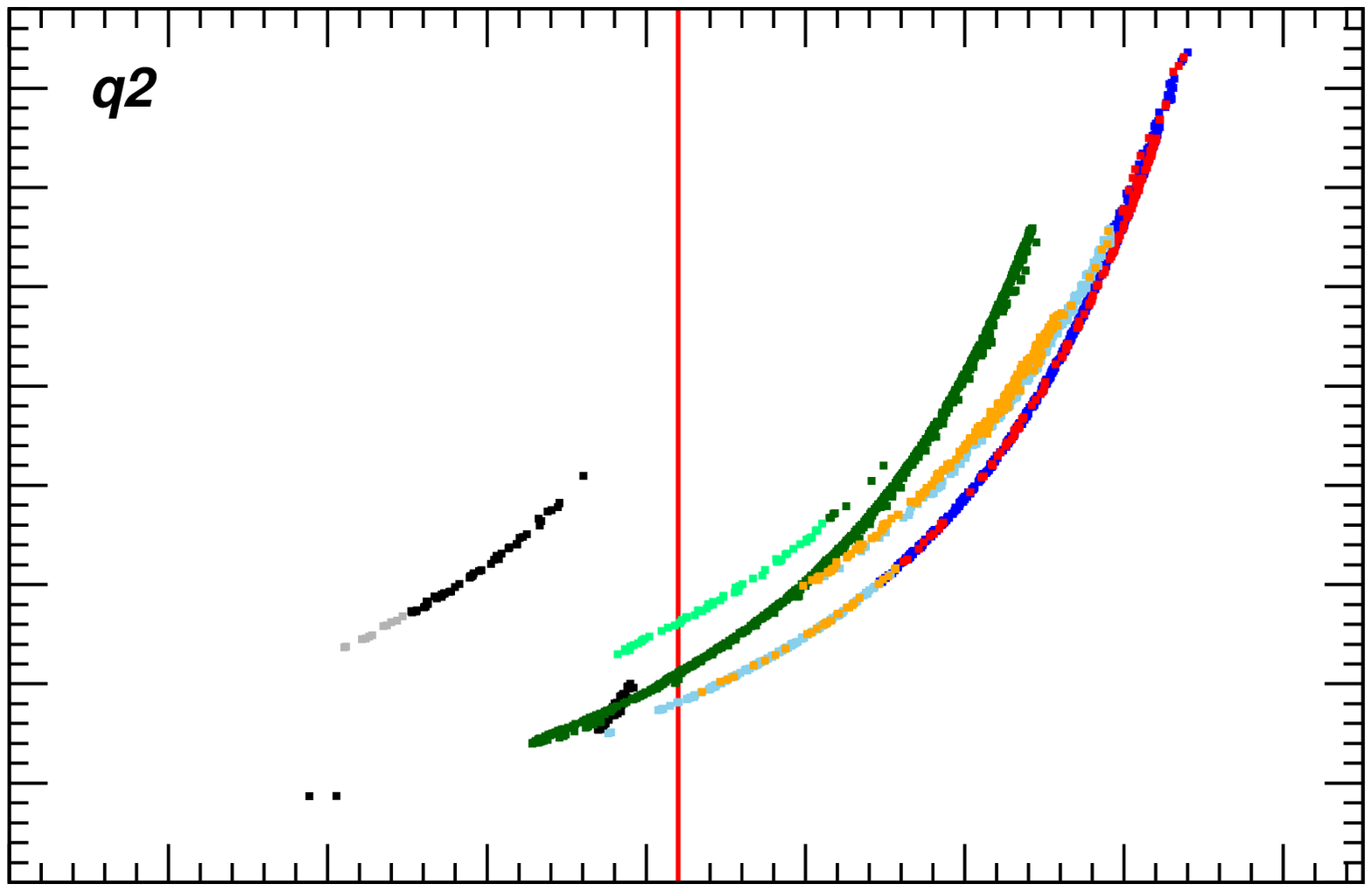}\vspace{-0.9cm}\\\hspace*{-0.2cm}%
\includegraphics[height=0.44\textwidth,angle=270]{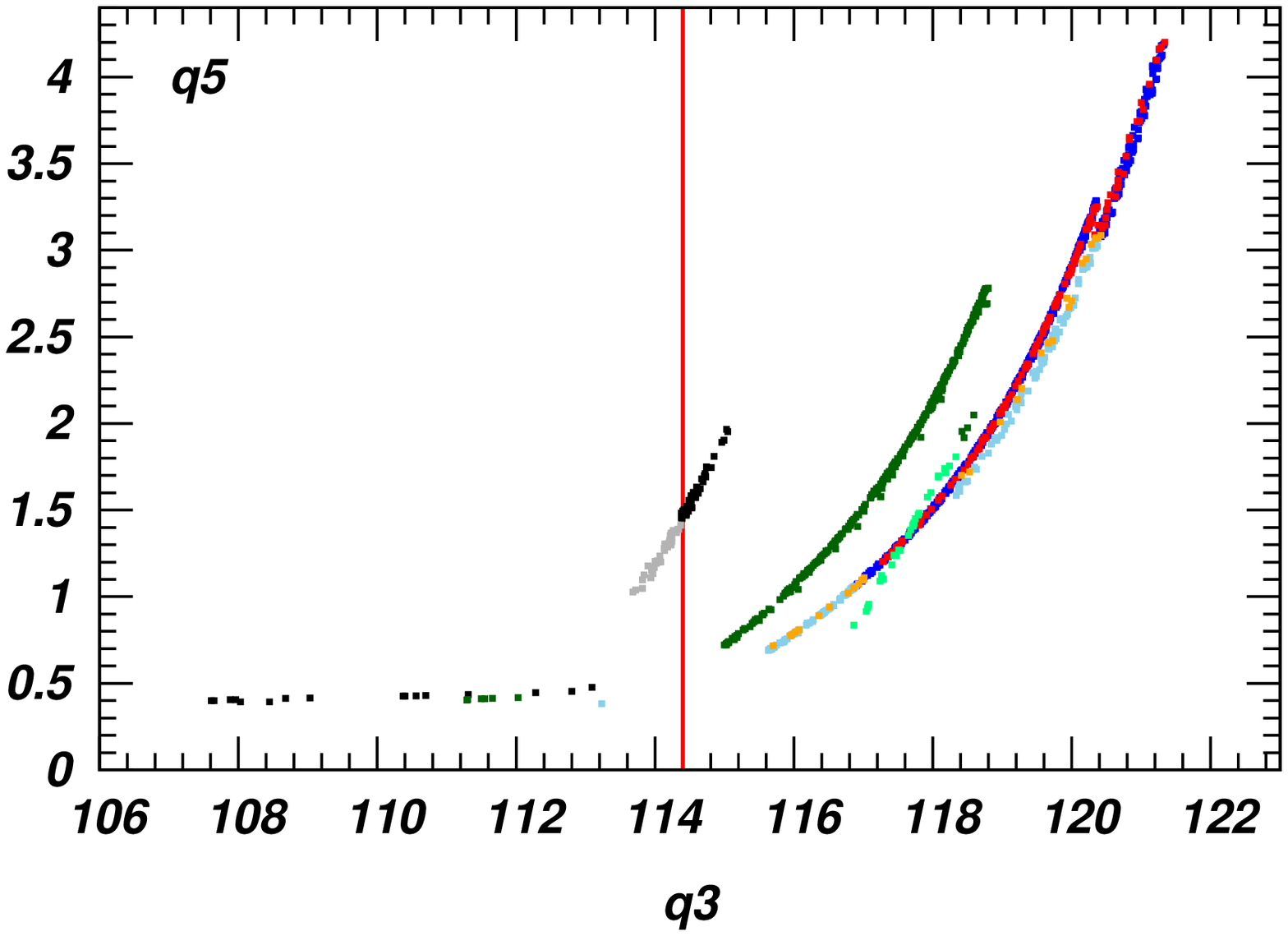}%
\includegraphics[height=0.44\textwidth,angle=270]{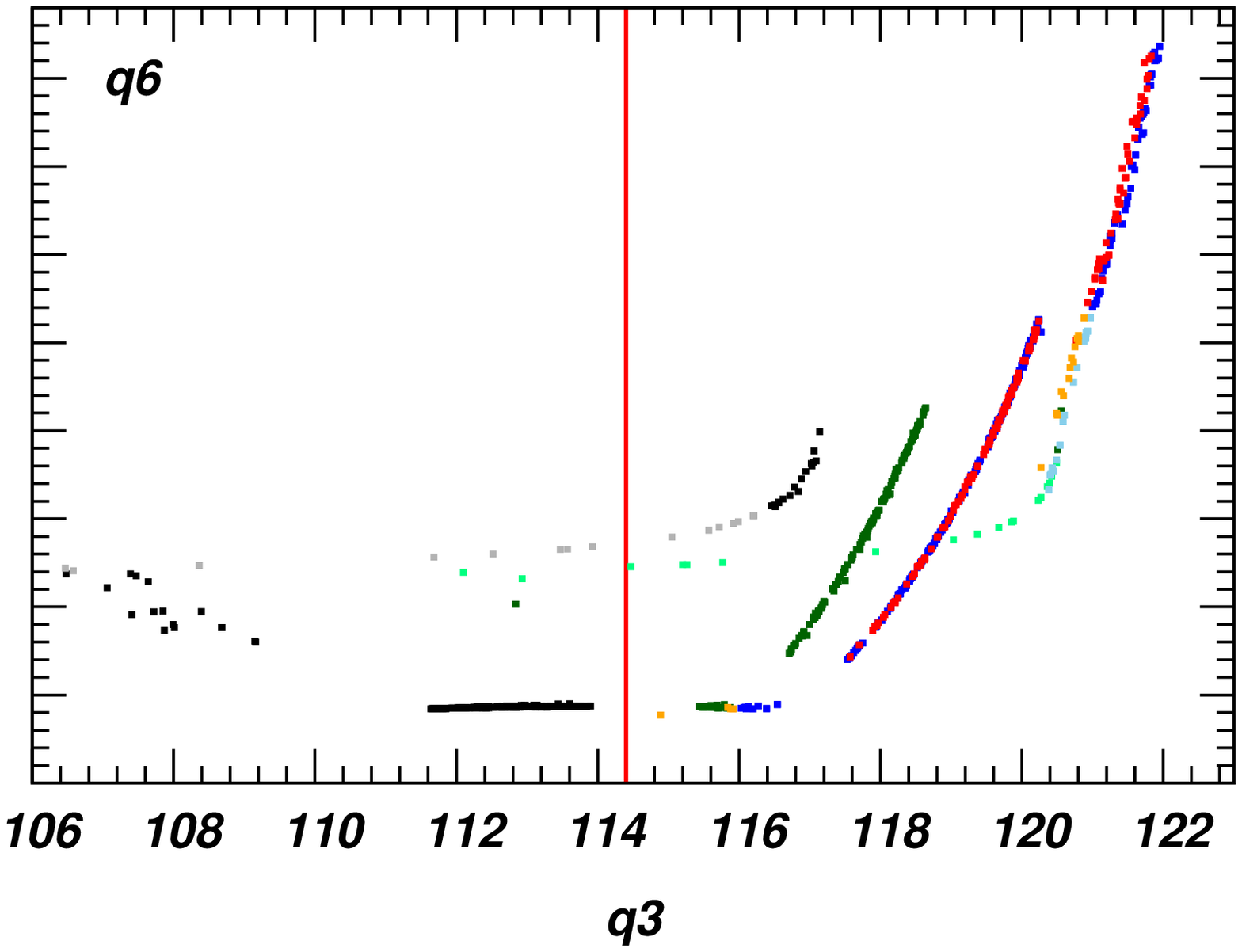}
\caption{Allowed region for gluino mass versus $m_h$. The color code is  the same as in Fig.~\ref{fig:ma-mh}.
\label{fig:mg-mh}}
\vspace*{-0.1cm}
}

Qualitatively similar patterns can be observed for other 
$A_0$ and $\tan\beta$ values and  are presented in Fig.~\ref{fig:ma-mh}. 
Here  gray, light green, light blue, and orange colors correspond to 
$\tan\beta =$5, 10, 50 and 53 respectively and satisfy the $3 \sigma$ bound on $\Delta a_{\mu}$ 
given by Eq.(\ref{eq:g-2}).
Black, dark green, blue and red colors correspond to the region outside the 
$\Delta a_{\mu}$ $3\sigma$ bound; frames a),b,)c),d) present results 
for $A_0 = 0.5,\ 0,\ -1$ and $-2$~TeV respectively. The upper and lower bands for each 
color represent  the \fp(+HF) and \stac(+\stoc+AF) regions respectively.
One finds that the CP-odd Higgs boson mass is always above $\sim 200$~GeV,  
corresponding to the decoupling regime and  SM-like nature of $h$.
One can see that for  $A_0=0.5$~TeV and 0 (frames a) and b) respectively) the parameter 
space for $\tan\beta=5$ is completely excluded by the LEP2 limit, $m_h \geq 114.4$~GeV. 
This is  due to the low tree-level Higgs mass and the absence of large enough radiative corrections. 
However, for $A_0=-1$~TeV and $-2$~TeV (frames c) and d) respectively), 
enhanced  radiative corrections to the Higgs boson mass open up some viable 
CMSSM parameter space even for $\tan\beta$ as low as 5. 
We note here that at the EW scale  $m_h$ depends, to a good approximation~\cite{Carena:2000dp}, 
on the absolute value of $A_0$.
Starting with negative values of $A_t$ at 
$M_{\rm GUT}$, it turns out that $A_t$ evolves to even larger negative values 
at the EW scale, thereby enhancing $m_h$ and  relaxing the lower bounds on 
the masses of the SUSY particles. 
This explains our choice for $A_0$ values used in the paper.

We stress that the light colors in Fig.~\ref{fig:ma-mh} 
and subsequent figures represent the parameter space which respect 
all of the constraints mentioned above {\it including} $(g-2)_\mu$, 
while the  dark colors represent the parameter space respecting all 
constraints {\it except} $(g-2)_\mu$. 
For low and intermediate values of $\tan\beta$ only some portion of the 
$\tilde\tau$-coannihilation regions (lower curves)  denoted by gray 
($\tan\beta=5$) and light green ($\tan\beta=10$) colors satisfy all of 
the constraints, while the focus point region is excluded. In fact, this 
known feature of CMSSM represents the tension between essentially 
two constraints: $b\to s\gamma$ and $(g-2)_\mu$. The root cause of 
this tension is the universality 
condition in CMSSM
which implies equal squark and slepton soft mass parameters at $M_{\rm GUT}$. 
On the other hand, $b\to s\gamma$ data agrees with the SM prediction 
(implying {\it heavy} third generation of squarks at the EW 
scale) whereas there seems to be  quite a significant deviation 
between the SM prediction and  $(g-2)_\mu$ measurement  (implying 
{\it light} second generation of sleptons at the EW scale). It has been 
shown~\cite{Baer:2004xx} that this tension can be resolved either 
with large  $\tan\beta$ values  or, alternatively, by relaxing the 
GUT-scale generation universality condition.  
Indeed, one can see that at large 
$\tan\beta$ there are allowed regions with both upper and lower 
(light blue ($\tan\beta=50$) and orange ($\tan\beta=53$)) curves. 
In the case of large $\tan\beta$, $m_A$ ($m_H$) up to
about 1~TeV can be accessible at the LHC and used for a 
consistency check of the CMSSM. 

In Fig.~\ref{fig:mg-mh} we present correlations between the 
gluino and the light Higgs boson masses with the same color coding as in 
the previous figure. 
The pattern of bands in the $(m_{\tilde{g}},m_h)$ plane is quite different from the 
one for $(m_A,m_h)$ in the previous figure.
Gluino mass is mainly defined by $m_{1/2}$ which, in turn, is similarly correlated with $m_h$
 along the \stac ~and \fp ~regions. In the $(m_{\tilde{g}},m_h)$ plane the bands are close 
to each other. The gluino mass 
increases in the `far' HB/FP 
region of large  $m_0$ and large $m_{1/2}$,
since in this region $m_{1/2}$ goes beyond 1~TeV
where the \stac ~region typically ends.
The `far' \fp ~region is characterized by the heaviest $m_h$ and heaviest gluino
mass in this particular plane.
Notice that the gluino mass 
is always above 400-500~GeV, so that the recent  Tevatron bounds on 
gluino mass~\cite{CDF:gluino,D0:gluino} do not affect the parameter 
space left after imposing  DM, LEP2 and $b\to s\gamma$ constraints.
On the other hand, the LHC with 100~fb$^{-1}$ luminosity will be able to probe the parameter space 
corresponding  to $m_{\tg}$ up to $\sim 3$~TeV
~\cite{Baer:1995nq,Baer:1998sz,Abdullin:1998nv,Allanach:2000ii,Baer:2003wx,Abdullin:1998pm}, 
covering almost the entire \stac ~band.
Also, the LHC with 100~fb$^{-1}$ can indirectly probe 
$m_{\tg}$ up to $\sim 2.5$~TeV in the \fp  ~region
via observation of 
$lepton+jets+\not\!\!{E_T}$ signal  from lighter gauginos~\cite{lhc-fp-montreal}. 
Distinguishing  \stac ~and \fp ~bands in  $(m_{\tilde{g}},m_h)$ plane
could be somewhat problematic, even though,
potentially, the knowledge of the light Higgs boson mass at the percentage level 
would allow one to  indirectly determine  the mass of very heavy 
gluinos with $5-10\%$ precision.

In Figs.~\ref{fig:mw-mh} and \ref{fig:mz-mh} we present plots for 
$m_{\tw_1}$ versus $m_h$ and $m_{\tz_1}$ versus $m_h$, respectively. 
%
\FIGURE{
\vspace*{-0.3cm}
\psfrag{q3}{{\normalsize {\boldmath $m_h~(GeV)$}}}
\psfrag{q4}{{\normalsize {\boldmath $m_{\tw_1}~(TeV)$}}}
\psfrag{q1}{{\small {\boldmath {\it (a)} $A_0=0.5~TeV$}}}
\psfrag{q2}{{\small {\boldmath {\it (b)} $A_0=0$}}}
\psfrag{q5}{{\small {\boldmath {\it (c)} $A_0=-1~TeV$}}}
\psfrag{q6}{{\small {\boldmath {\it (d)} $A_0=-2~TeV$}}}
\includegraphics[height=0.43\textwidth,angle=270]{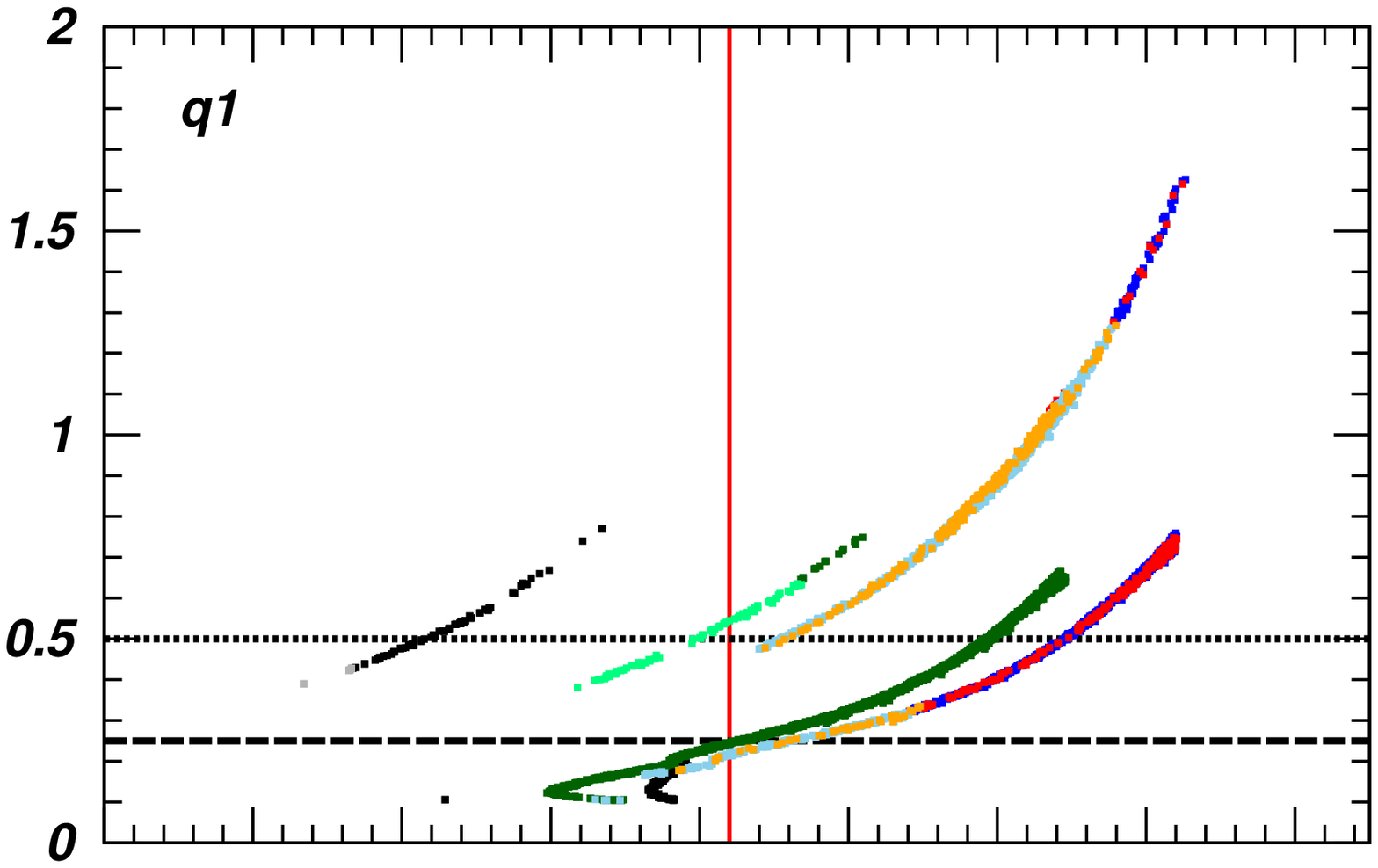}%
\includegraphics[height=0.43\textwidth,angle=270]{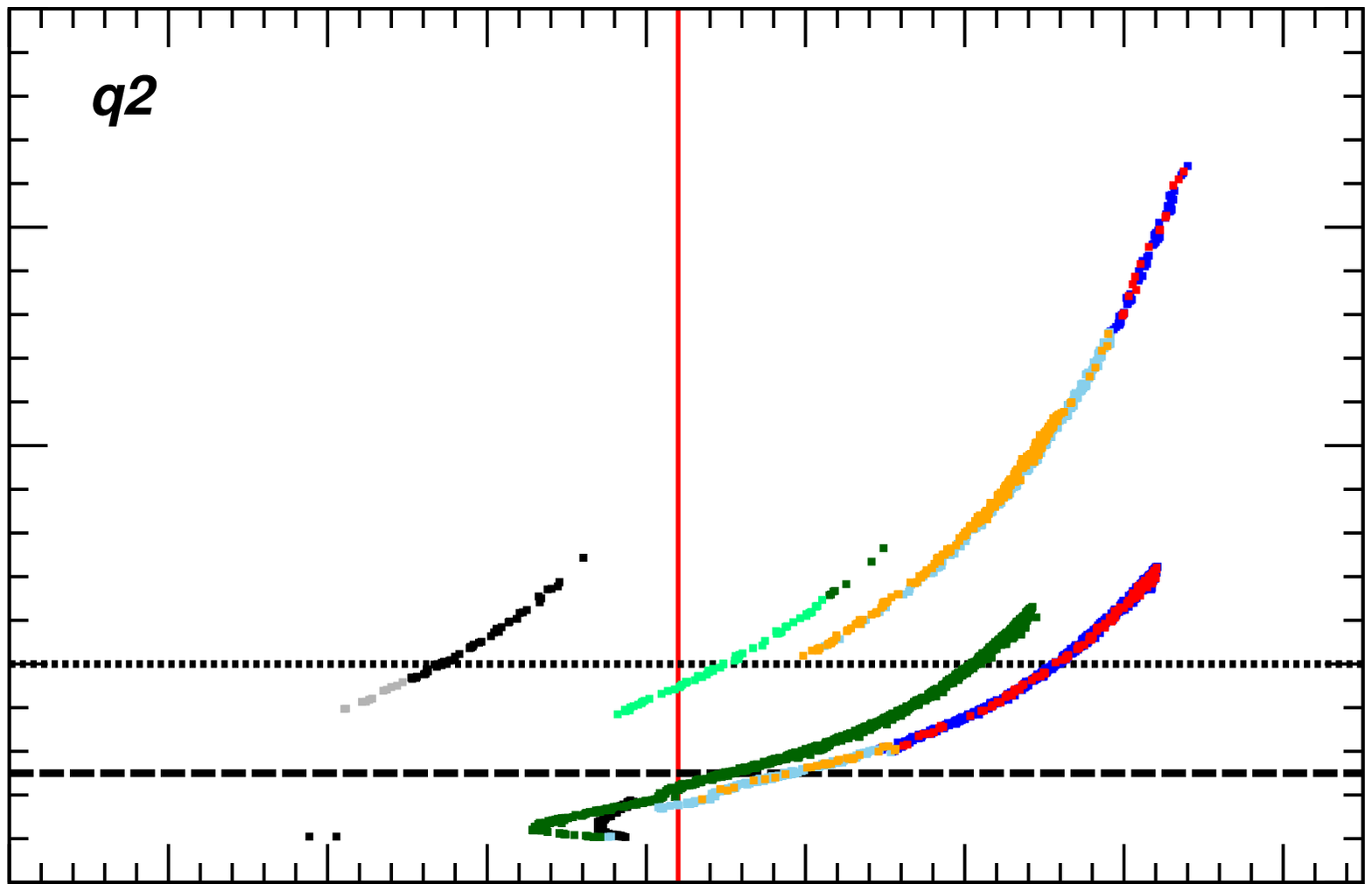}\vspace{-0.9cm}\\\hspace*{-0.3cm}%
\includegraphics[height=0.43\textwidth,angle=270]{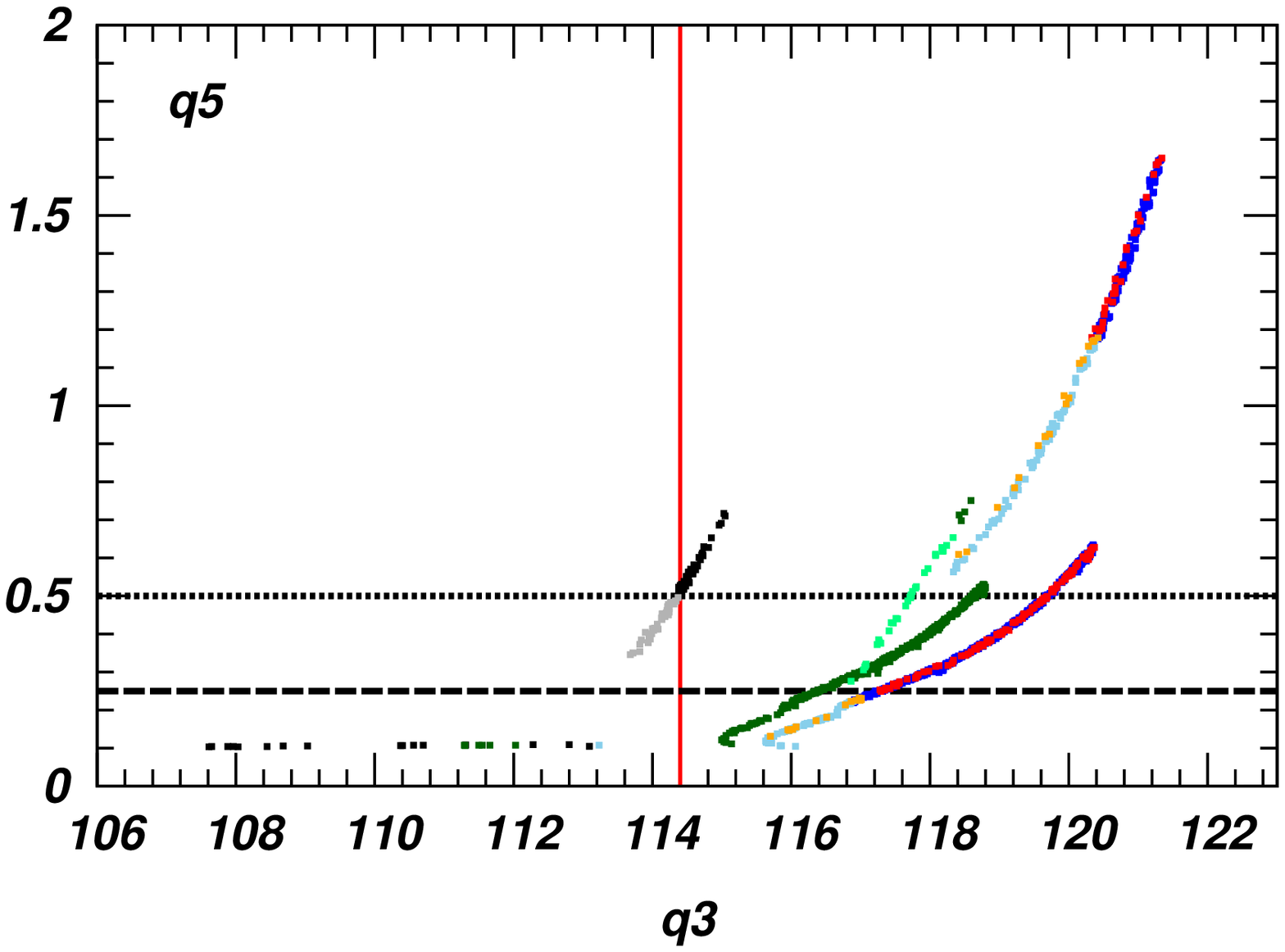}%
\includegraphics[height=0.43\textwidth,angle=270]{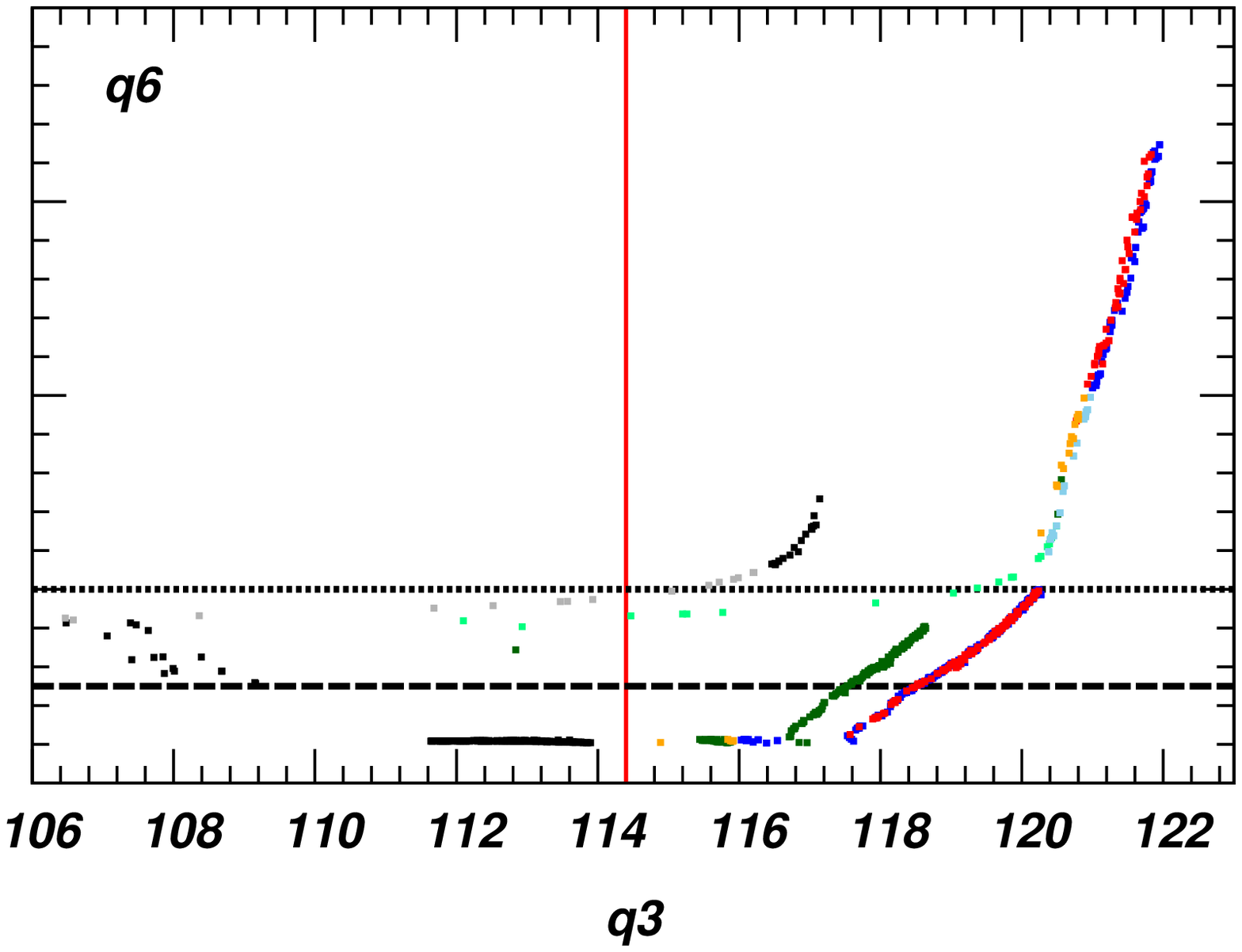}
\caption{\label{fig:mw-mh} Allowed region for the lightest chargino mass versus $m_h$.
The color code is  the same as in Fig.~\ref{fig:ma-mh}.
Dashed (dotted) horizontal lines represent approximate reach of ILC500 (ILC1000).}
}
\FIGURE{
\vspace*{-0.3cm}
\psfrag{q3}{{\normalsize {\boldmath $m_h~(GeV)$}}}
\psfrag{q4}{{\normalsize {\boldmath $m_{\tz_1}~(GeV)$}}}
\psfrag{q1}{{\small {\boldmath {\it (a)} $A_0=0.5~TeV$}}}
\psfrag{q2}{{\small {\boldmath {\it (b)} $A_0=0$}}}
\psfrag{q5}{{\small {\boldmath {\it (c)} $A_0=-1~TeV$}}}
\psfrag{q6}{{\small {\boldmath {\it (d)} $A_0=-2~TeV$}}}
\includegraphics[height=0.43\textwidth,angle=270]{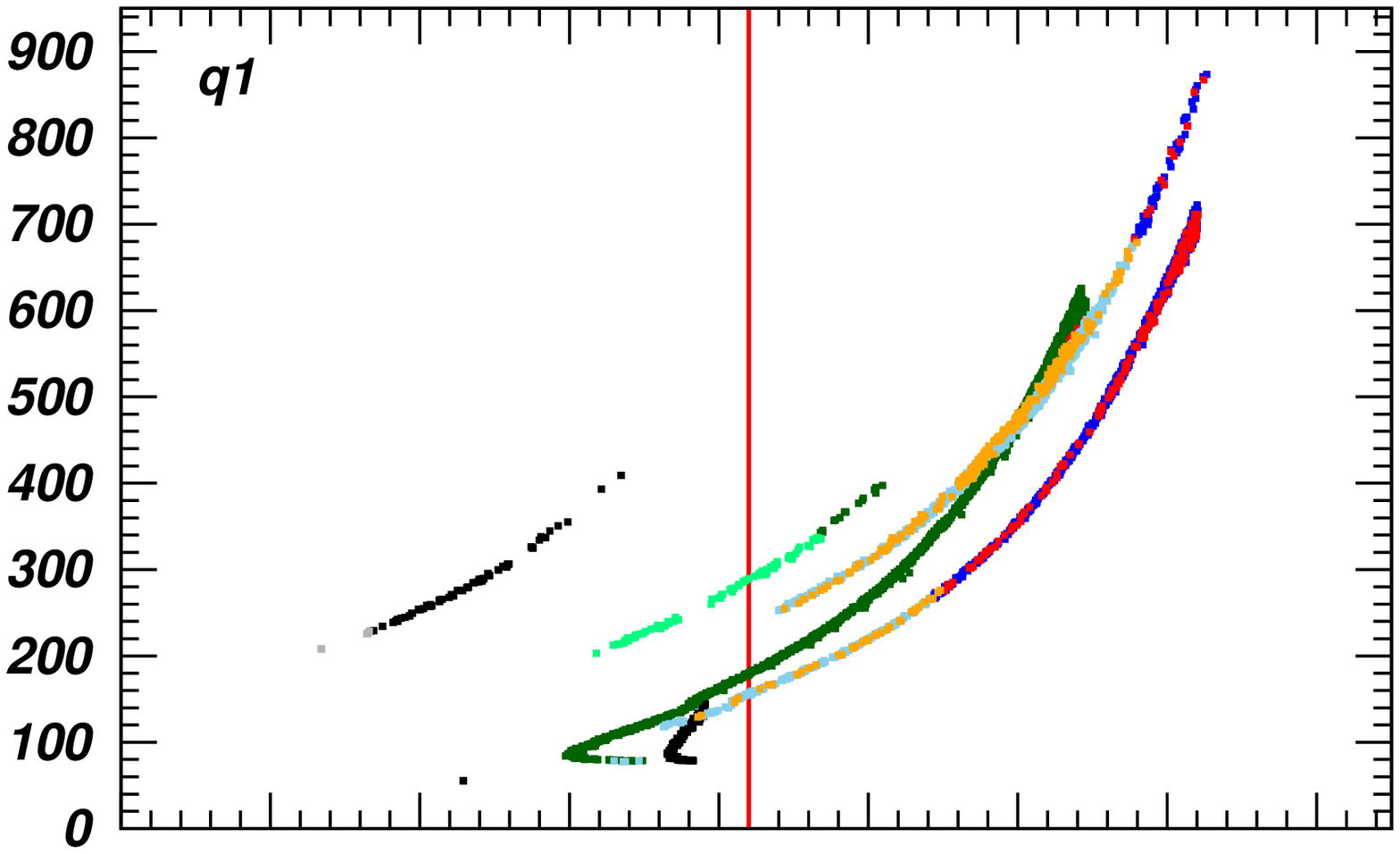}%
\includegraphics[height=0.43\textwidth,angle=270]{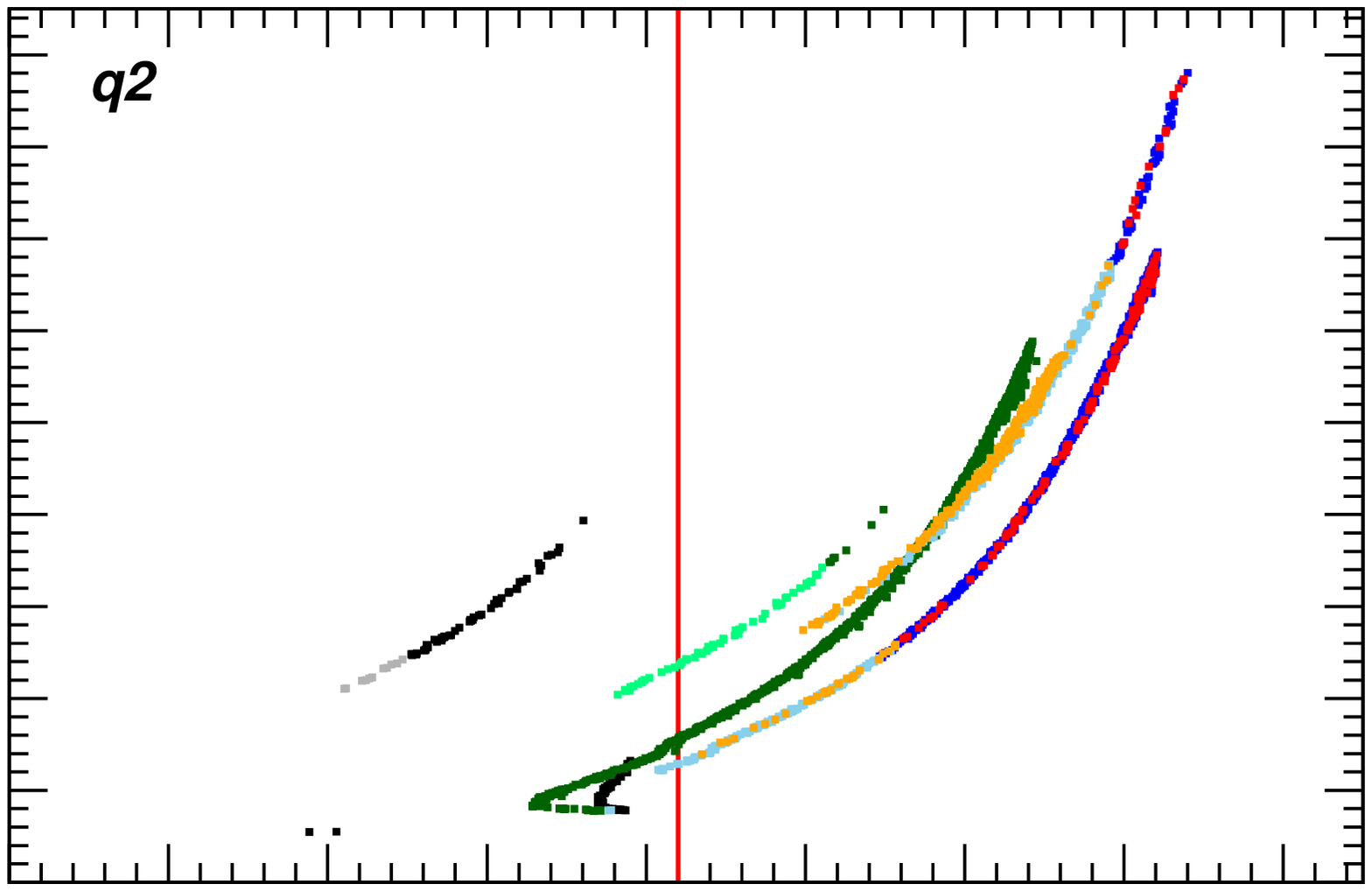}\vspace{-0.9cm}\\\hspace*{-0.3cm}%
\includegraphics[height=0.43\textwidth,angle=270]{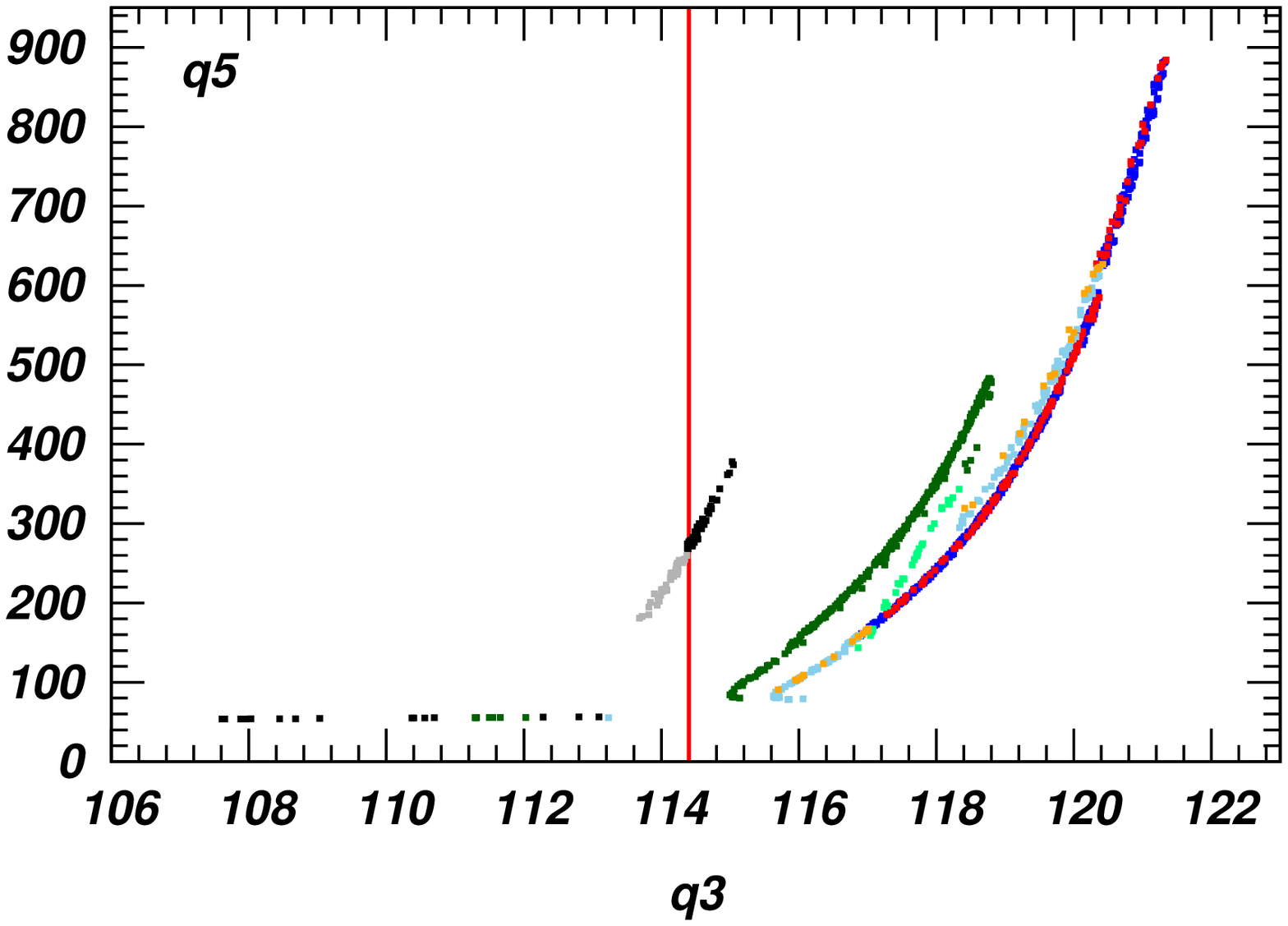}%
\includegraphics[height=0.43\textwidth,angle=270]{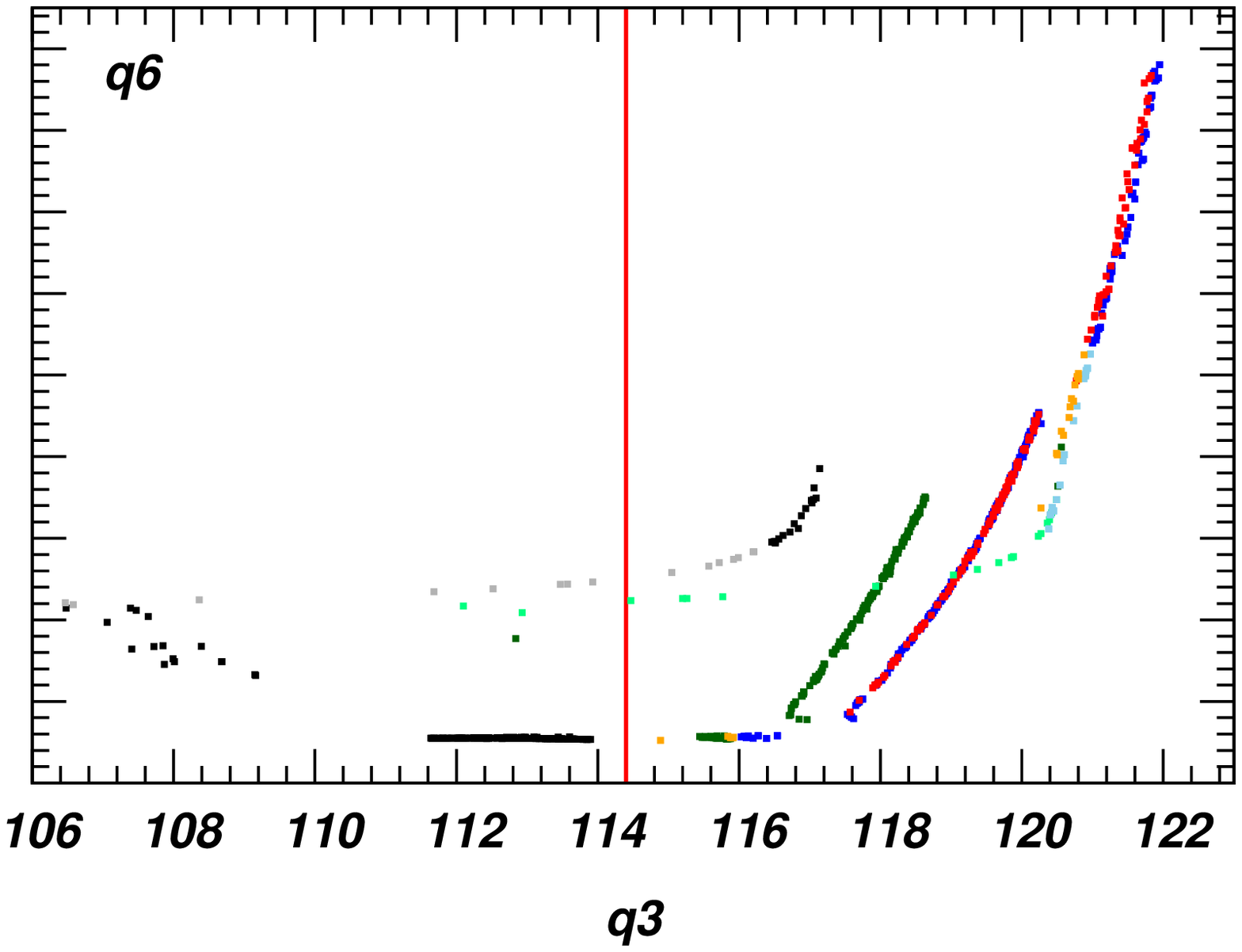}
\caption{\label{fig:mz-mh}Allowed region for the lightest neutralino mass versus $m_h$.
The color code is the same as in Fig.~\ref{fig:ma-mh}.}
}
%
The upper  band of the allowed  parameter space for each particular 
value of $\tan\beta$ corresponds to the  \stac(+\stoc+AF) region. 
In this region, the mass of the chargino and the neutralino is being 
driven up with the rise of $\tan\beta$ up to about $m_{\tw_1} \simeq 
1.7$~TeV and  $m_{\tz_1} \simeq 850$~GeV for $\tan\beta=53$. 
The \fp  ~and HF regions are characterized by small $|\mu|$ 
values resulting in light quasi-degenerate $\tz_1$ and 
$\tw_1$ states. This happens as a result of the following non-trivial 
interplay of the RG equations. With $m_{1/2}$ fixed, increasing $m_0$
leads to the suppression of terms with top and bottom Yukawa couplings 
in the RG equations for $m_{H_u}$ and $m_{H_d}$. This effect drives the higgs SSB 
masses towards the ``no REWSB'' condition and respectively lowers the value of $|\mu|$. 
These regions form the lower band of the allowed parameter space. The upper tip
of this band can reach the mass region as large as $\sim 700$~GeV.
This limit is defined  by the maximal value of 
$m_0=5$~TeV in our scan and practically does not depend on 
$\tan\beta$. One can see that the slope of the coannihilation 
band is typically larger than the slope of the allowed band of the 
parameter space corresponding to the focus point region. Therefore, 
in  the coannihilation region one can expect better accuracy in the 
chargino/neutralino mass determination through the correlations 
presented in Figs.~\ref{fig:mw-mh} and \ref{fig:mz-mh}. Also, using 
$m_{\tilde g} - m_h$ correlation, when the gluino is observed and 
reconstructed, one can actually predict the chargino and the neutralino 
masses, whose observation in EW production processes could be problematic.
To put things in perspective, we show in Fig.~\ref{fig:mw-mh} 
the kinematic limits for $\tw_1 \tw_1$ pair production, which 
can serve as good approximations for the reach of 
ILC500 ($\sqrt{s}=500$~GeV) and ILC1000 ($\sqrt{s}=1000$~GeV)
machines~\cite{Baer:2003ru,Baer:2004zk,Baer:1996vd}.
We see that the linear collider can access only the lower end of the coannihilation band
(corresponding to low $m_{1/2}$ values), while the HB/FP region can be probed almost entirely. 
This is in contrast to the LHC, which has large coverage in the coannihilation region, but can only
probe the lower half of the HB/FP region, as was shown in Fig.~\ref{fig:mg-mh}.

%
\FIGURE{
\vspace*{-0.3cm}
\psfrag{q3}{{\normalsize {\boldmath $m_h~(GeV)$}}}
\psfrag{q4}{{\normalsize {\boldmath $m_{\tst_1}~(TeV)$}}}
\psfrag{q1}{{\small {\boldmath {\it (a)} $A_0=0.5~TeV$}}}
\psfrag{q2}{{\small {\boldmath {\it (b)} $A_0=0$}}}
\psfrag{q5}{{\small {\boldmath {\it (c)} $A_0=-1~TeV$}}}
\psfrag{q6}{{\small {\boldmath {\it (d)} $A_0=-2~TeV$}}}
\includegraphics[height=0.44\textwidth,angle=270]{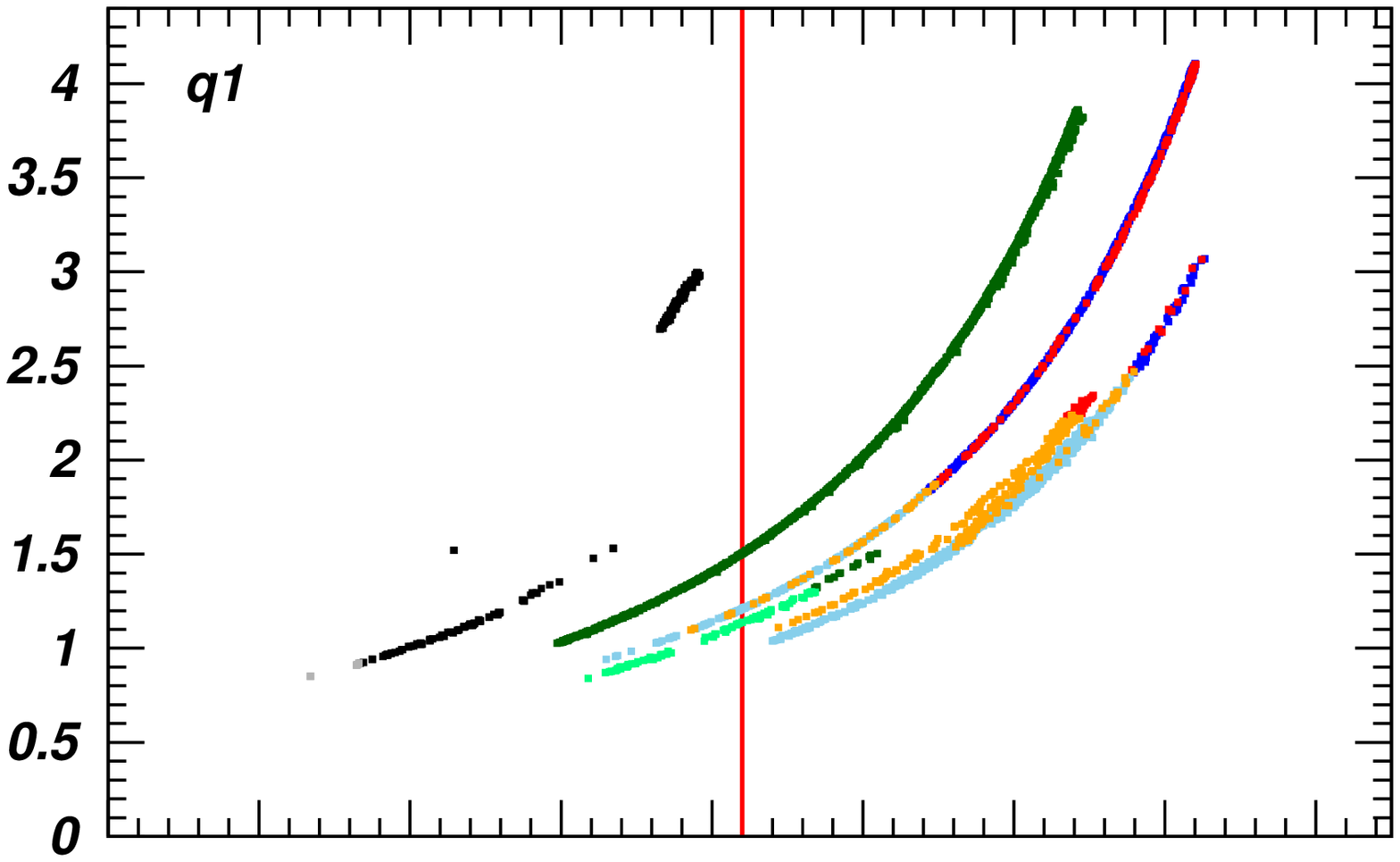}%
\includegraphics[height=0.44\textwidth,angle=270]{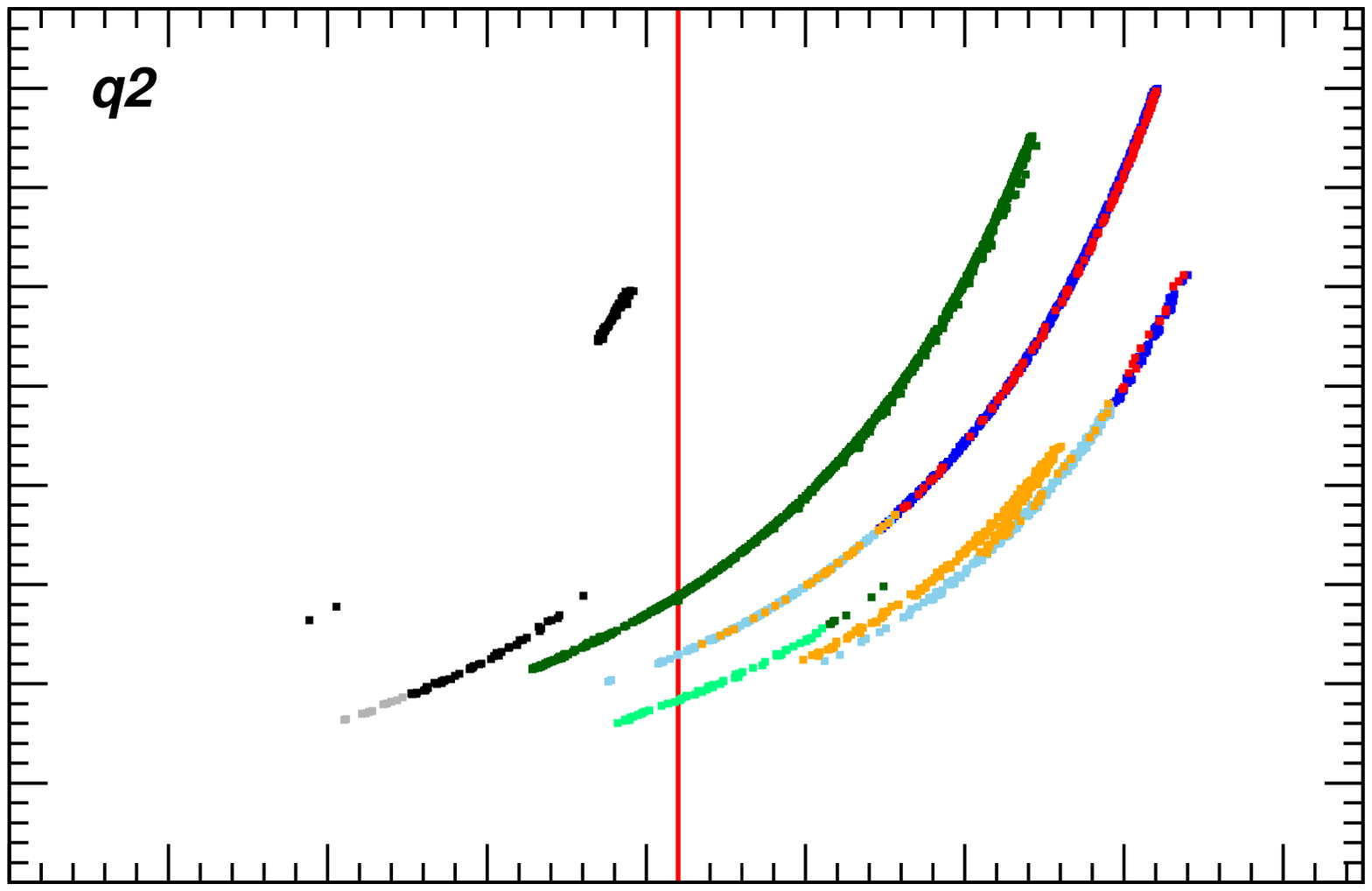}\vspace{-0.9cm}\\\hspace*{-0.2cm}%
\includegraphics[height=0.44\textwidth,angle=270]{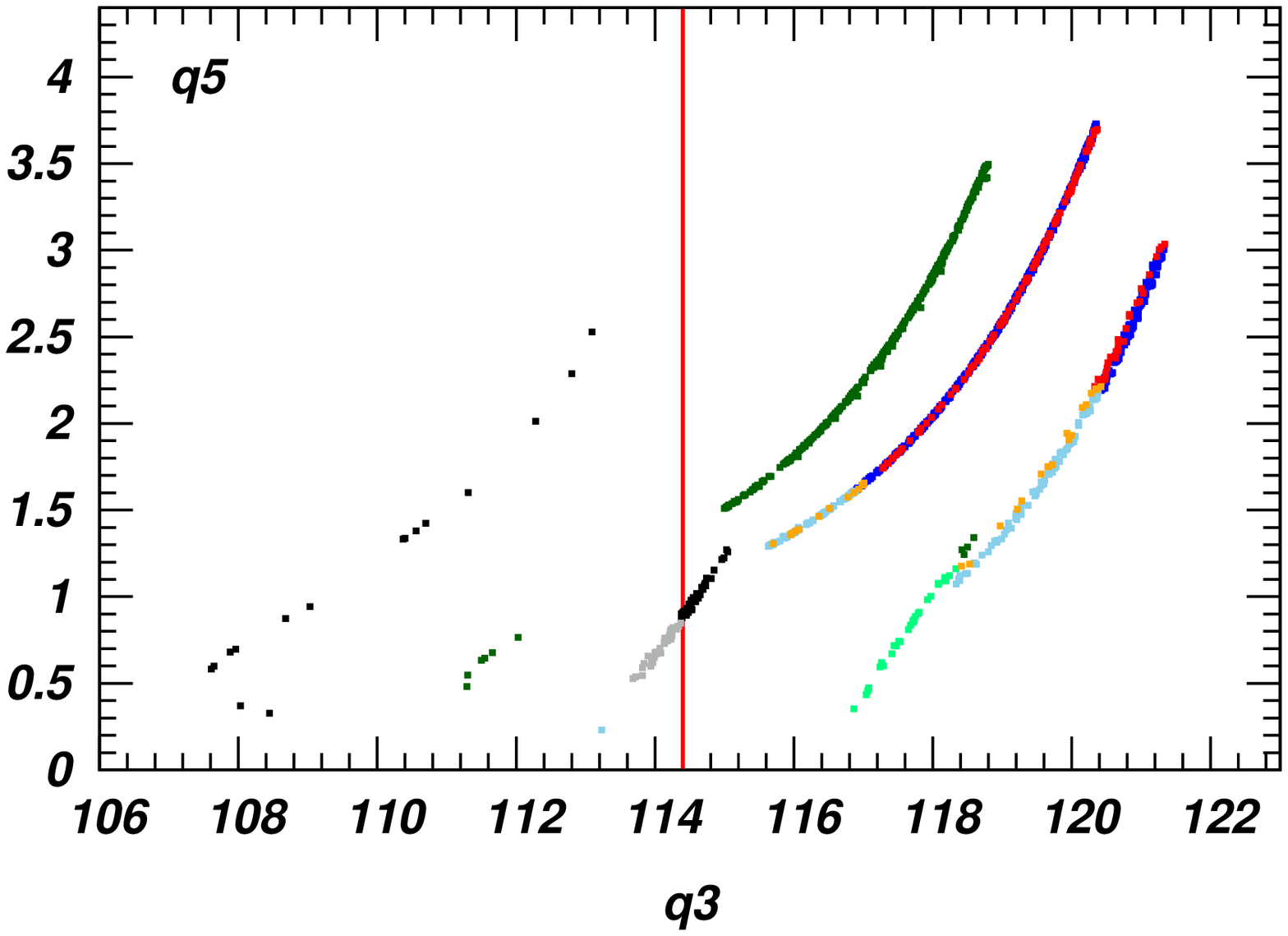}%
\includegraphics[height=0.44\textwidth,angle=270]{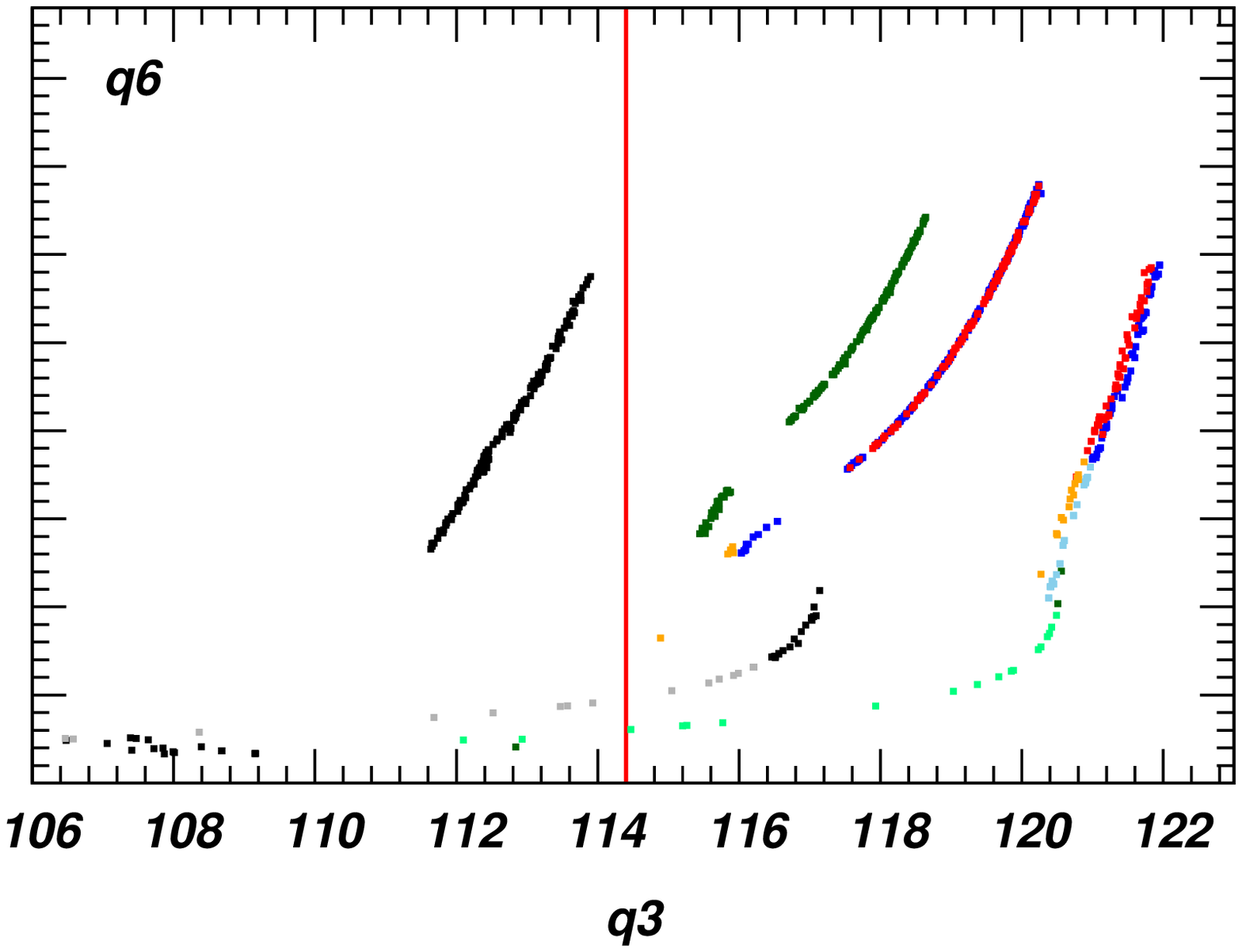}
\caption{\label{fig:mt-mh} Allowed region for top squark mass mass versus $m_h$.
              The color code is  the same as in Fig.~\ref{fig:ma-mh}.}
}
In Fig.~\ref{fig:mt-mh} we present correlations 
between the lightest top-squark mass and $m_h$. 
In this figure, the 
upper curves for each value of $\tan\beta$ correspond to the 
\fp(+HF)  region and  the lower curve  to the \stac(+\stoc+AF) region. At 
$\tan\beta=53$ and $A_0=0.5$~TeV and $A_0=0$ in frames a) and b) 
respectively, one can see that the orange curve is widened in its upper part. 
This region corresponds to the AF region, 
where the large range of radiative corrections to the 
top squark mass
and its stronger dependence on various model parameters 
spoil somewhat the $m_{\tst_1}-m_h$ correlation. 
In this region, $m_{\tst_1} \gtrsim 500$~GeV, thus exceeding the Tevatron
bound~\cite{CDF:stop,D0:stop}.
One can also notice 
that for $A_0=-1$~TeV and $ A_0=-2$~TeV, presented in frames c) and d), 
respectively, the top squark
exhibiting stop-coannihilation region 
is allowed to be as light as 200~GeV. This region is 
represented by the lower tip of the light-green line 
in frame c), and the lower tips of gray and light-green line in frame d). 
As $A_0$ becomes increasingly negative, the mixing between $\tst_L$ and $\tst_R$ 
increases (thereby reducing the mass of $\tst_1$) which, in turn, drives $m_h$ to 
larger values through radiative corrections. This explains why the 
curves shift to the lower right corner, as we go from frame a) to d). 

Let's take a closer look at the `anatomy' of the stop co-annihilation region.
In Fig.~\ref{fig:mst1-mh-tb30-n1} we present constraints similar to 
Figs.~\ref{fig:m0-mhf-tb10-n2} and \ref{fig:ma-mh-tb10-n2},
but in the $(m_{\tilde{t}},m_h)$ plane with the same color 
coding for $\tan{\beta}=30$ and $A_0=-1$~TeV.
The DM-allowed area consists of HB/FP region (upper branch),
\stac ~(lower branch),
and \stoc ~(the very bottom tip of the DM allowed region) regions.
From the frame b) one can clearly notice that the most serious constraint for 
the stop-coannihilation region  comes from the $b\to s \gamma$.
Indeed, for a light top squark, the contribution from the stop-chargino loop
increases $Br(b\to s \gamma)$ beyond the acceptable level.
The reason for the appearance of the very small allowed island 
surrounded by the $b\to s \gamma$ excluded region is the non-trivial cancellations
between  stop-chargino and top-charged Higgs loops.
Thus, the \stoc ~region does survive in the CMSSM  framework,
but this scenario is highly constrained.
\FIGURE{
\vspace*{-1cm}
\includegraphics[width=1.0\textwidth]{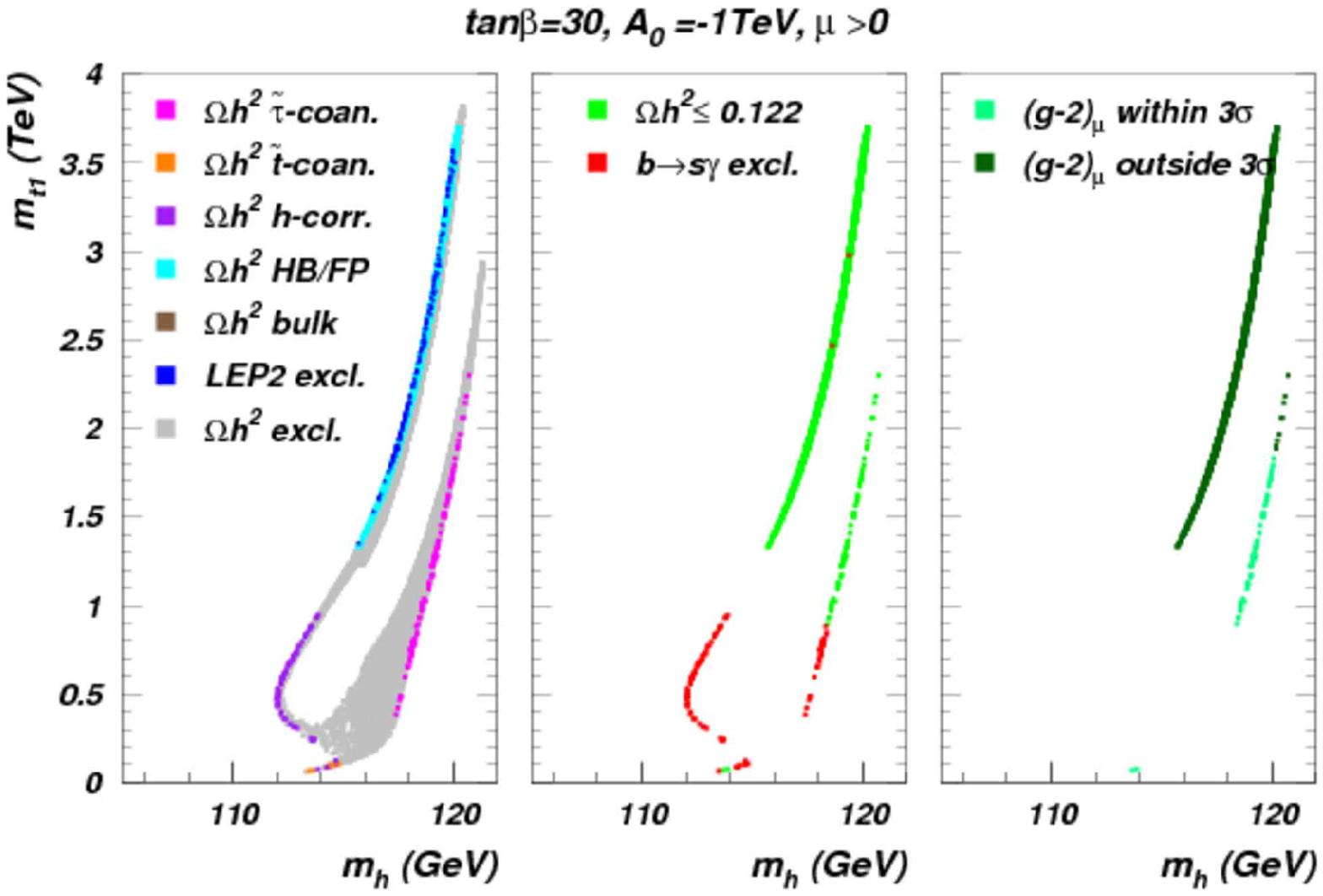}
\caption{\label{fig:mst1-mh-tb30-n1} 
Allowed region for top squark mass mass versus $m_h$.
The color code is  the same as in Fig.~\ref{fig:m0-mhf-tb10-n2}.}
}

\FIGURE{
\vspace*{-0.3cm}
\psfrag{q3}{{\normalsize {\boldmath $m_h~(GeV)$}}}
\psfrag{q4}{{\normalsize {\boldmath $m_{\tb_1}~(TeV)$}}}
\psfrag{q1}{{\small {\boldmath {\it (a)} $A_0=0.5~TeV$}}}
\psfrag{q2}{{\small {\boldmath {\it (b)} $A_0=0$}}}
\psfrag{q5}{{\small {\boldmath {\it (c)} $A_0=-1~TeV$}}}
\psfrag{q6}{{\small {\boldmath {\it (d)} $A_0=-2~TeV$}}}
\includegraphics[height=0.43\textwidth,angle=270]{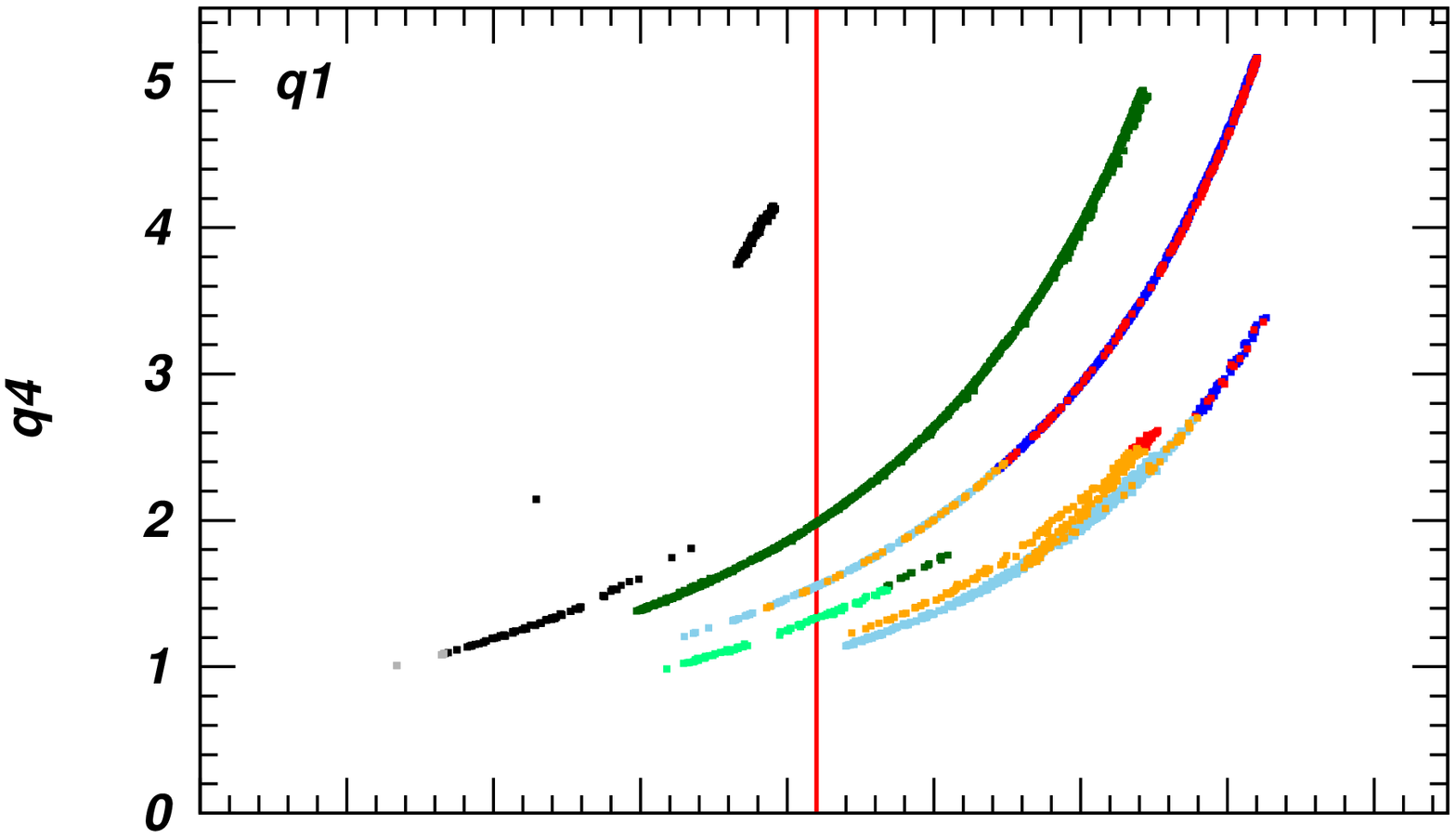}%
\includegraphics[height=0.43\textwidth,angle=270]{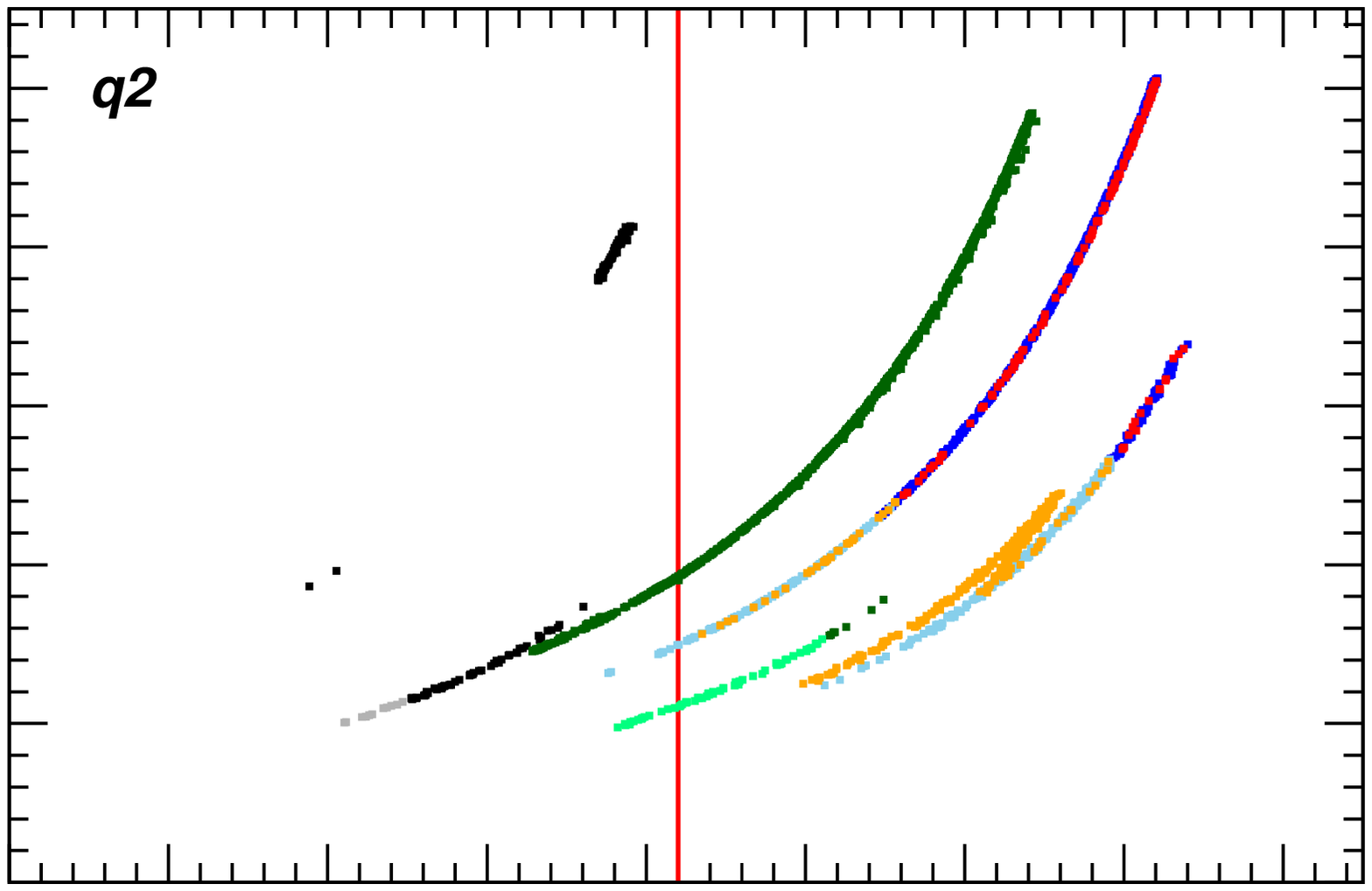}\vspace{-0.9cm}\\\hspace*{-0.3cm}%
\includegraphics[height=0.43\textwidth,angle=270]{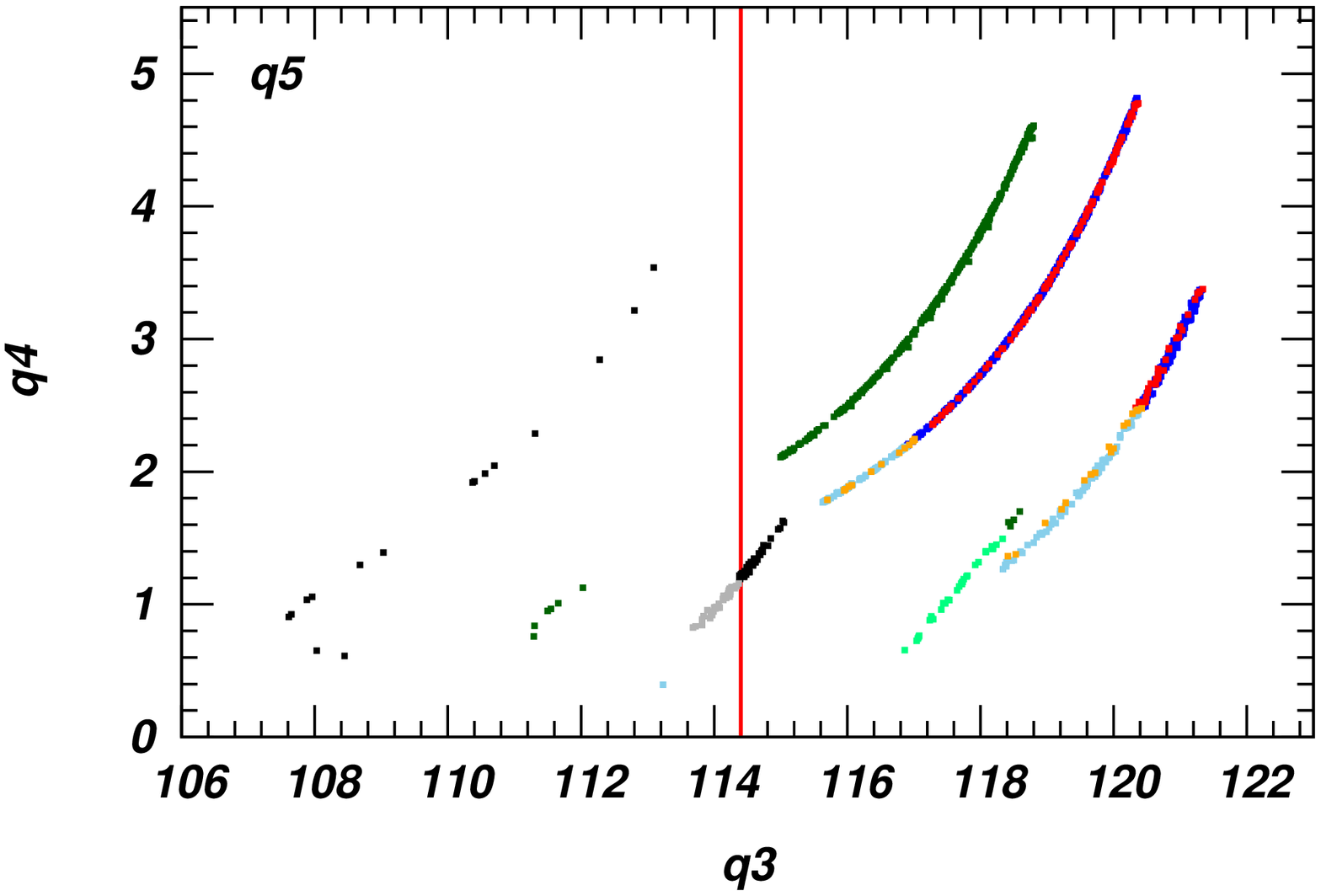}%
\includegraphics[height=0.43\textwidth,angle=270]{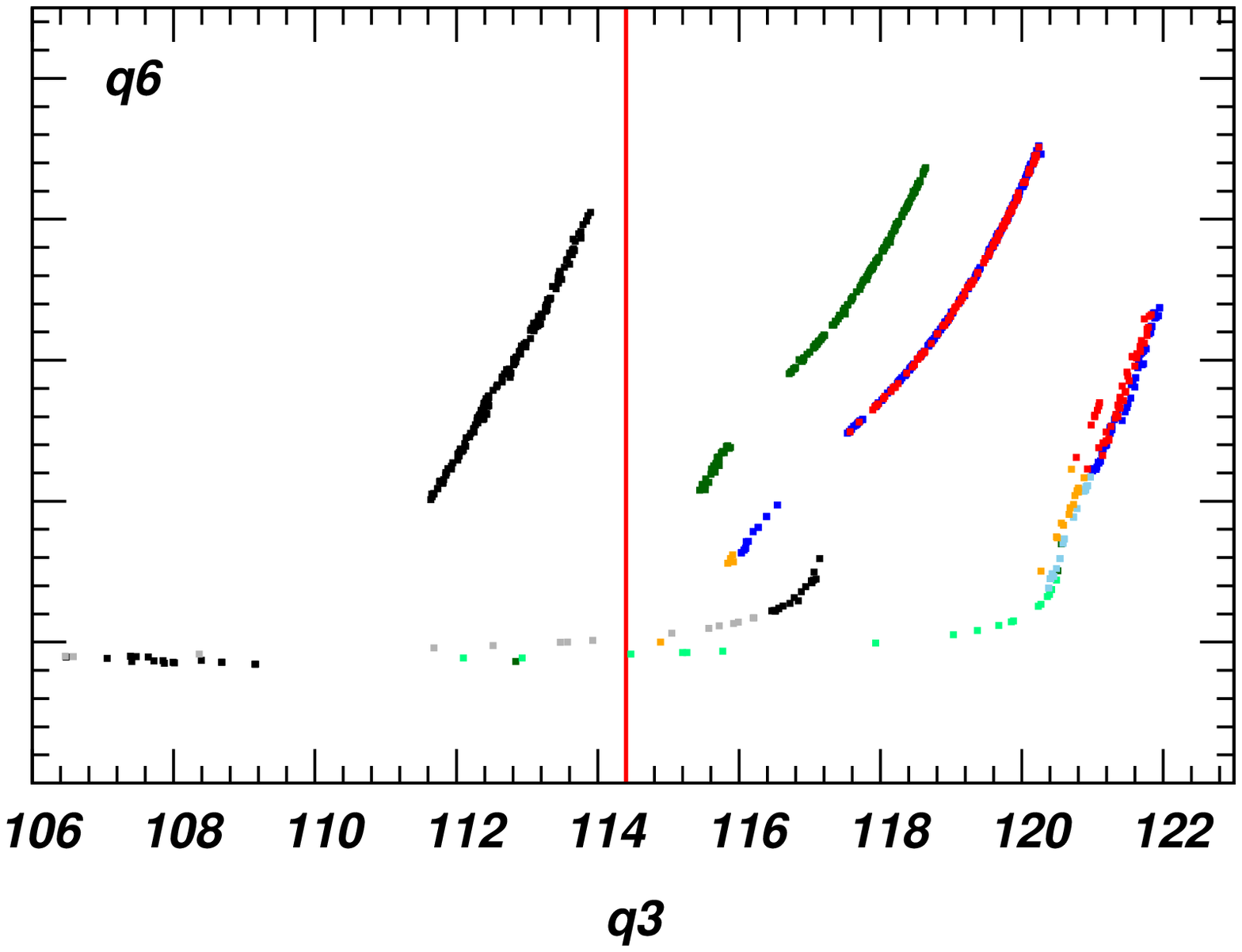}
\caption{\label{fig:mb-mh} Allowed region for bottom squark mass mass versus $m_h$.
The color code is  the same as in Fig.~\ref{fig:ma-mh}.}
}
\FIGURE{
\vspace*{-0.3cm}
\psfrag{q3}{{\normalsize {\boldmath $m_h~(GeV)$}}}
\psfrag{q4}{{\normalsize {\boldmath $m_{\ttau_1}~(TeV)$}}}
\psfrag{q1}{{\small {\boldmath {\it (a)} $A_0=0.5~TeV$}}}
\psfrag{q2}{{\small {\boldmath {\it (b)} $A_0=0$}}}
\psfrag{q5}{{\small {\boldmath {\it (c)} $A_0=-1~TeV$}}}
\psfrag{q6}{{\small {\boldmath {\it (d)} $A_0=-2~TeV$}}}
\includegraphics[height=0.43\textwidth,angle=270]{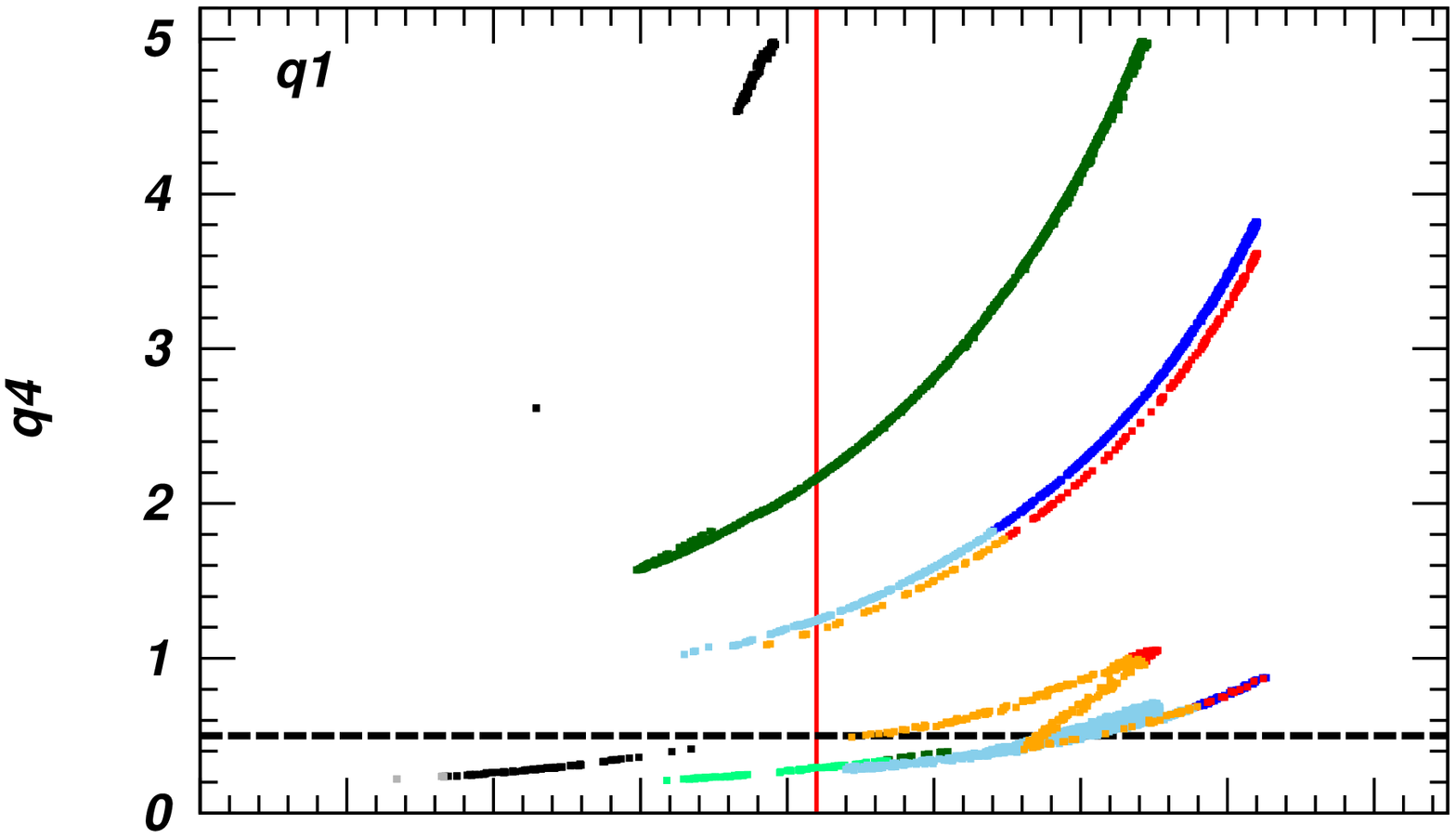}%
\includegraphics[height=0.43\textwidth,angle=270]{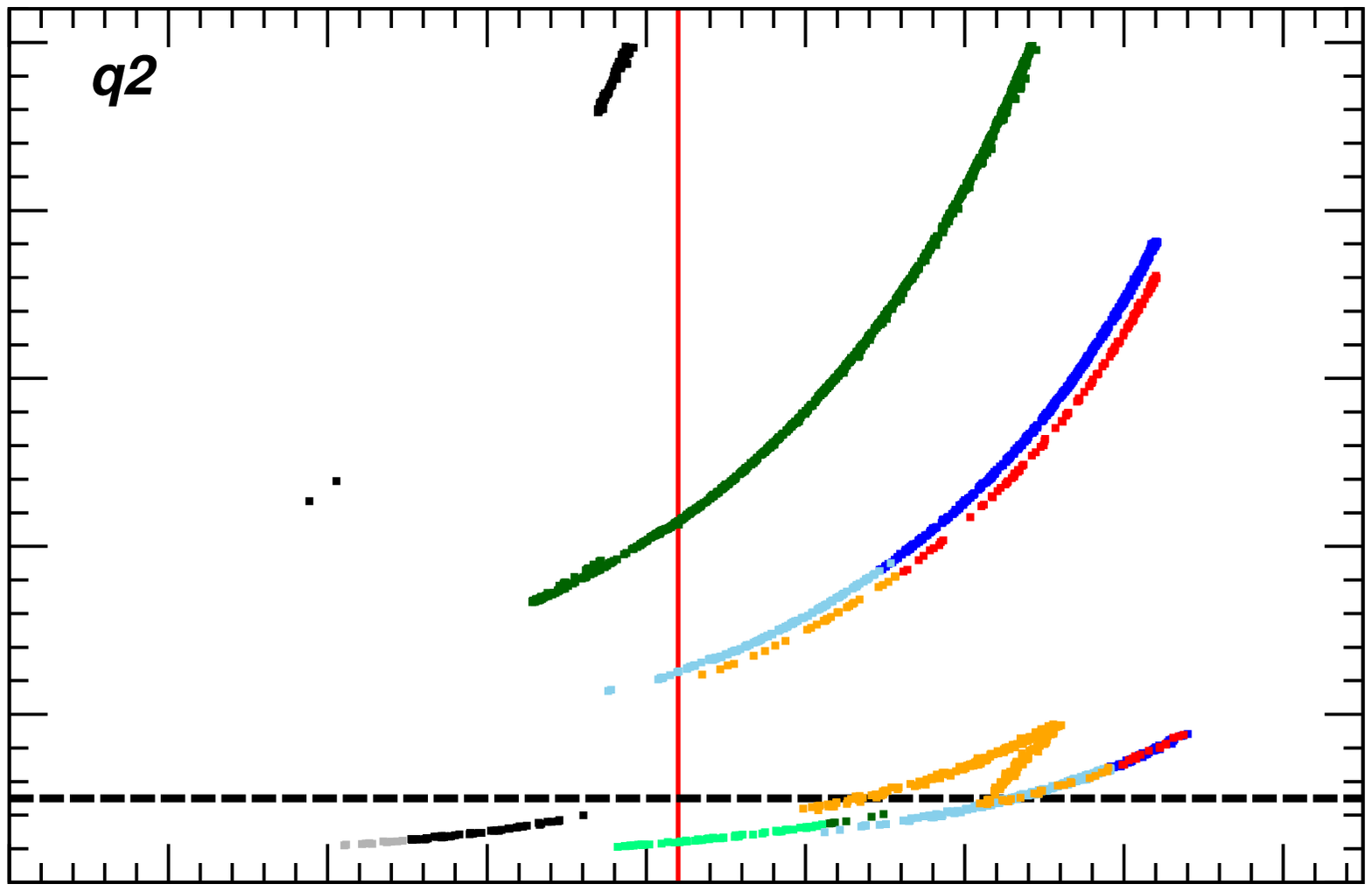}\vspace{-0.9cm}\\\hspace*{-0.3cm}%
\includegraphics[height=0.43\textwidth,angle=270]{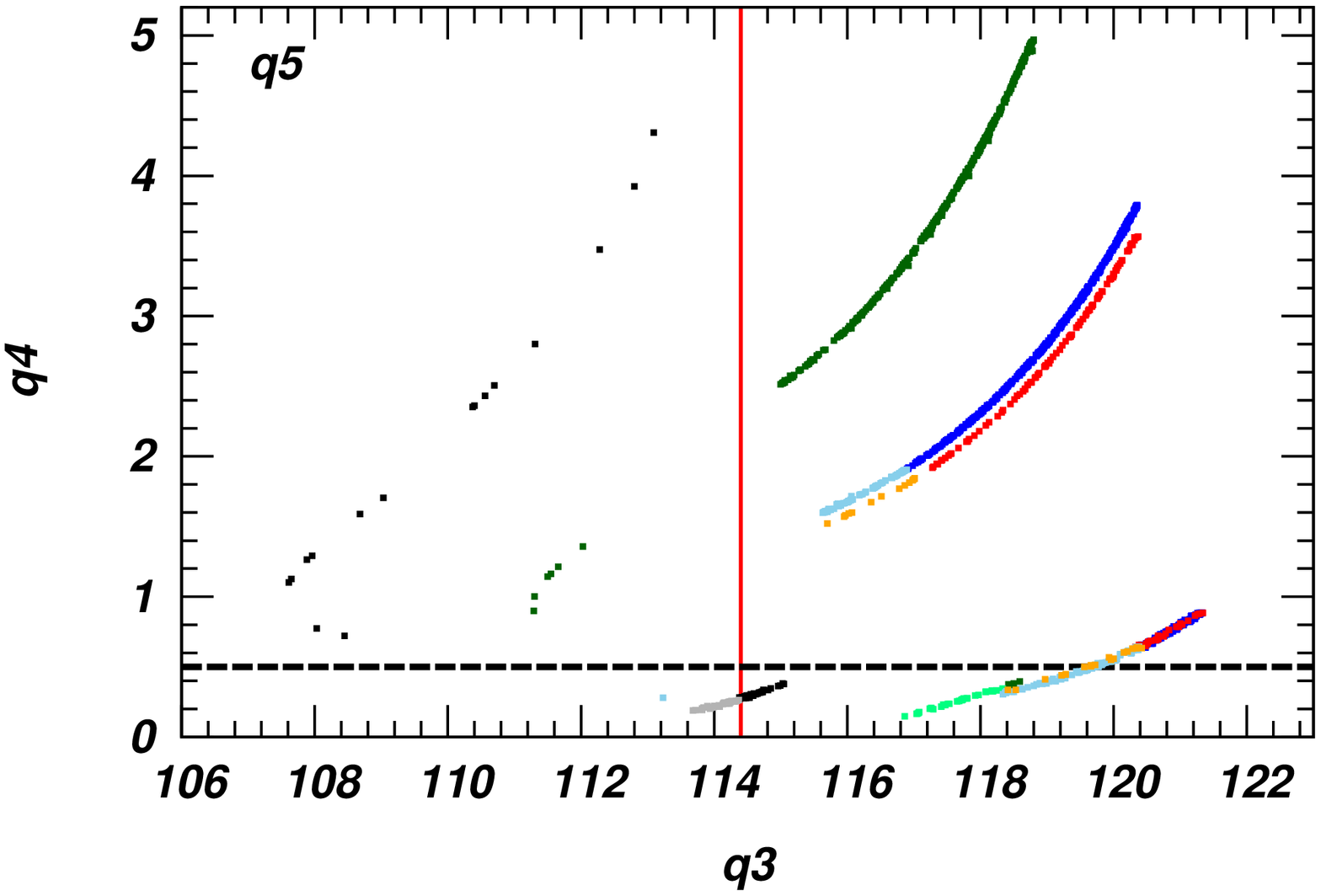}%
\includegraphics[height=0.43\textwidth,angle=270]{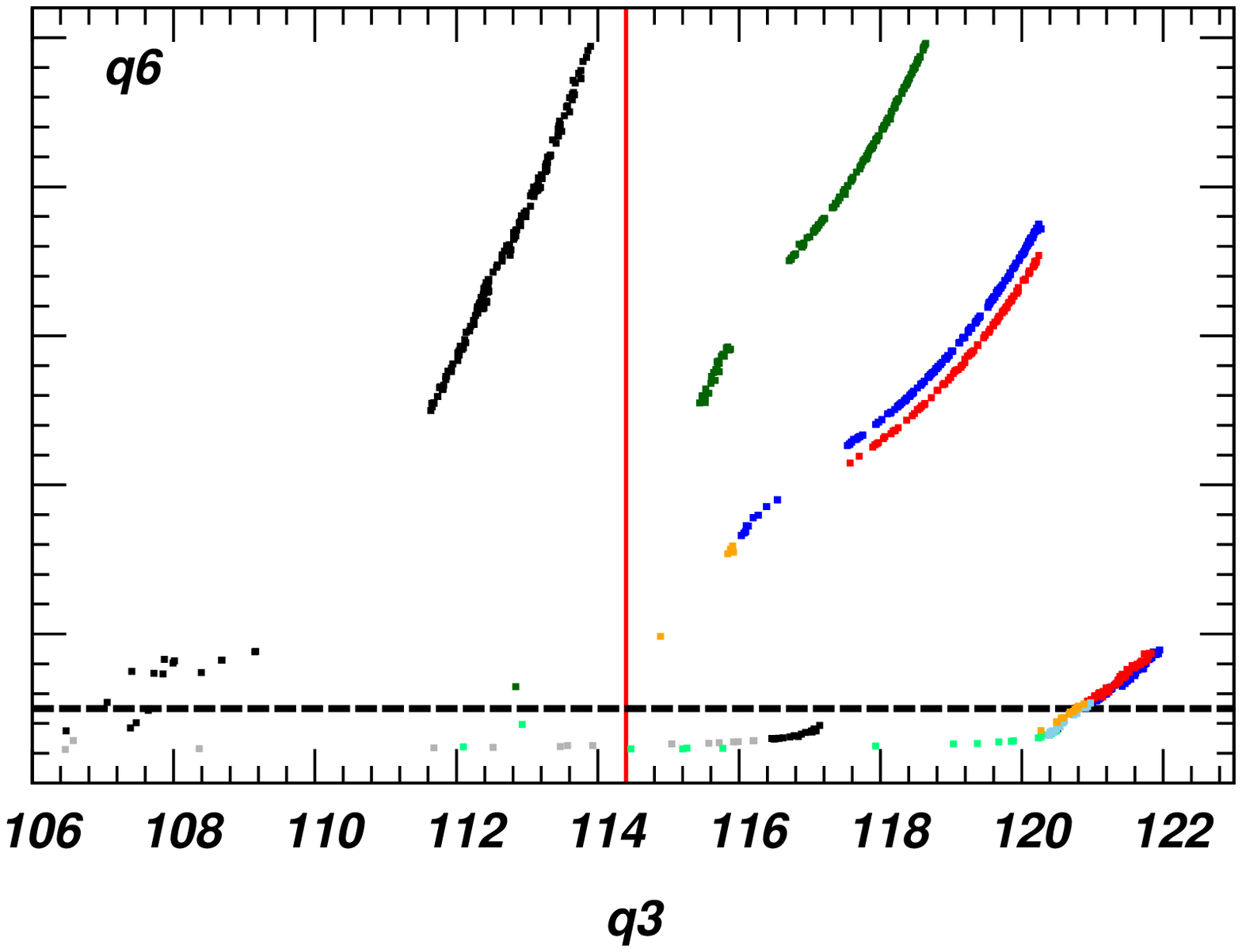}
\caption{\label{fig:mta-mh} Allowed region for stau mass versus $m_h$. 
The color code is  the same as in Fig.~\ref{fig:ma-mh}.
Dashed horizontal lines represent approximate reach of ILC1000.}
}

Next, in Figs.~\ref{fig:mb-mh} and ~\ref{fig:mta-mh} we present 
correlations between the bottom squark mass and $m_h$, and stau mass and
$m_h$,  respectively. One can observe qualitatively similar pattern of 
correlations in these two figures, where again, as in 
Fig.~\ref{fig:mt-mh}, the upper curve corresponds to the \fp  ~region, while the lower curve corresponds to the stau-coannihilation one. 
At large $\tan\beta$ and $A_0=0.5$~TeV and  $A_0=0$, the $A-$funnel region
opens up, which is reflected in  the behavior of the orange curve. 
Increasing $\tan\beta$ boosts up the bottom and tau Yukawa couplings, which
reduces the sbottom and stau SSB masses through RGE effects, and increases L-R
mixing; 
both effects tend to reduce $\tb_1$ and $\ttau_1$ masses. This
effect is more noticeable for staus, as seen from the wider separation of the
\fp ~bands in Fig.~\ref{fig:mta-mh}. Notice that in Fig.~\ref{fig:mb-mh}, the minimal value of
bottom squark mass is $\sim 500$~GeV, which is significantly above the
current Tevatron bound~\cite{CDF:stop,D0:sbottom}. Kinematical limit for
$\tw_1 \tw_1$ pair production at $\sqrt{s}=1000$~GeV, shown by the dashed
horizontal line, indicates an approximate reach of ILC1000~\cite{Baer:2003ru,Baer:2004zk,Baer:1996vd}.

It is quite informative to also  consider the correlation plots for a  {\it
fixed  value of $\tan\beta$} and different $A_0$ values in the same frame. 
%
\FIGURE{
\vspace*{-0.3cm}
\psfrag{p3}{{\normalsize {\boldmath $m_h~(GeV)$}}}
\psfrag{p4}{{\normalsize {\boldmath $m_A~(TeV)$}}}
\psfrag{p1}{{\small {\boldmath {\it (a)} $\tan\beta=5$}}}
\psfrag{p2}{{\small {\boldmath {\it (b)} $\tan\beta=10$}}}
\psfrag{p5}{{\small {\boldmath {\it (c)} $\tan\beta=50$}}}
\psfrag{p6}{{\small {\boldmath {\it (d)} $\tan\beta=53$}}}
\includegraphics[height=0.45\textwidth,angle=270]{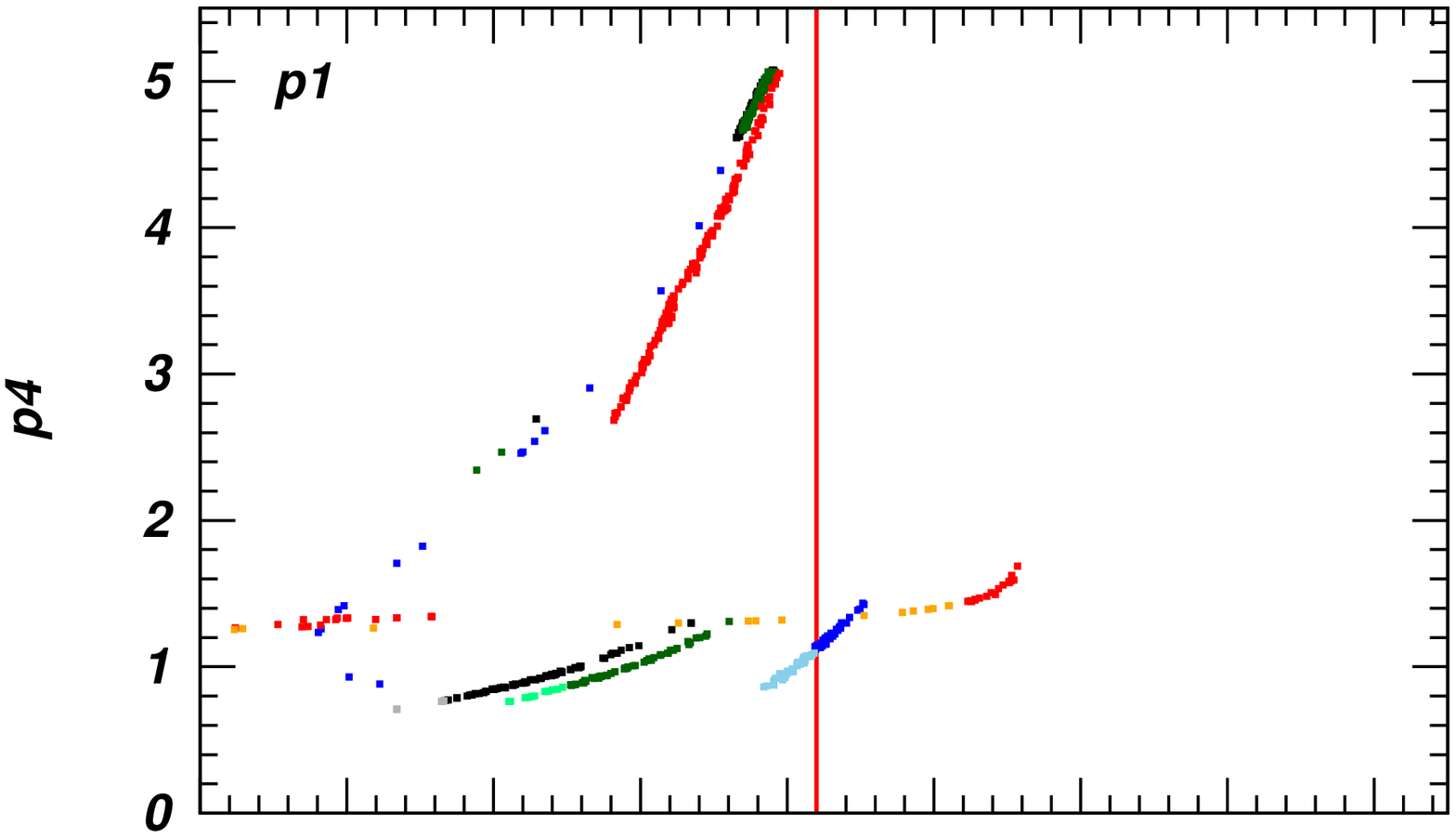}%
\includegraphics[height=0.45\textwidth,angle=270]{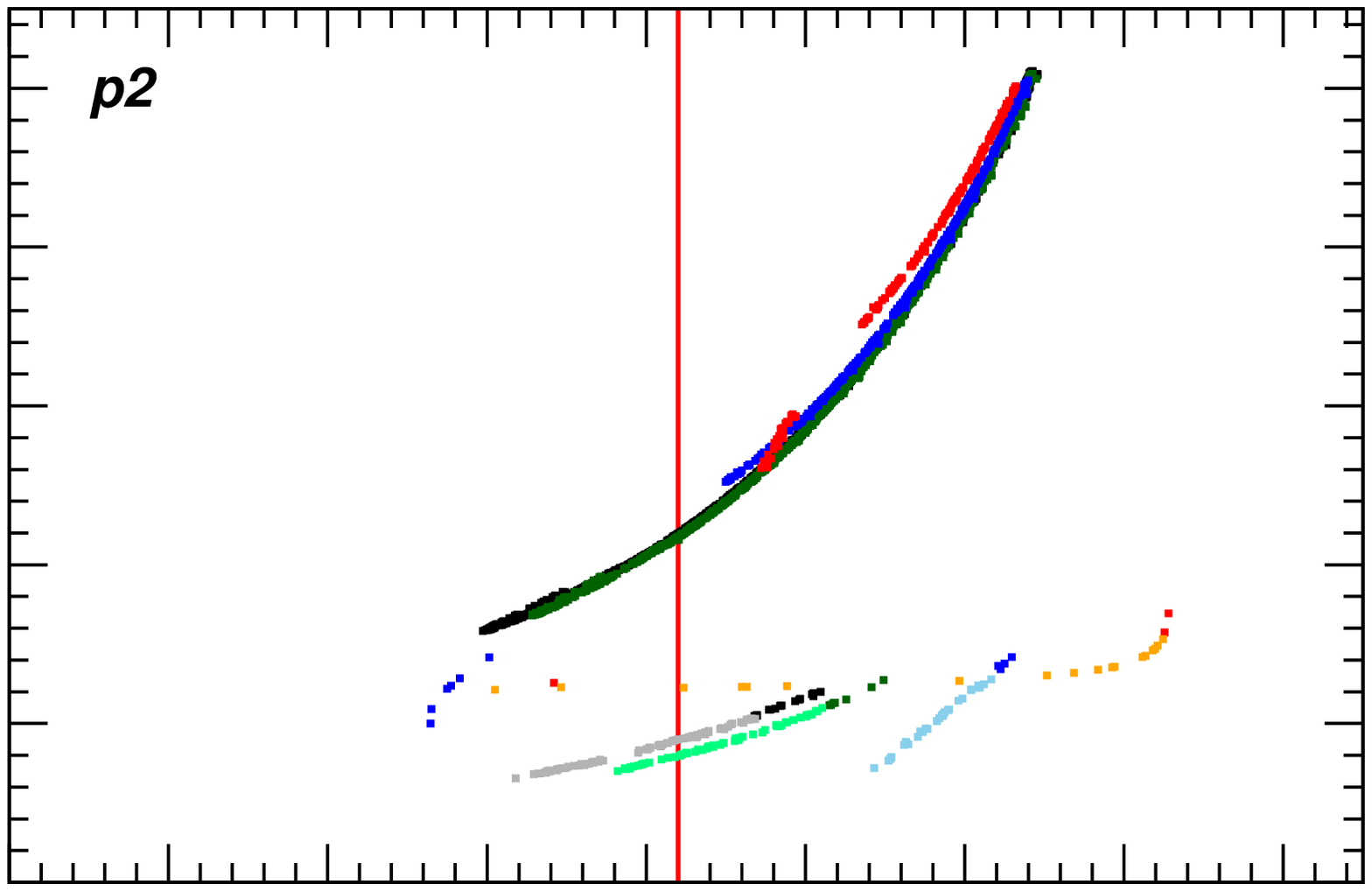}\vspace{-0.9cm}\\\hspace*{-0.1cm}%
\includegraphics[height=0.45\textwidth,angle=270]{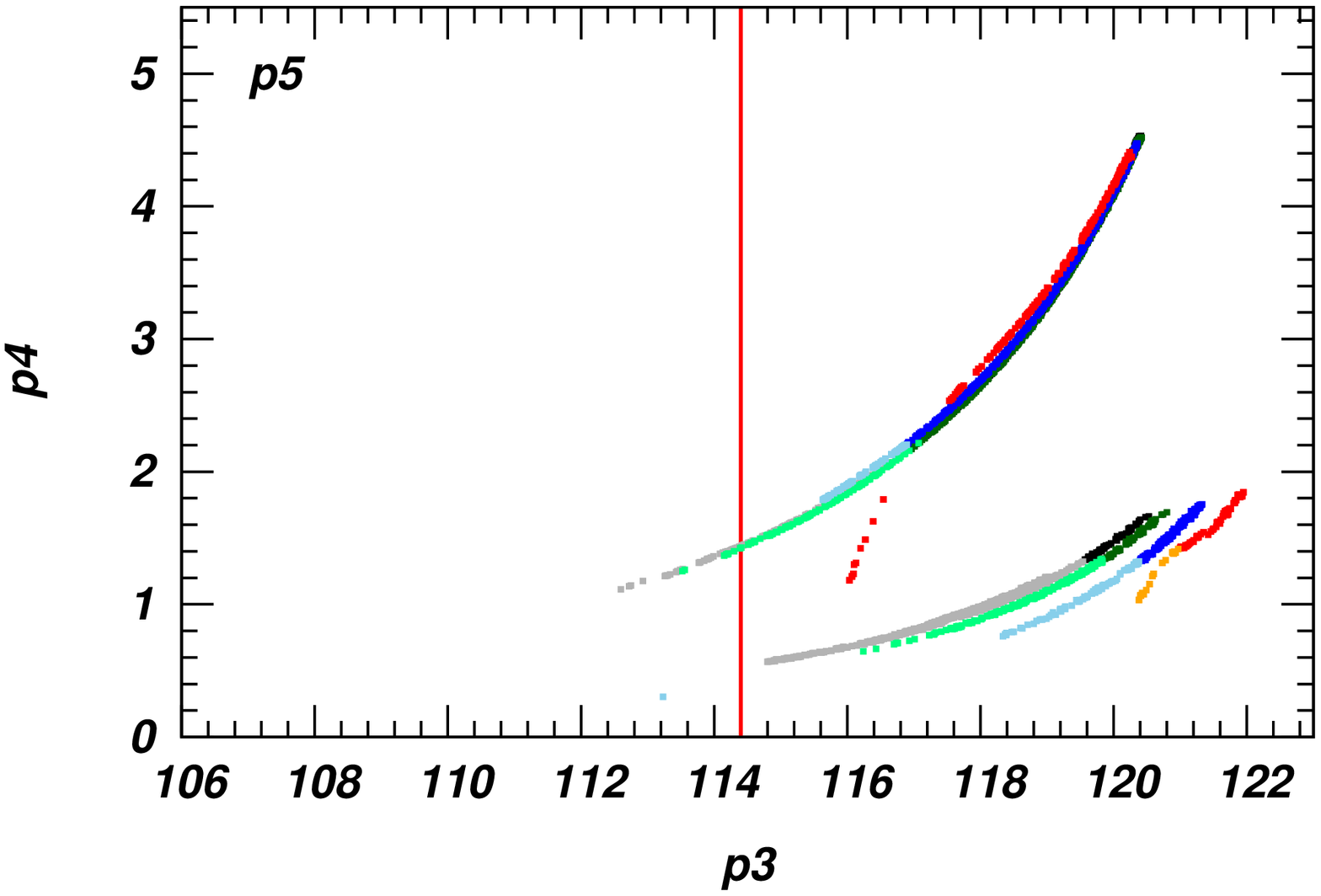}%
\includegraphics[height=0.45\textwidth,angle=270]{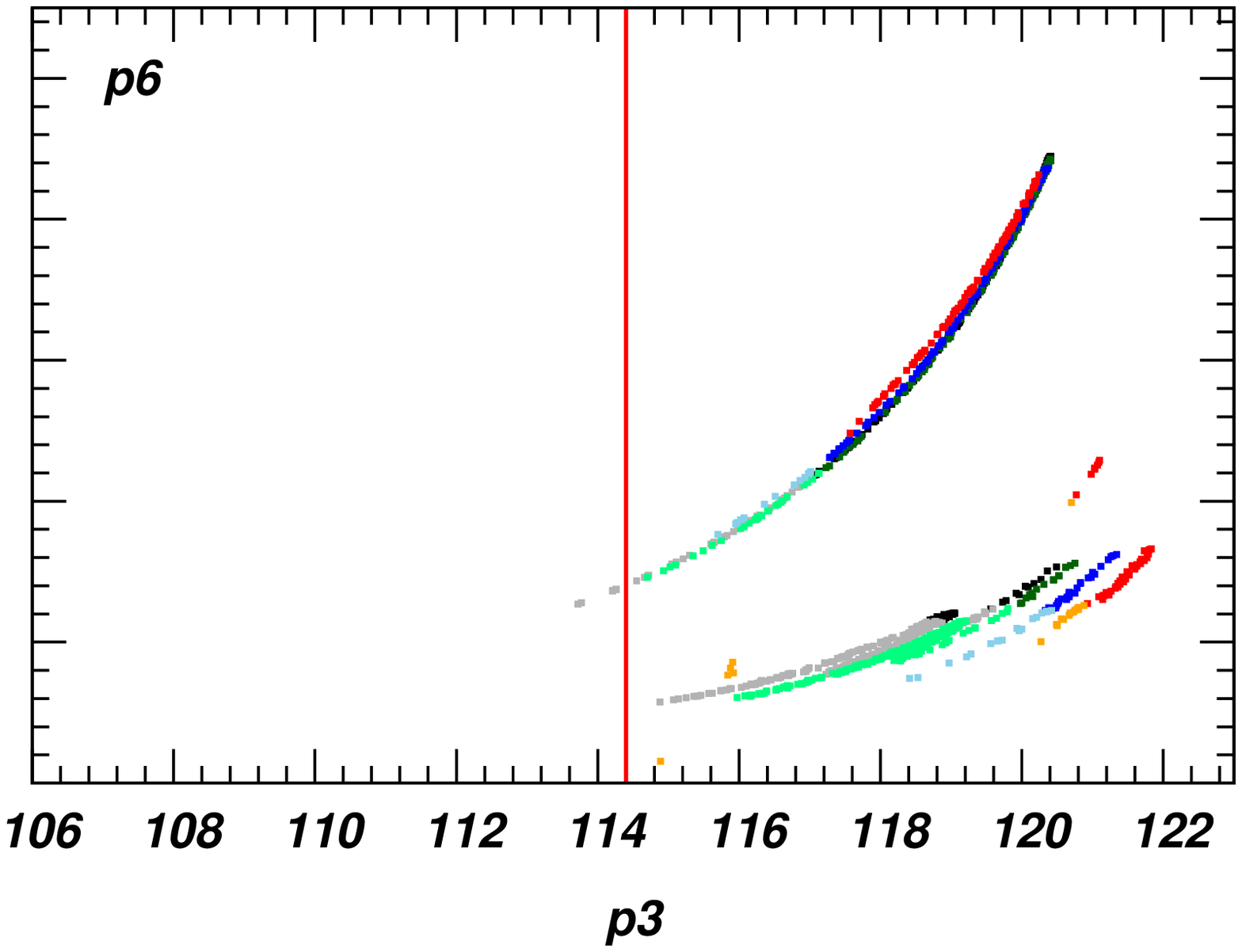}
\caption{\label{ma-mh-fix_tanb}Allowed region for CP-odd Higgs boson mass versus $m_h$.
Gray, light green, light blue, and orange correspond to  
$A_0=0.5,\ 0,\ -1$ and $-2$~TeV, 
respectively and satisfy $3 \sigma$ bound on $\Delta a_{\mu}$ (\ref{eq:g-2}).
 Black, dark green, blue and red correspond to 
$A_0=0.5,\ 0,\ -1$ and $-2$~TeV and have $\Delta a_{\mu}$ outside the $3 \sigma$ range.} 
 }
For example, in Fig.~\ref{ma-mh-fix_tanb} we present the allowed region for the CP-odd
Higgs mass versus $m_h$. Gray, light green, light blue, and orange regions 
correspond to $A_0=0.5,\ 0,\ -1$ and $-2$~TeV  respectively and
satisfy the $3 \sigma$ bound (\ref{eq:g-2}) on $\Delta a_{\mu}$; black, dark green, 
blue and red  colors correspond to regions outside the $\Delta a_{\mu}$ bound. 
If we assume that $\tan\beta$ is measured, which can be done at the LHC with an 
accuracy better than 20\% 
for low to intermediate values of $m_A$~\cite{ATLAS_TDR},
then the strong $m_A-m_h$ correlations above can be used for an indirect prediction of $m_A$,
in case the CP-odd Higgs boson is too heavy to be directly accessible. 
A 1$\%$ accuracy in the $m_h$ 
measurement would allow one  to estimate $m_A$  with a precision of 5-10$\%$.

One can also use these correlations to determine (or at least strongly
constrain) the trilinear coupling $A_0$, whose direct measurements are quite
problematic. In Fig.~\ref{mg-mh-fixtanb}, we present correlations  between the gluino
and light higgs boson masses, where we group 
various  $A_0$ values for fixed $\tan\beta$ in each frame. For
each $A_0$ value, the upper band corresponds to the coannihilation region
and the lower one to the HB/FP region. The two groups of bands are very
close, but if we discover that we are in the HB/FP or in the 
coannihilation region, half of the curves will  be  removed and the remaining
ones are well separated for $\tan\beta \gtrsim 10$, which can be used to extract the $A_0$ value.
%
\FIGURE{
\vspace*{-0.3cm}
\psfrag{p3}{{\normalsize {\boldmath $m_{h}~(GeV)$}}}
\psfrag{p4}{{\normalsize {\boldmath $m_{\tg}~(TeV)$}}}
\psfrag{p1}{{\small {\boldmath {\it (a)} $\tan\beta=5$}}}
\psfrag{p2}{{\small {\boldmath {\it (b)} $\tan\beta=10$}}}
\psfrag{p5}{{\small {\boldmath {\it (c)} $\tan\beta=50$}}}
\psfrag{p6}{{\small {\boldmath {\it (d)} $\tan\beta=53$}}}
\includegraphics[height=0.44\textwidth,angle=270]{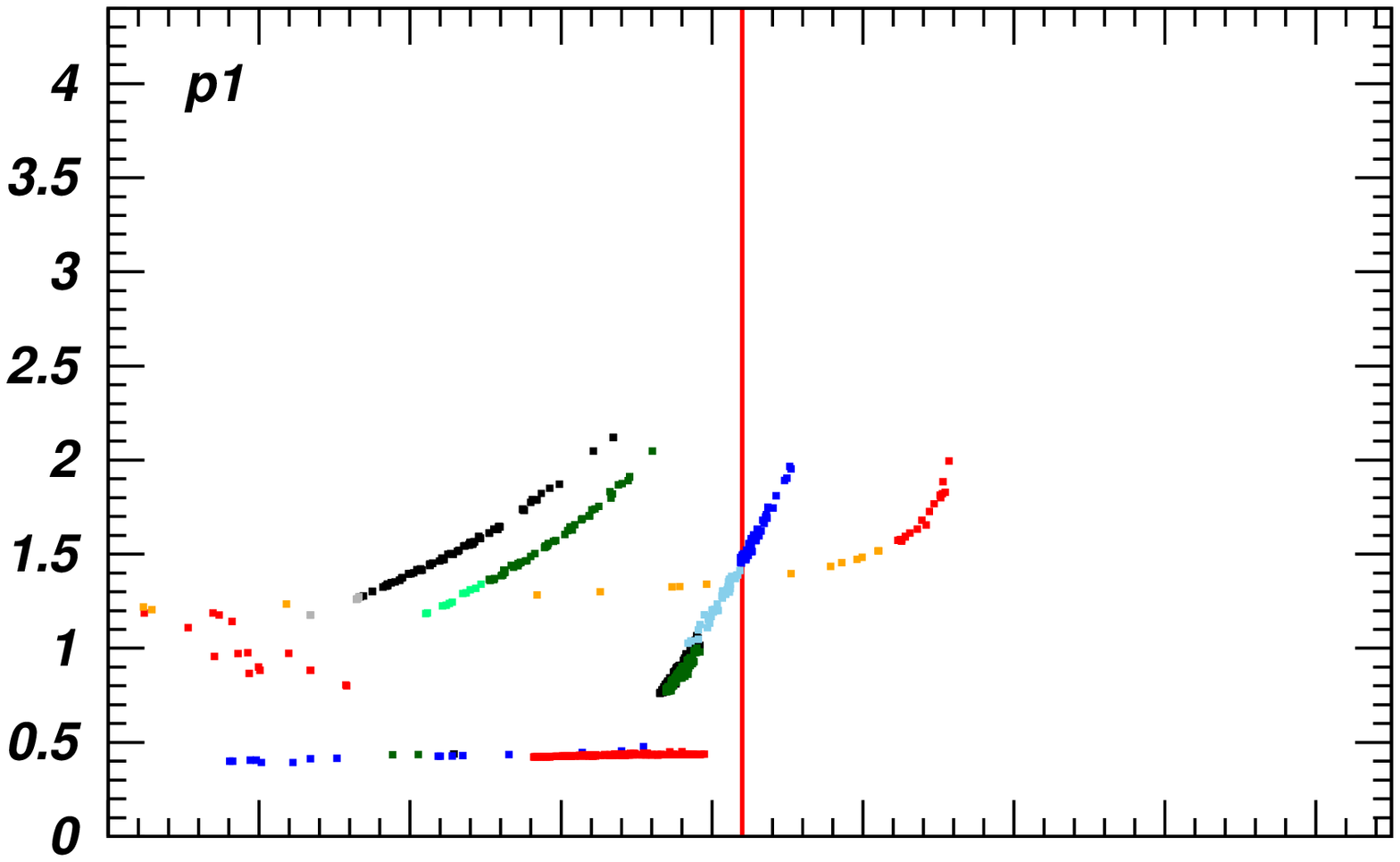}%
\includegraphics[height=0.44\textwidth,angle=270]{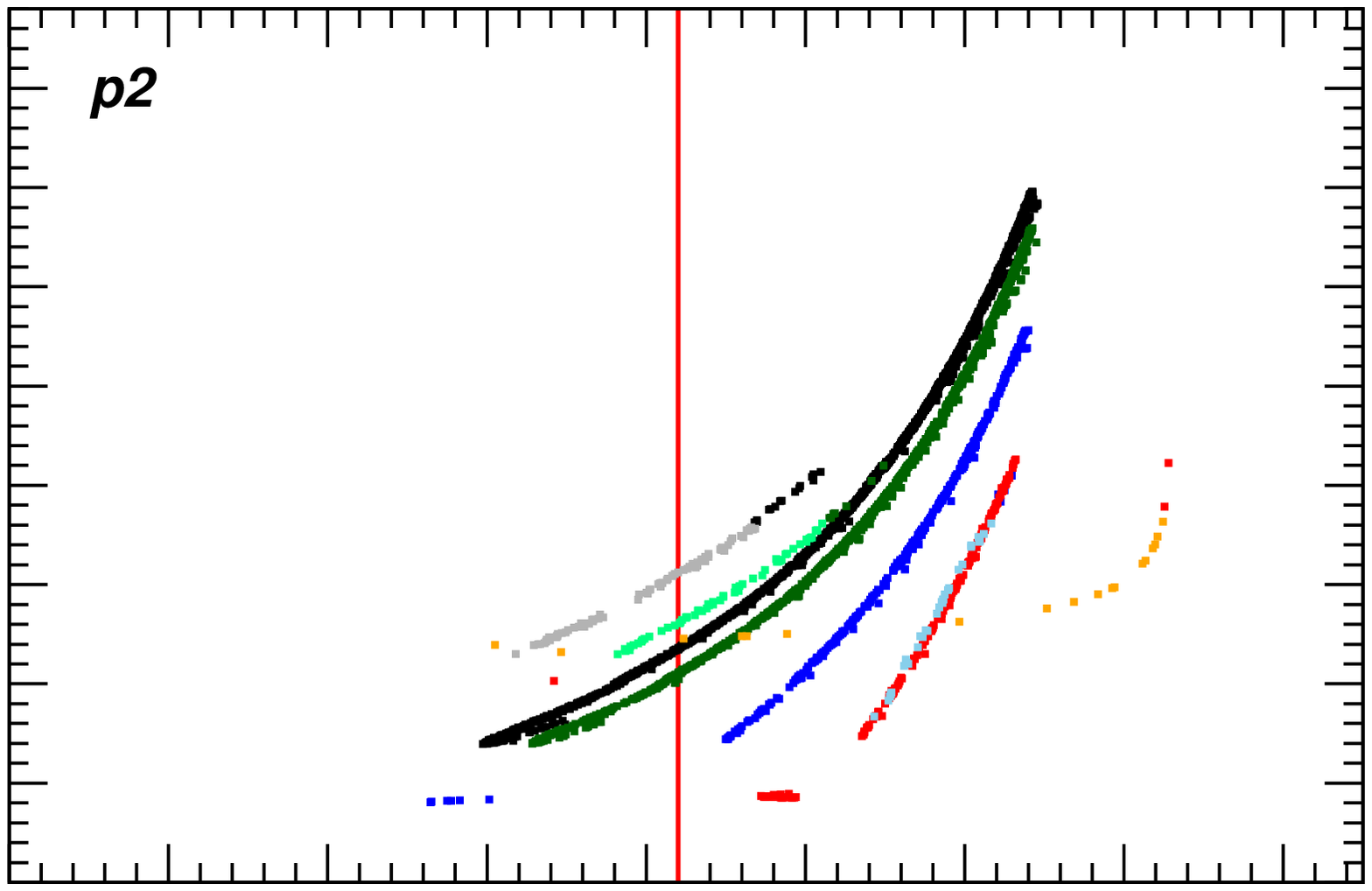}\vspace{-0.9cm}\\\hspace*{-0.2cm}%
\includegraphics[height=0.44\textwidth,angle=270]{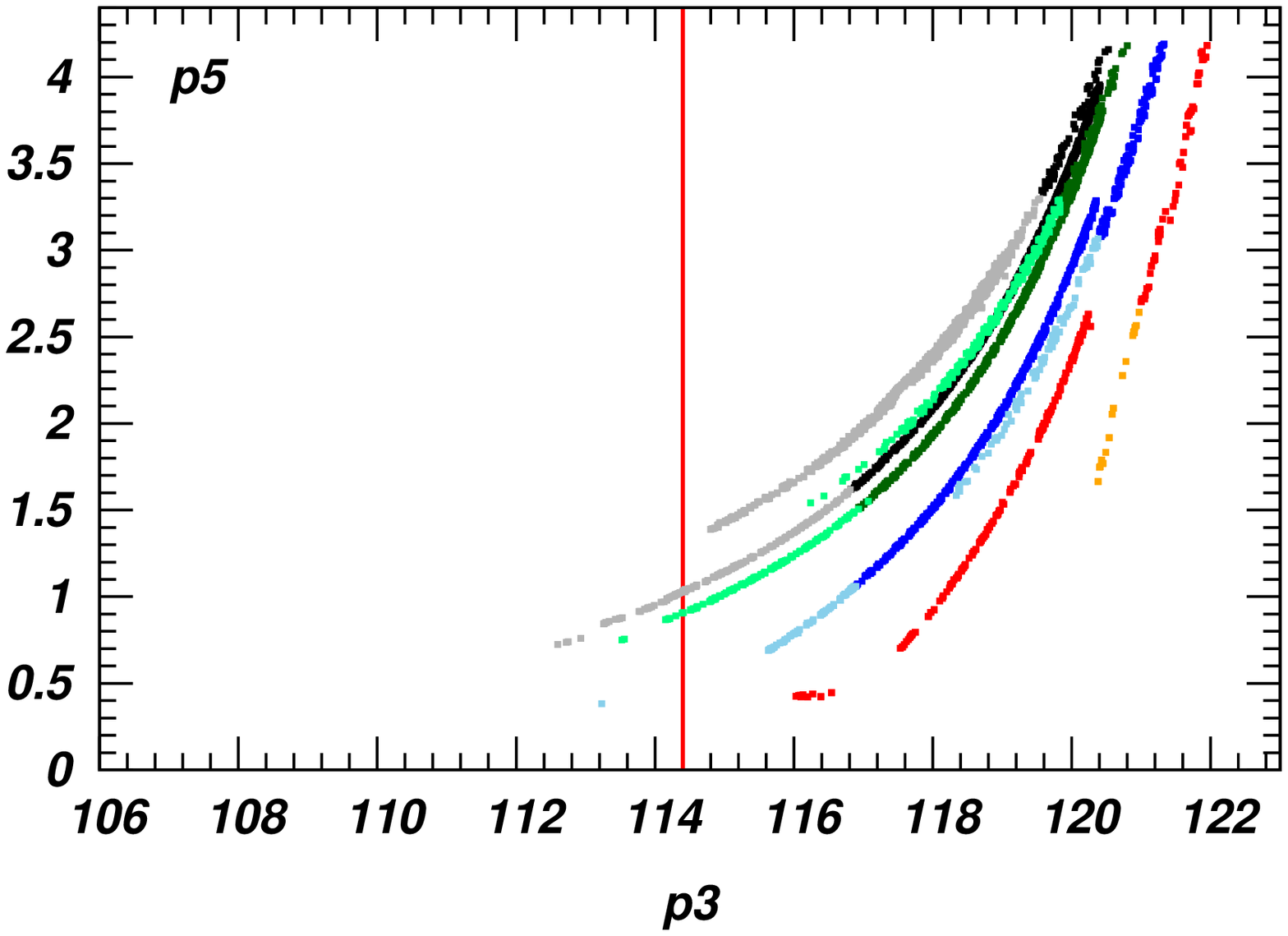}%
\includegraphics[height=0.44\textwidth,angle=270]{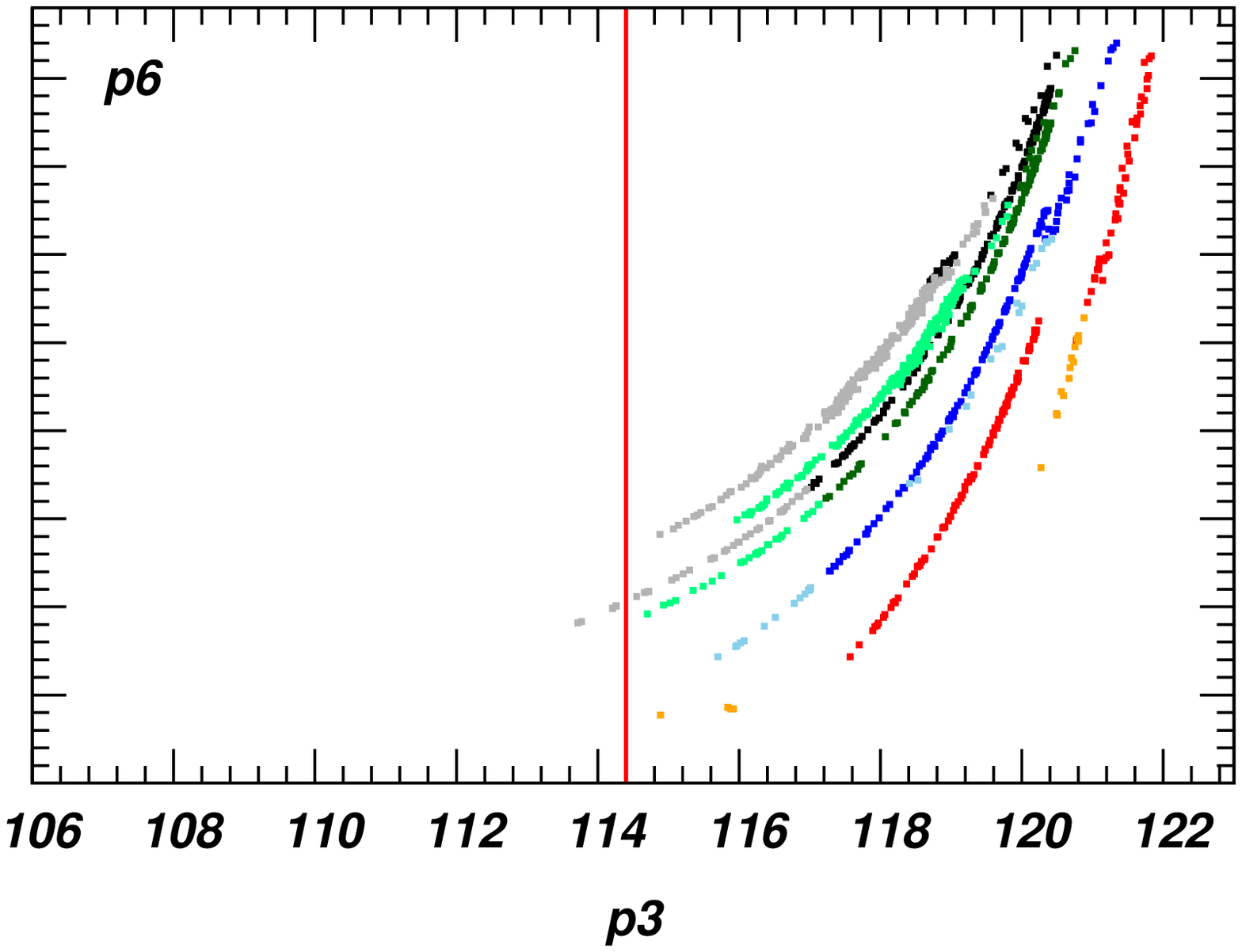}
\caption{Allowed region for gluino mass versus $m_h$.
The color code is  the same as in Fig.~\ref{ma-mh-fix_tanb}.
\label{mg-mh-fixtanb}}
}
\FIGURE{
\vspace*{-0.3cm}
\psfrag{p3}{{\normalsize {\boldmath $m_{\tz_1}~(TeV)$}}}
\psfrag{p4}{{\normalsize {\boldmath $m_{A}~(TeV)$}}}
\psfrag{p1}{{\small {\boldmath {\it (a)} $\tan\beta=5$}}}
\psfrag{p2}{{\small {\boldmath {\it (b)} $\tan\beta=10$}}}
\psfrag{p5}{{\small {\boldmath {\it (c)} $\tan\beta=50$}}}
\psfrag{p6}{{\small {\boldmath {\it (d)} $\tan\beta=53$}}}
\includegraphics[height=0.44\textwidth,angle=270]{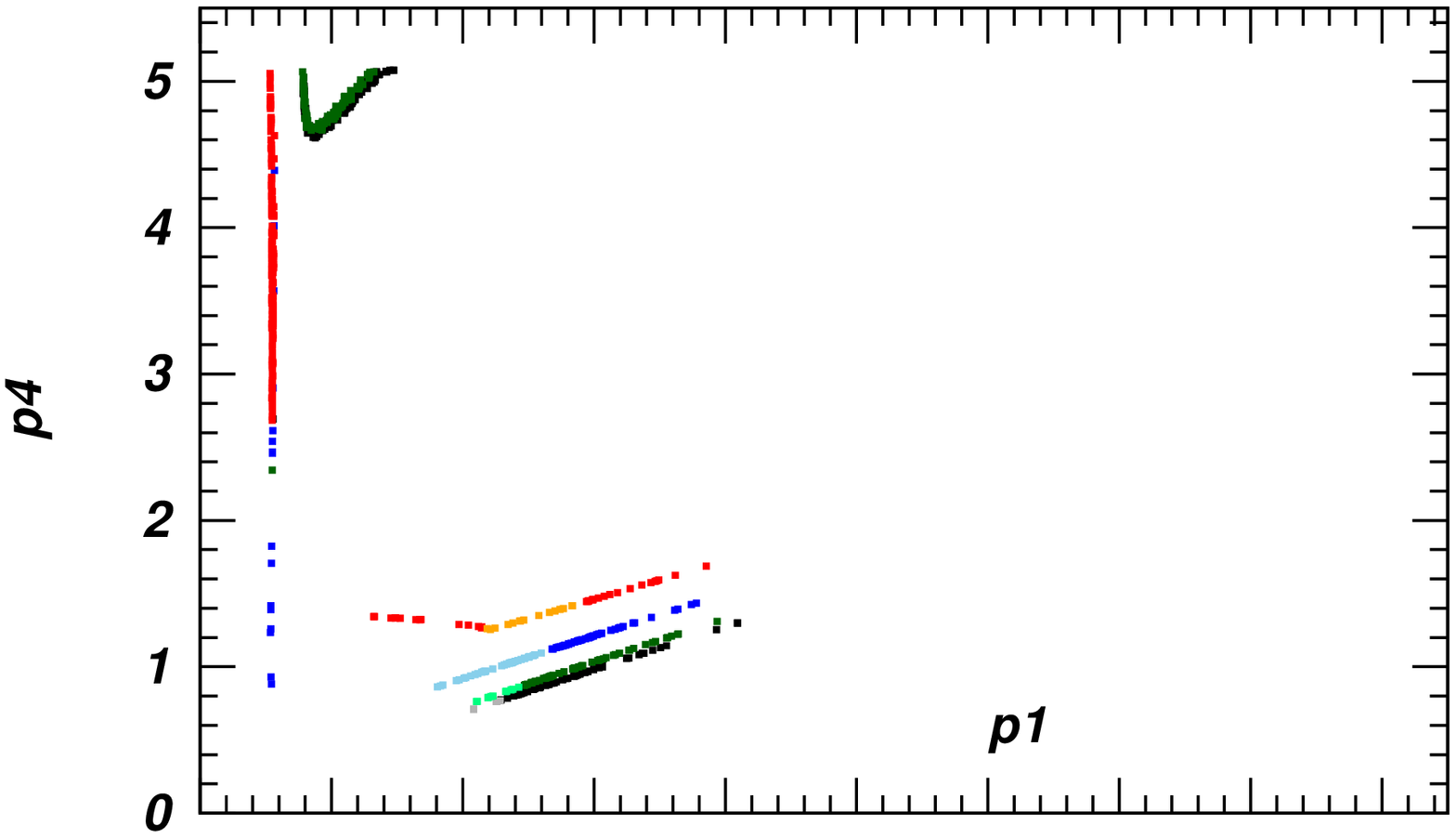}%
\includegraphics[height=0.44\textwidth,angle=270]{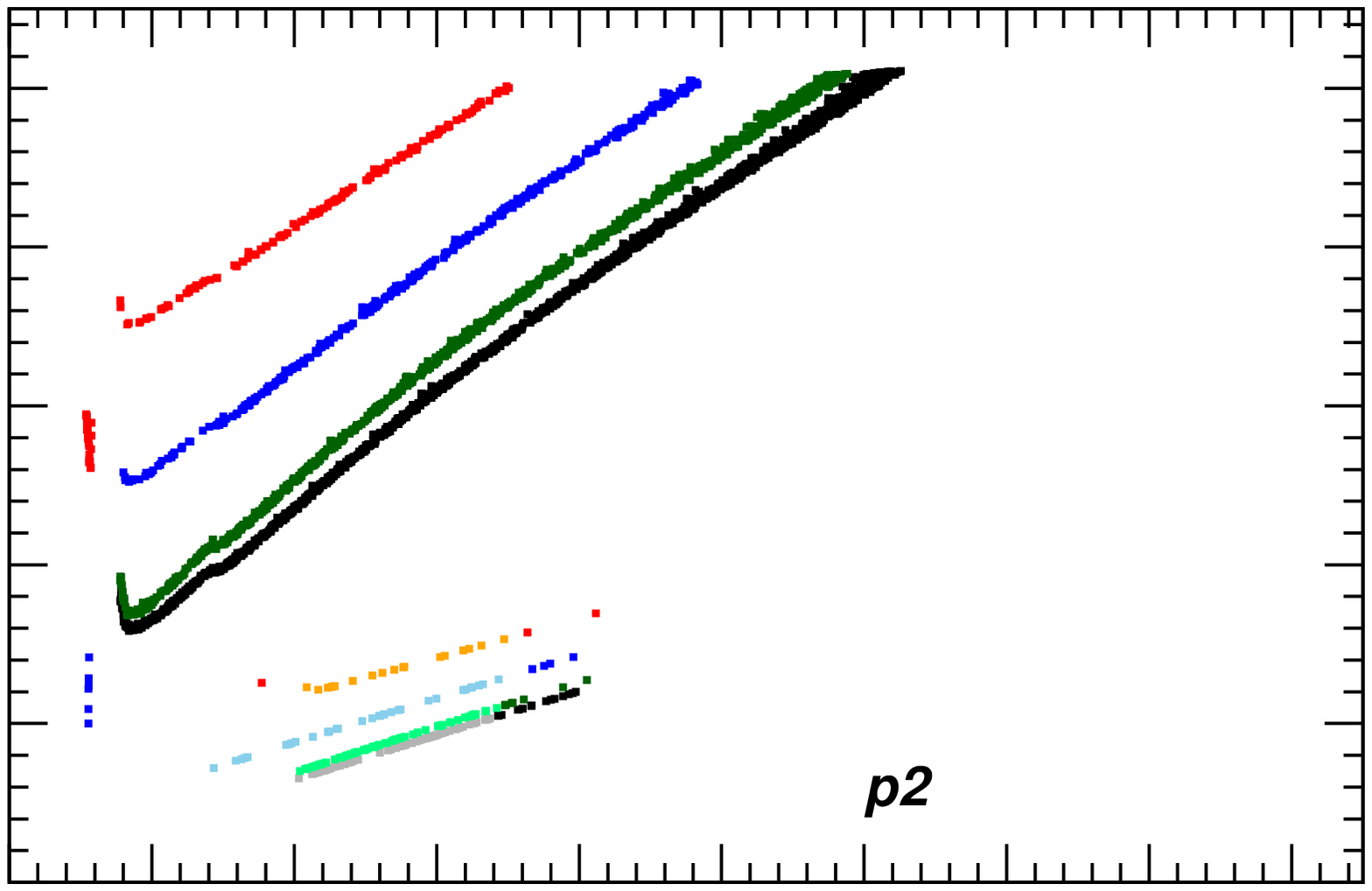}\vspace{-0.9cm}\\\hspace*{-0.2cm}%
\includegraphics[height=0.44\textwidth,angle=270]{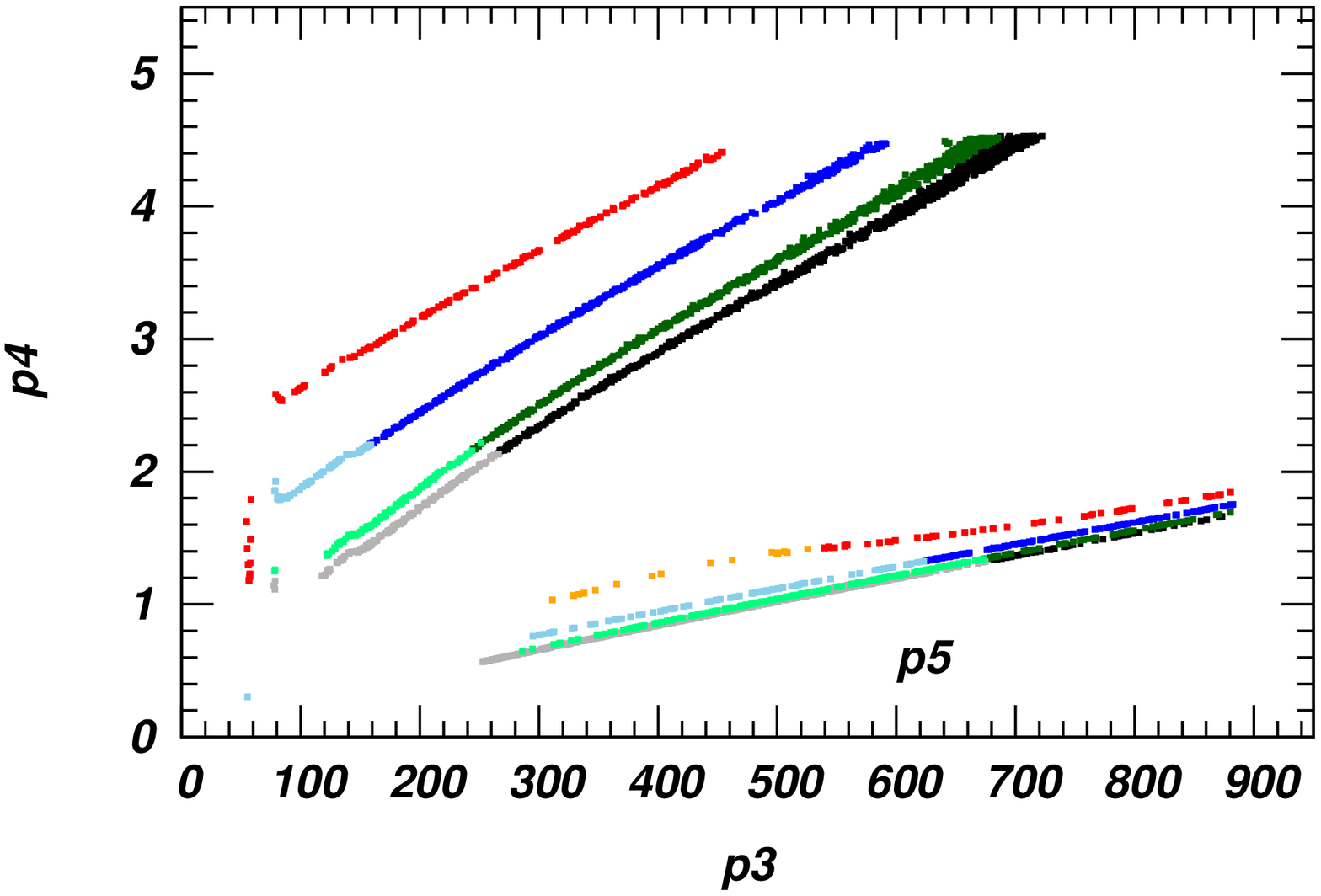}%
\includegraphics[height=0.44\textwidth,angle=270]{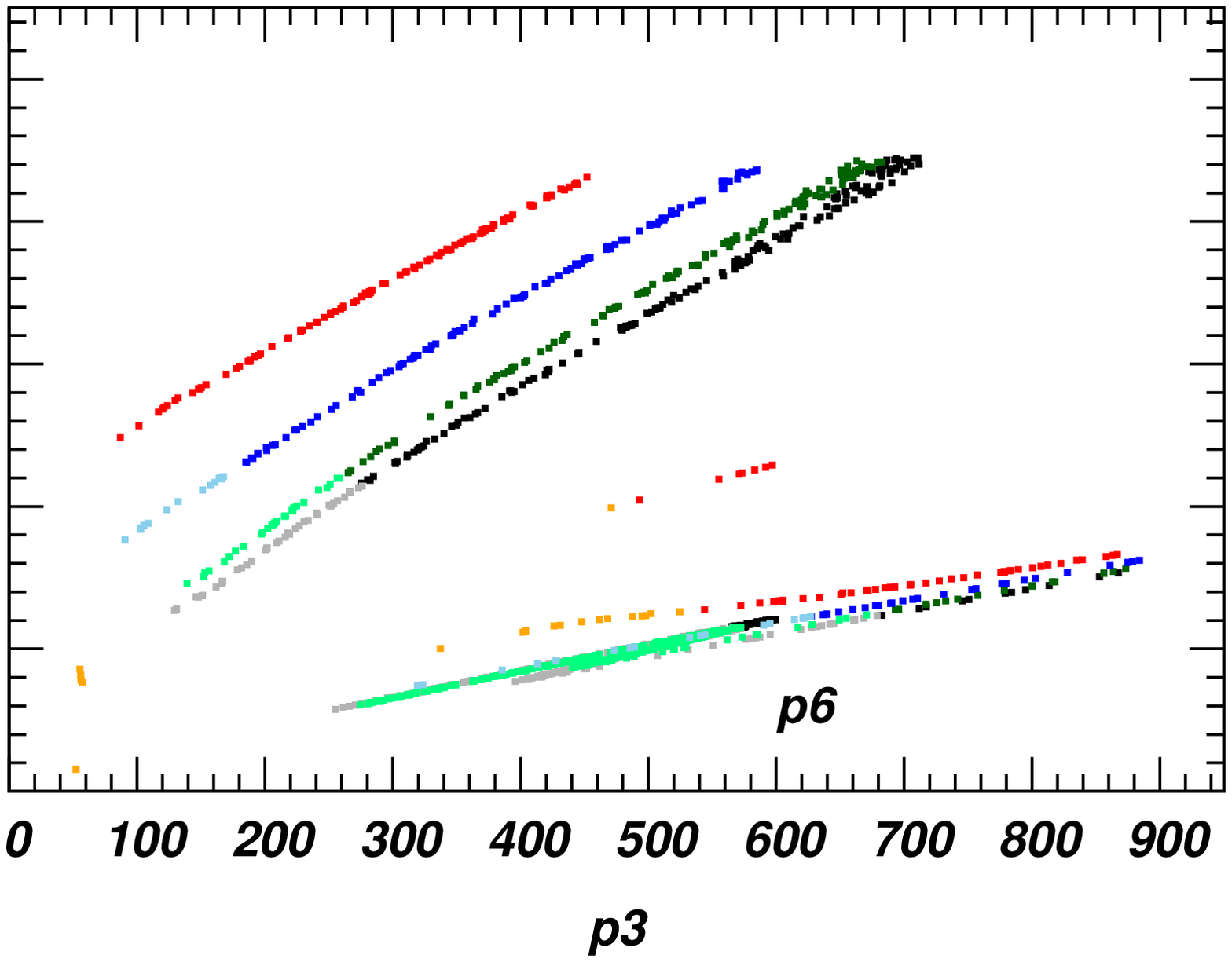}
\caption{Allowed region for $m_A$ versus $m_{\tz_1}$.
The color code is  the same as in Fig.~\ref{ma-mh-fix_tanb}.
\label{ma-mz1-fixtanb}}
}

Furthermore, the study of the correlations  between various SUSY masses
 would allow one to further delineate the SUSY parameter space. 
For example, in Fig.~\ref{ma-mz1-fixtanb} we plot
the allowed region in $(m_{\tz_1},m_A)$ plane. We see that the two-fold
solution bands form two well separated groups -- the lower one
representing the coannihilation region and the upper one corresponding
to the \fp  ~region. One should also notice a third type of bands, namely the vertical one,
located at low $m_{\tz_1}$ values and corresponding to the HF region.
Given $m_A$ (either from direct measurements or
extracted from Fig.~\ref{ma-mh-fix_tanb}) and $\tan\beta$ values, one
can deduce $A_0$ from the reconstructed neutralino mass. 
One should point out, however, that the LHC potential can be quite 
limited in the reconstruction of the heavy neutralino masses
in certain regions of the parameter space~\cite{Abdullin:1998pm}. 
Analyzing the direct DM detection (DD) rates, we have found
an important complementarity of DM detection experiments in this respect, 
since these experiments can cover a much  wider range of
neutralino masses.

\FIGURE{
\psfrag{p3}{{\normalsize {\boldmath $m_{\tz_1}~(GeV)$}}}
\psfrag{p4}{{\normalsize {\boldmath $\sigma_{SI}(\tz_1 p)~~(pb)$}}}
\psfrag{p1}{{\small {\boldmath {\it (a)} $\tan\beta=5$}}}
\psfrag{p2}{{\small {\boldmath {\it (b)} $\tan\beta=10$}}}
\psfrag{p5}{{\small {\boldmath {\it (c)} $\tan\beta=50$}}}
\psfrag{p6}{{\small {\boldmath {\it (d)} $\tan\beta=53$}}}
\includegraphics[height=0.45\textwidth,angle=270]{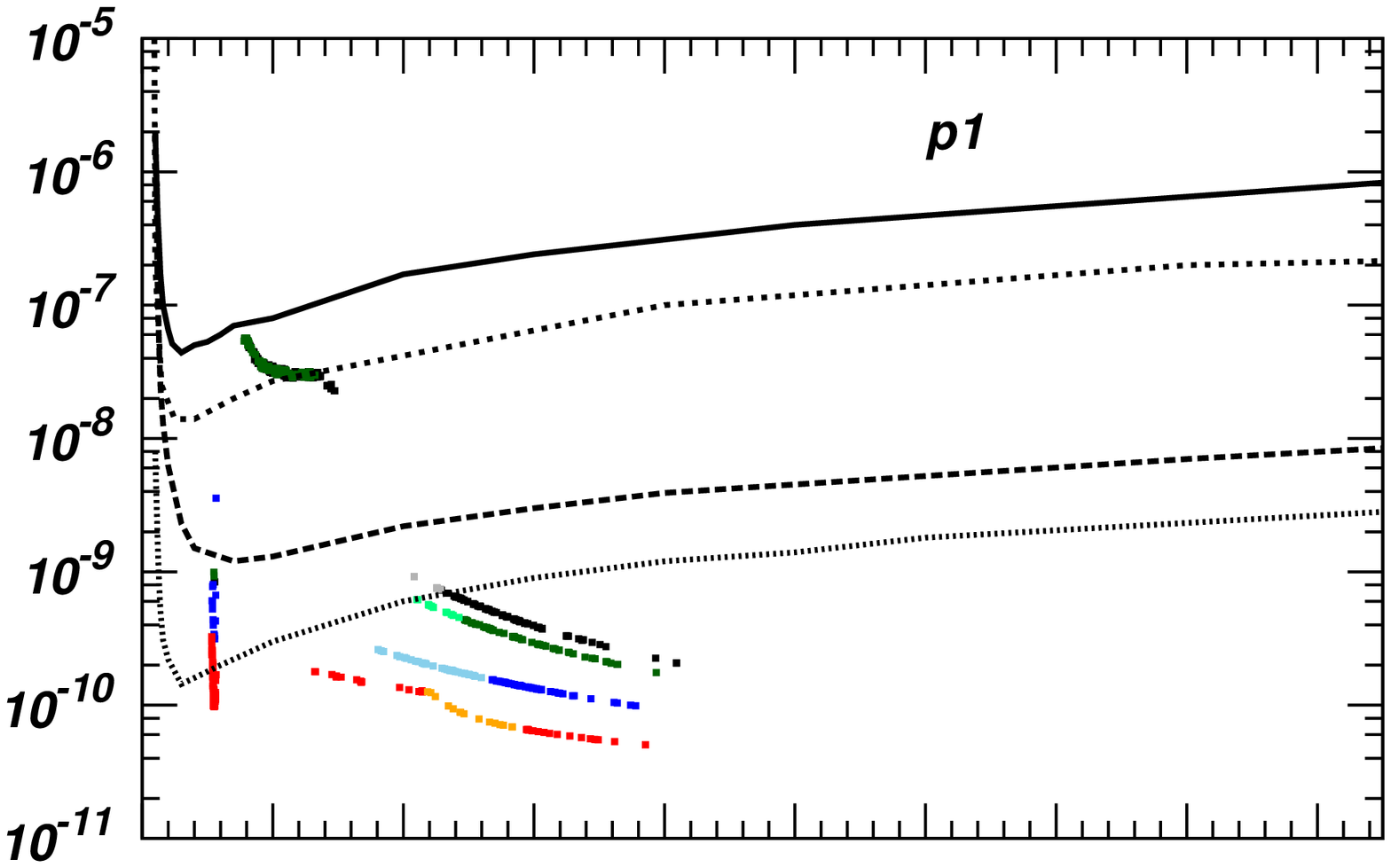}%
\includegraphics[height=0.45\textwidth,angle=270]{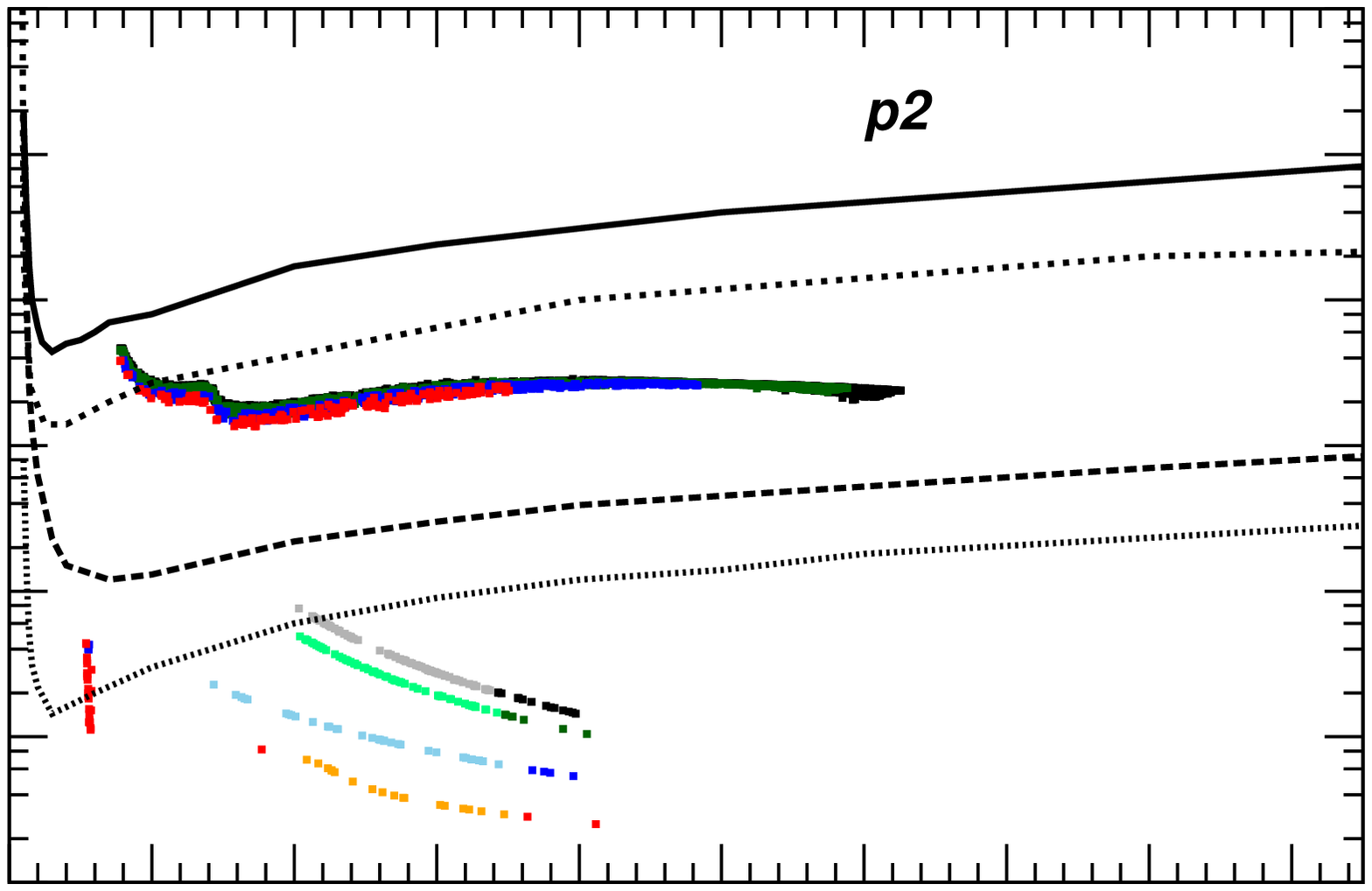}\vspace{-1.0cm}\\
\includegraphics[height=0.45\textwidth,angle=270]{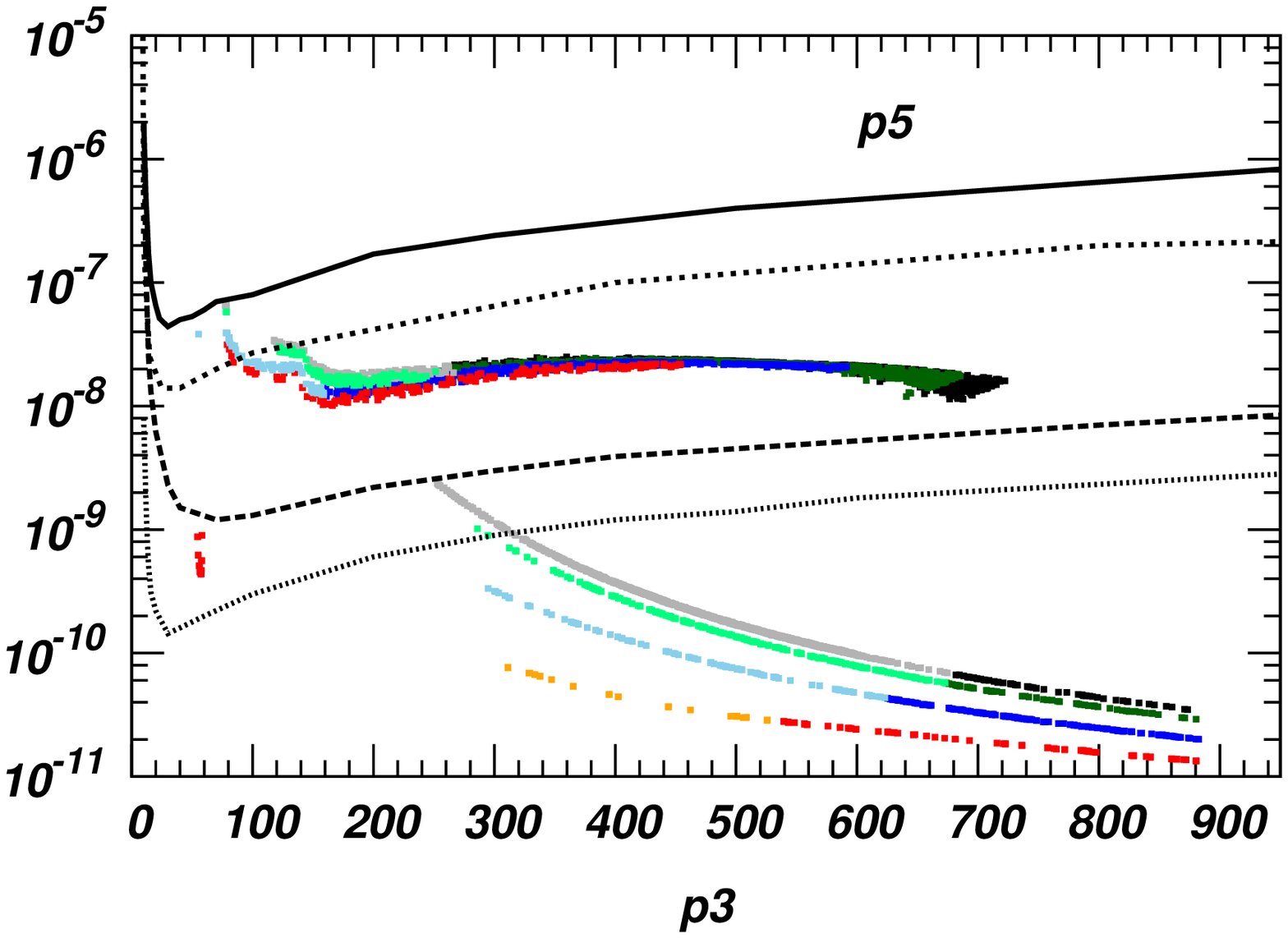}%
\includegraphics[height=0.45\textwidth,angle=270]{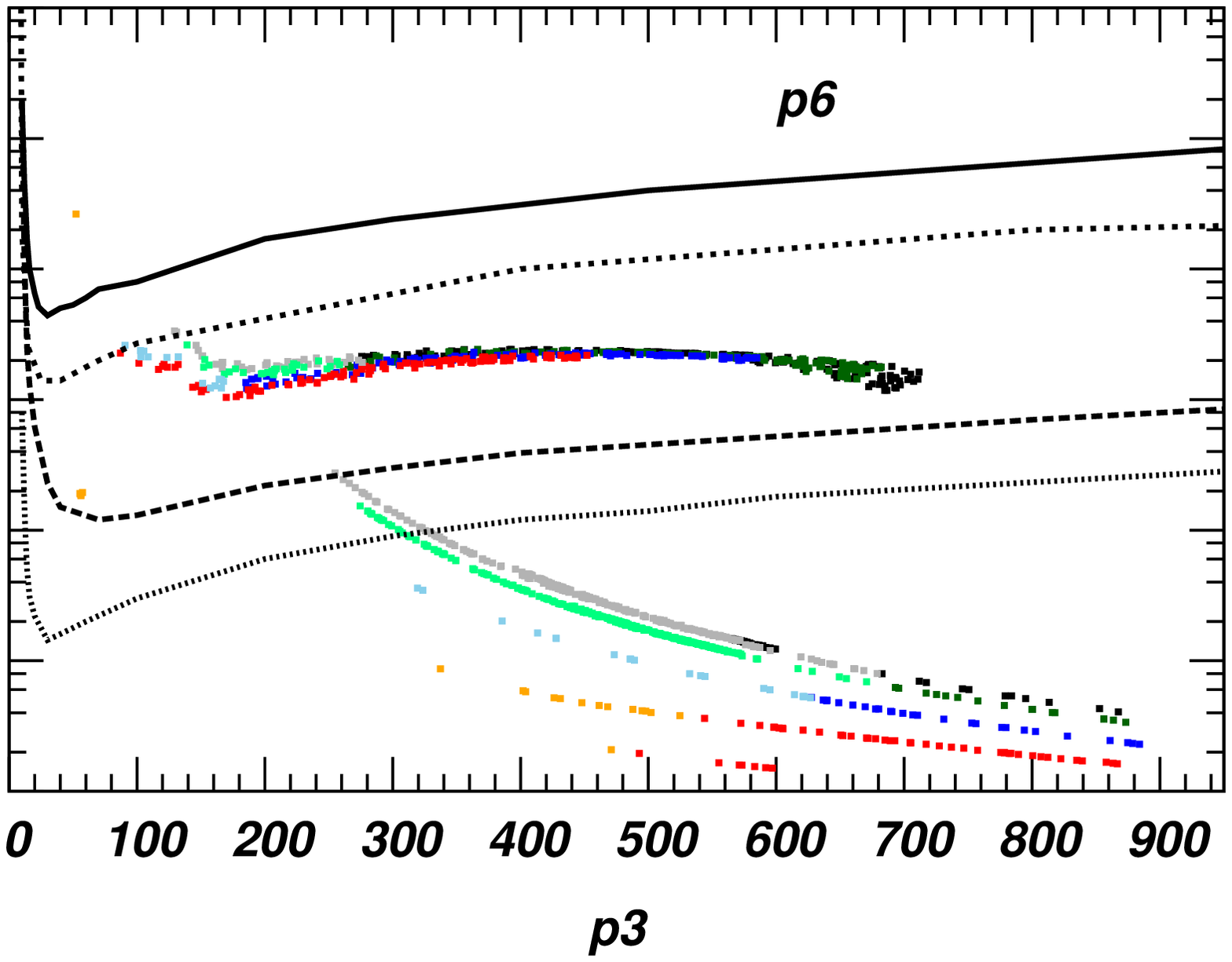}
\caption{Allowed region for spin-independent neutralino scattering cross section on proton plotted 
versus $m_{\tz_1}$. The color code is  the same as in Fig.~\ref{ma-mh-fix_tanb}.
We also show the reach and projected reach of XENON-10, CDMS-II, SuperCDMS and XENON-1~ton detectors by
solid, dashed, long-dashed and dotted lines, respectively.
\label{fig:si}}
}
%
In Fig.~\ref{fig:si}, we present the DD rates,  in
detector-invariant way, as spin-independent neutralino-proton elastic
scattering  cross section versus the neutralino mass. We used IsaReS
code~\cite{isares}, a part of IsaTools package, with pion-nucleon $\Sigma$
term assumed to be 45~MeV
~\footnote{ Recent experimental results suggest a somewhat different value of the 
$\Sigma$ term than the canonical value we assumed in this work. 
This affects the contribution from s-quark diagrams and can change our predictions for
$\sigma_{SI}(\tz_1 p)$ by about a factor of three~\cite{Ellis:2005mb}.}.
One can clearly read patterns coming from different dark matter motivated regions. 
The upper band corresponds to HB/FP region,
where $\tz_1$ has a substantial higgsino component thereby enhancing its
scattering cross section off the proton.  The lower pattern comes
from the stau coannihilation region  (merging with A-funnel for
$\tan\beta=53$). 
We see that the HB/FP and coannihilation regions are always separated in terms of
spin-independent cross section. 
One should also note the h-funnel region, which
is represented by narrow vertical band at $m_{\tz_1} \sim 55$~GeV. 
This region is well separated from HB/FP and \stac ~regions and
could be almost completely covered by XENON 1~ton detector.

The current best limit comes from the XENON-10
collaboration~\cite{xenon10}, which obtained an upper limit
$\sigma_{SI}(\tz_1 p)\lesssim 8\times 10^{-8}$~pb for $m_{\tz_1}\sim
100$~GeV.  We also show the projection for CDMS2~\cite{cdms2}, which is
expected to release results next year, and for its planned upgrade,
SuperCDMS~\cite{supercdms}. The reach of 
XENON-1~ton detector
as representative of many planned large noble
gas detectors~\cite{xenon1t,warp,zeplin4,Nikkel:2005qj}
aiming  for a sensitivity of $\sim 10^{-10}$~pb
is presented as well.
One can see that Stage-2 detectors, like
SuperCDMS, and eventually Stage-3 detectors, like  XENON-1~ton, will be able
to observe the signal from neutralino scattering off the nuclei in the entire
HB/FP region and measure the respective neutralino mass
exhibiting the prominent complementarity to LHC. Observation of the signal in
CDMS2 or a Stage-2 detector will clearly be an indication of the HB/FP region.
These results can be used in conjunction,  for example, with the 
LHC results for $m_{\tg}$ and $m_h$ to determine $A_0$ from Fig.~\ref{mg-mh-fixtanb}, 
since the ambiguity of the HB/FP and the coannihilation region curves could be resolved
using DD results. 
Also, positive result from DM searches will provide us with the neutralino mass,  which, in turn, can be
employed to extract $A_0$ from the correlations of Fig.~\ref{ma-mz1-fixtanb}.
If the sensitivity of DD experiments would be further increased to 
$\sim 10^{-10}-10^{-11}$~pb level, then individual curves from the 
stau-coannihilation pattern  in Fig.~\ref{fig:si} could be resolved 
and one could probe the $A_0$ parameter in this region.

\section{Conclusion}
\label{ch:conclusions}

We have performed an
updated scan of the CMSSM parameter space, taking into account the revised (lower) 
value of top quark mass,
updated SUSY constraints from collider and low energy physics
as well as crucial constraint on dark matter abundance from WMAP3. The scan was 
 performed in the CMSSM  parameter space for $\mu>0$, and for a plausible range of 
values for $m_0$, $m_{1/2}$ and $|A_0|$ 
and for $\tan\beta = $5, 10, 50 and 53.

We have demonstrated that  taking account of the SUSY constraints, especially the dark matter abundance, 
strong correlations  occur between the sparticle and Higgs masses. 
The correlations between the light CP-even Higgs boson mass and SUSY particles
could potentially allow determination of  the sparticle spectra with a few percent accuracy, 
assuming a theoretical control of $m_h$ at the percent level.
All correlations we have found are represented by separated narrow bands, 
two-fold solutions which exhibit  focus point and 
h-funnel regions, or co-annihilation and A-funnel regions respectively.

The correlations found among the sparticle and Higgs masses 
would also allow one to delineate the value of the trilinear coupling 
which can potentially be large. 
The large value of the trilinear coupling $A_0$
is especially motivated by the present light Higgs mass constraints
and the latest top-quark mass measurements. Since the top quark 
world-averaged mass went down to 170.9~GeV,
the additional contributions from $A_t$ and $A_b$ 
are even more important to increase $m_h$  
and thereby overcome the LEP2 limit.  
We have found that even for a parameter space  
with $m_0$ and $|A_0|$ as large as 5~TeV and 2~TeV  respectively at $M_{GUT}$, 
the lightest Higgs boson mass 
is limited by 122-123~GeV from above.

We have found that some correlations, like between 
$(m_A,m_h)$, $(m_A,m_{\tilde{g}})$ and $(m_A,m_{\tz_1})$ could 
allow one to reveal the value of $A_0$ 
and help to distinguish the DM motivated band of the parameter space mentioned above.
Moreover, we have demonstrated  striking complementarity between 
the LHC and direct dark matter detection experiments
in resolving the SUSY mass spectrum and determination of SUSY
parameters.  
Stage~2 experiments like SuperCDMS 
will be able to cover the entire HB/FP region 
and resolve the ambiguity of HB/FP and non-HB/FP
bands for several important correlations.

The correlations we have presented in this paper can be a useful
input for SUSY global analysis fit as well as for experimental 
delineation of the SUSY parameter space.
We would also like to emphasize that including the experimental
input both from collider physics and from dark matter detection experiments
would allow one to significantly improve the understanding of 
the SUSY spectrum and the underlying
parameter space.


\section*{Acknowledgments}

We thank H.~Baer, X.~Tata and A.~Djouadi for useful discussions.
This work is supported in part by grants from the US Department of Energy, 
S.D., I.G., and Q.S. (Grant \# DE-FG02-91ER40626) and A.M. (Grant \# DE-FG02-04ER41308). 
S.D. was also supported by the University of Delaware competitive fellowship.
A.B.  acknowledges partial support from  PPARC Rolling Grant PPA/G/S/2003/00096.


\bibliographystyle{JHEP}
\bibliography{SALADIN}

\end{document}